\documentclass[11pt,a4paper]{article}

\usepackage{booktabs}
\usepackage{jheppub}
\usepackage{capt-of}
\usepackage{comment}
\usepackage{siunitx}
\usepackage{amssymb}
\usepackage{enumitem}   
\usepackage[english]{babel}
\usepackage{url}
\usepackage{mathtools}
\usepackage{soul}
\usepackage{bbold}
\usepackage{slashed}
\usepackage{multirow}
\usepackage{slashed}
\usepackage{multirow}

\usepackage[usenames,dvipsnames,svgnames]{xcolor}
\usepackage{appendix}
\newcommand{\s}{s_{\rm w}}
\newcommand{\be}{\begin{equation}}
\newcommand{\ee}{\end{equation}}

\newcommand{\tev}{\,\textrm{TeV}}
\newcommand{\gev}{\,\textrm{GeV}}

\newcommand\aNLO{{\sc\small MadGraph5\_aMC@NLO}}
\newcommand\aNLOs{{\sc\small MG5\_aMC}}

\newcommand\UFO{{\sc\small UFO}}

\newcommand\TO{\rightarrow}

\begin{document}



\title{Top-philic ALP phenomenology at the LHC:\\ the elusive mass-window}

\author[a,b]{Simone Blasi,}
\author[c,d,e]{Fabio Maltoni,}
\author[b]{Alberto Mariotti,}
\author[f,g]{Ken Mimasu,}
\author[e]{Davide Pagani,}
\author[c]{Simone Tentori}

\affiliation[a]{Deutsches Elektronen-Synchrotron DESY, Notkestr.~85, 22607 Hamburg, Germany}
\affiliation[b]{Theoretische Natuurkunde and IIHE/ELEM, Vrije Universiteit Brussel, \& The International
Solvay Institutes, Pleinlaan 2, B-1050 Brussels, Belgium}
\affiliation[c]{Centre for Cosmology, Particle Physics and Phenomenology (CP3),
Universit\'{e} Catholique de Louvain, B-1348 Louvain-la-Neuve, Belgium}
\affiliation[d]{Dipartimento di Fisica e Astronomia, Universit\`{a} di Bologna, Via Irnerio 46, 40126 Bologna, Italy}
\affiliation[e]{INFN, Sezione di Bologna, Via Irnerio 46, 40126 Bologna, Italy}
\affiliation[f]{School of Physics and Astronomy, University of Southampton, \\Highfield, Southampton S017 1BJ, United Kingdom}
\affiliation[g]{Theoretical Particle Physics and Cosmology Group, Department of Physics,\\
King's~College~London, London WC2R 2LS, United Kingdom}

\emailAdd{simone.blasi@desy.de}
\emailAdd{fabio.maltoni@uclouvain.be}
\emailAdd{alberto.mariotti@vub.be}
\emailAdd{ken.mimasu@soton.ac.uk}
\emailAdd{davide.pagani@bo.infn.it}
\emailAdd{simone.tentori@uclouvain.be}


\abstract{
We study the LHC phenomenology of an Axion Like Particle (ALP) that couples only derivatively with the top quark at tree level.
We inspect the radiatively induced couplings with the SM fermions and (gauge) bosons and the associated production and decay mechanisms of the ALP. We focus on the most challenging mass window that remains open for a top-philic ALP, {\it i.e.}, the range between tens and hundreds of GeV. 
Not only ALP production processes but also virtual  ALP contributions to final states with top quarks are considered in detail. We show how searches through resonant production, such as ALP production in association with a $t\bar t$ pair, are complementary to  precision measurements of $t \bar t$  and $t\bar t t \bar t$ final states, the latter being competitive or even more powerful for a top-philic ALP in this mass window. Finally, we  explore the scenario where the top-philic ALP acts as a mediator to a dark-matter sector, resulting in missing energy signatures at the LHC. 
We find that the LHC constraints from $t \bar t$,  $t\bar t t \bar t$ and ALP + jet production, together with those from $t \bar t$ + ALP production, can already exclude a large fraction of the parameter space leading to the correct relic abundance.
}

\preprint{
\begin{flushright}
IRMP-CP3-23-69\\
DESY-23-198\\
KCL-PH-TH/2023-68\\
\end{flushright}
}

\maketitle


\section{Introduction}
\label{sec:introduction}
Axion-like Particles (ALPs) represent a compelling benchmark scenario for physics beyond the Standard Model (BSM) in which a new, singlet pseudoscalar state ($a$) is introduced in the low-energy spectrum of our theory  of fundamental interactions. In particular,
the ALP interactions with the Standard Model (SM) particles can be described within an effective field theory (EFT) that begins at canonical dimension-five, and respects an approximate shift symmetry, $a\to a+\textit{c}$  where $c$ is a constant, reflecting the fact that $a$ is the pseudo-Nambu Goldstone boson (pNGB) of a spontaneously broken global symmetry.
As such, the ALP is naturally light compared to other BSM states, and can in fact be considered as the sole additional, dynamical degree of freedom at low energy. 

Beyond the original motivation of the axion solution to the strong CP problem~\cite{Peccei:1977hh,Peccei:1977ur,Wilczek:1977pj,PhysRevLett.48.11}, ALPs can 
arise in many BSM scenarios such as string compactifications~\cite{Cicoli:2013ana,Ringwald:2012cu}, supersymmetric theories \,\cite{Bellazzini:2017neg}, neutrino-mass-generation mechanisms~\cite{Alexandre:2020tba,Mavromatos:2020hfy}, as well as composite-Higgs realisations featuring additional pNGBs\,\cite{Gripaios:2009pe}. In addition, ALPs can be viable dark matter candidates as well as mediators to the dark-matter sector, see {\it e.g.}~Ref.~\cite{Preskill:1982cy,ABBOTT1983133,Dine:1982ah,Dror:2023fyd,Fitzpatrick:2023xks}.

The ALP mass ($m_a$) 
and its decay constant ($f_a$) are  independent, free parameters  as are the interaction strengths of the ALP with the various SM particles. 
In this paper we will focus on a specific type of ALP that is \emph{top-philic}.
Our set-up is defined by assuming that the ALP couples only to the top quark at the  tree level
in a derivative form that is consistent with the ALP shift-symmetry:
\begin{align}
   \mathcal{L}_{\text{int.}}= c_t \frac{\partial^{\mu} a}{f_a} \bar t_R \gamma_{\mu} t_R.
\end{align}
We will present an in-depth study of the phenomenology of the coupling $c_t$ and how it can be probed in a particular ALP-mass range where existing bounds are found to be relatively weak.

A top-philic ALP,  at  tree level, has no
anomaly-induced couplings with the gauge bosons of the SM. It therefore differs significantly from scenarios where the axion solves the strong CP problem via the anomalous coupling to gluons. 
Nevertheless, the model that we consider can, for instance, emerge when the ALP couples to the SM only via fermion mixing. A possible realisation is indeed the case of an ALP that is coupled to heavy (vector-like) top partners, such as those emerging from the strong sector of a composite Higgs model, as discussed, {\it e.g.}, in Ref.~\cite{Esser:2023fdo}.

Without committing to a specific UV completion, we will study the phenomenology of the top-philic ALP at the LHC,
focusing on the model-independent features that primarily depend on the low-energy physics described by the EFT (see \cite{Ebadi:2019gij,Carmona:2021seb,Carmona:2022jid,Esser:2023fdo,Rygaard:2023dlx,Bruggisser:2023npd} for recent related studies).
The first observation is that the top-philic nature of the ALP is not preserved beyond  tree level. In fact $c_t$ induces, through radiative corrections involving SM fields, couplings to all SM fermions and gauge bosons.
As we will explore and discuss in detail, this has important implications for the possible discovery channels of this light BSM state.
For example, for ALP masses below the bottom threshold, the constraints on the model are dominated by flavour observables (see, \emph{e.g.}, Ref.~\cite{Bauer:2021mvw}) from the induced fermion couplings. At even lighter, keV-scale masses, the induced coupling to electrons implies extremely strong bounds from stellar cooling via ALP emission, beyond any conceivable collider sensitivity~\cite{Chala:2020wvs}. Above the top-quark pair threshold, a top-philic ALP can be probed efficiently by resonant $t\bar{t}$ searches. However, by this point the ALP can no longer be considered light and shift-symmetry-breaking effects start to become significant.

For these reasons, in this paper we will focus on a specific mass range, between 10 GeV and 200 GeV, where the top-philic ALP proves to be rather elusive, as we will discuss. The upper limit of 200 GeV is chosen as a maximum value for which $m^2_a \ll 4 m^2_t$ approximately holds and the salient features of the top-philic ALP couplings are retained.
As we will show, resonance searches are quite ineffective in probing the decay constant of the top-philic ALP in this mass window, given that the gluon-fusion production cross section is naturally suppressed,
and its dominant decay channel is into bottom-quark pairs.
This motivates the investigation of alternative strategies to close this gap in the top-philic-ALP parameter space.

To this end, we will explore in detail the various processes through which a top-philic ALP in this mass window can be directly produced  at the LHC, or contribute non-resonantly to top-enriched final states that do not involve the ALP itself. 
In particular, we determine the constraints on this model that arise from precision measurements of 
$t\bar{t} b\bar{b}$, $t\bar{t}t\bar{t}$ and  $t\bar{t}$ production cross-sections.
ALPs contribute non-resonantly to the latter two processes, such that we can combine multiple inclusive and differential LHC measurements to probe the top-philic coupling. 
For $t\bar{t}b\bar{b}$, ALPs contribute resonantly via the $pp\to t\bar{t}a$ process and subsequent $a\to b\bar{b}$ decay but, since no such direct searches exist, we equally make use of total cross section measurements at the LHC to derive our constraints.
We will show that these datasets can imply stronger bounds on the ALP decay constant than tailored, resonant searches that probe the on-shell, direct production via gluon fusion ($pp\to a$), production in association with a pair of tops ($pp\to t\bar{t} a$) or an additional jet ($pp\to a+j$). See Refs.~\cite{Gavela:2019cmq,Carra:2021ycg,Bonilla:2022pxu} for other works considering related, non-resonant probes of ALPs at colliders.
We have produced a {\UFO}~\cite{Degrande:2011ua, Darme:2023jdn} implementation of the top-philic model capable of generating predictions in {\aNLO}  for the 
aforementioned processes, which in some cases require the evaluation of one-loop amplitudes. 

We will also explore the possibility that the top-philic ALP acts as a mediator to a dark matter (DM) sector.
We identify the viable regions in the parameter space where the correct relic abundance can be obtained via freeze-out, which typically requires a large ALP--top coupling and a large ALP--DM coupling.
Then, we study the constraining power of the LHC on this scenario in the regime where the top-philic ALP decays predominantly invisibly (see Refs.~\cite{Brivio:2017ije,Esser:2023fdo} for previous studies).
In addition to the ALP-mediated processes involving the top quarks, the ALP can be directly produced (in isolation or in associated with a top-quark pair) and decay into DM, giving rise to missing energy signatures. We will re-interpret existing LHC analyses, and also employ our {\aNLO} implementation to study the reach of mono-jet searches on the ALP decay constant. We find that, although an invisibly decaying top-philic ALP makes for quite distinctive channels at the LHC, the sensitivity of precision SM cross section measurements involving top quarks in the final state remains comparable to that of missing energy based searches.
Moreover, we will show that the LHC limits strongly constrain the regions leading to a correct dark matter relic density, concluding that a top-philic ALP as portal to DM is viable only in cases where the dark matter annihilation is resonantly enhanced.

The paper is structured as follows. In Sec.~\ref{sec:topphilicALP} we briefly recap the main features of the general ALP effective theory before specialising to the top-philic case of interest. We discuss the ALP couplings generated by $c_t$ beyond tree level, namely the induced couplings to fermions and to SM gauge bosons. We discuss the impact of shift symmetry breaking effects on the induced couplings and motivate the importance of two-loop effects for the gauge boson interactions, studying the associated form factors in phenomenologically relevant kinematical limits. We also discuss the conversion to an equivalent basis in which derivative ALP-fermion interactions are traded for those without derivatives. This elucidates the relation between our top-philic ALP and a generic top-philic pseudoscalar, explaining why the two models yield very different predictions for certain LHC processes. In Sec.~\ref{sec:basics} we 
explore the LHC phenomenology of the top-philic ALP in detail, computing its decay branching fractions and production cross sections for the aforementioned processes. We also derive bounds on $f_a/c_t$ based on the interpretation of resonance searches at the LHC, quantifying the gap in the parameter space in our mass window. Sec.~\ref{sec:newprobes} considers new, alternative probes of the top-philic ALP using SM cross section measurements in top-quark only final states, which we find to yield the strongest limits to date in this parameter region. Finally, Sec.~\ref{sec:DM} presents our study of the invisible ALP as a portal to a DM sector before summarising and concluding in Sec.~\ref{sec:conclusions}.

\section{The Top-philic ALP}
\label{sec:topphilicALP}

\subsection{The ALP effective Lagrangian}
\label{subsec:ALPgeneral}

In this section we summarise the general structure of the effective Lagrangian for an ALP, denoted as $a$, at dimension-five, and introduce the notation and conventions used in this paper. The specific features of a top-philic ALP are discussed in Sec.~\ref{sec:top-philic}, while in this section the ALP is not yet assumed to be top-philic.

The effective Lagrangian reads \cite{Bauer:2020jbp}:
\begin{equation}
\begin{split}
\label{eq:full_Lag_a}
\mathcal{L}_a =
\frac{1}{2}(\partial_{\mu} a)^2 -\frac{1}{2} m_a^2 a^2 & +
\frac{\partial^{\mu}a}{f_a} \sum_f \bar \psi_f \mathbf{c}_f \gamma_{\mu} \psi_f + c_H \frac{\partial^{\mu} a}{f_a} \phi^{\dagger} i D_{\mu}\phi\\
&
+c_{GG} \frac{\alpha_S}{4 \pi} \frac{a}{f_a} G \tilde G + 
c_{WW} \frac{\alpha_2}{4 \pi} \frac{a}{f_a} W \tilde W + 
c_{BB} \frac{\alpha_1}{4 \pi} \frac{a}{f_a} B \tilde B\,, 
\end{split}
\end{equation}
where $f_a$ is the ALP decay constant, $m_{a}$ is the ALP mass and the $c$ parameters are the Wilson coefficients associated to the effective operators, with $\mathbf{c}_f$ being a $3\times3$ matrix in flavor space. $\phi$ is the $SU(2)$ Higgs-doublet  field and $G, W$ and $B$ are the gauge fields associated to the $SU(3)$, $SU(2)$ and $U(1)$ symmetries, with coupling strengths parameterised by $\alpha_{S}$, $\alpha_{2}$ and $\alpha_{1}$, respectively, and $\alpha_{i}\equiv g_{i}^{2}/(4\pi)$. 
The sum over $f$ runs over all of the chiral SM fermions, namely $f= e_R,u_R,d_R,Q_L,L_L$. In the following, we shall consider only flavour-diagonal ALP couplings to the SM fermions, and neglect all effects related to a non-diagonal CKM matrix.

At the classical level, the SM Lagrangian is invariant under hypercharge, baryon- and lepton-number transformations. These symmetries can be exploited in order to identify the combinations of ALP couplings that are physically meaningful, thus removing redundancies in the set of parameters presented in Eq.~\eqref{eq:full_Lag_a}, see {\it e.g.}~Refs.~\cite{Bauer:2020jbp,Bauer:2021mvw,Bonilla:2021ufe}.
In particular, we can apply the following ALP-dependent transformations to the SM fields:
\begin{equation}
\label{eq:rephasing}
    \psi_f \TO {\rm exp}\left( i c \frac{a(x)}{f_a} Q_f\right) \psi_f, \quad \phi \TO {\rm exp}\left( i c \frac{a(x)}{f_a} Q_\phi\right) \phi.
\end{equation} 
By imposing the charges $Q_{f,\phi}$   to match the symmetries of the SM mentioned above, these transformations redefine the couplings in Eq.~\eqref{eq:full_Lag_a} in terms of a new set of parameters $c$, while leaving the SM part of the action invariant. Thus, they can be used to remove some of the ALP interactions in Eq.~\eqref{eq:full_Lag_a},  reducing the number of free parameters that are needed to fully characterise the ALP theory. In other words, only some combinations of the ALP couplings will have physical meaning, namely those that are invariant under the possible rephasings in Eq.~\eqref{eq:rephasing}. Note that, when the global symmetry corresponding to the transformation in \eqref{eq:rephasing} is anomalous, the Lagrangian acquires extra terms from the non-invariance of the path integral measure, that modify the axion-vector boson couplings in Eq.~\eqref{eq:full_Lag_a}.

Assuming a diagonal CKM matrix, the classical symmetries of the SM at our disposal include not only the hypercharge and the generation-specific lepton numbers, but also the generation-specific baryon numbers.
If we further assume that the ALP couplings to fermions are flavour-diagonal in the basis in which the SM up-quark Yukawa matrix is diagonal, the $\mathcal{O}(1/f_a)$ EFT in Eq.~\eqref{eq:full_Lag_a} includes 20 free parameters. By using the 7 aforementioned  symmetries only 13 combinations thereof will be physically relevant.

It is customary to use the freedom given by the hypercharge symmetry to eliminate the Higgs operator, $c_H=0$. Then, together with the ALP mass $m_a$ and $c_{GG}$ (which is not modified by baryon and lepton number transformations), one can take the following set of
reparameterisation invariant quantities: \footnote{These differ from the quantities defined in Ref.\,\cite{Bauer:2020jbp} as here we only consider transformations that are classical symmetries of the SM, so that pseudo-Yukawa couplings are not generated and the operators basis in Eq.~\eqref{eq:full_Lag_a} is mapped onto itself.}
\be
\label{eq:EFTdata}
c_Q^{ii} - c_u^{ii}, \quad c_Q^{ii} - c_d^{ii}, \quad c_L^{ii} - c_e^{ii}, \quad c_{WW} + c_{BB}, \quad c_{WW} - \frac{1}{2}\, \text{Tr}\, \left(3 c_Q + c_L\right),
\ee
where $i=1,2,3$ runs over the flavour generations. Notice that $c_{WW} + c_{BB}$ is related to the ALP coupling to photons, see {\it e.g.}~Refs.~\cite{Bauer:2021mvw,Craig:2018kne} (and Eq.~\eqref{eq:aFF} later on).

\subsection{Top-philic scenario and induced ALP couplings}
\label{sec:top-philic}
In this section we restrict our model to the top-philic case, quantifying the relevant interactions of the ALP.
As we will discuss in detail in the following, while at tree level one can impose an ALP that only interacts with the top quark, loop corrections will generate all of the other interactions in Eq.~\eqref{eq:full_Lag_a} (see also Refs.~\cite{Bonilla:2021ufe,Craig:2018kne,Bauer:2021mvw}). In particular, in the next section we discuss the ALP couplings with fermions, which are logarithmically enhanced and proportional to $m_t^{2}$. The ALP interactions with gluons and photons, which with on-shell legs are suppressed by ($m_{a}/m_{t})^{2}$ when $m_{a}\ll m_{t}$, are discussed in Sec.~\ref{subsec:gluphot}. We discuss the case of ALP interactions with $W$ and $Z$ bosons, which are instead not suppressed when $m_{a}\ll m_{t}$ in Sec.~\ref{sec:weak_gauge}. 

Let us now define our model of a top-philic ALP, whose couplings are generated at some UV scale $\Lambda\sim f_a$. Imposing tree-level interactions of the ALP with top quarks only, all reparameterisation invariant quantities in Eq.~\eqref{eq:EFTdata}  vanish with the exception of 
\be
c_t \equiv c_u^{33} - c_Q^{33} \, . \label{eq:ctdef}
\ee 
The simplest way of obtaining this condition is by setting all coefficients apart from $c_u^{33}$ to zero, obtaining the Lagrangian
\be
\label{eq:top-philic}
\mathcal{L}_{\text{top-philic}} = \frac{1}{2} (\partial_\mu a)^2 -\frac{1}{2} m_a^2 a^2 + c_t \frac{\partial^{\mu} a}{f_a} \bar t_R \gamma_{\mu} t_R\, .
\ee
where with this convention $c_t = c_u^{33}$.\footnote{Note that while Eq.~\eqref{eq:ctdef} is always true, this condition is valid only at  tree level. As soon as loop corrections induce further interactions, it is not valid anymore.}

Top-philic EFT scenarios can emerge, for instance, when the top quark mixes with heavy partners charged under the $U(1)$ Peccei-Quinn-like (PQ) symmetry in the underlying UV model, as in the case we discuss in Appendix~\ref{app:UVmodel} or also in models inspired by Higgs compositeness~\cite{Esser:2023fdo}.

We notice that the Lagrangian in Eq.~\eqref{eq:top-philic} is actually consistent with approximate minimal-flavour-violation\,\cite{DAmbrosio:2002vsn}, which imposes the following structure on the ALP couplings to right-handed quarks, $c_{u}$, in the basis where the SM Yukawa matrices are diagonal\,\cite{Bauer:2020jbp}:
\be
\label{eq:MFV}
c_{u} = c_0^u \mathbb{1} + \epsilon c_1^u (Y_u^{\rm diag})^2 + \mathcal{O}(\epsilon^2)\,,
\ee
where $Y_u^{\rm diag} =(y_u, y_c, y_t)$. As we can see, our top-philic model is a particular case of Eq.~\eqref{eq:MFV} when $c_0^u=0$ and the light quark Yukawas are neglected, $y_{u,c}=0$.

An important assumption behind Eq.~\eqref{eq:top-philic} is not only that the SM matter fields are not charged under the $U(1)$ PQ symmetry, but also that this symmetry has no mixed anomaly with the SM gauge group. In other words, in terms of EFT data at  tree level, $c_{WW}=c_{BB}=c_{GG}=0$.  As we shall see, this implies a qualitatively different behaviour compared to models featuring, in particular, contact interactions between the ALP and the gluons ($c_{GG} \neq 0$). On the other hand, under our assumptions, the shift symmetry  enjoyed by the ALP, $a \TO a + \rm c$ with c a constant, is only softly broken in the Lagrangian Eq.~\eqref{eq:top-philic} by the ALP mass, $m_a^2$, which we treat as an additional free parameter.

Our main goal is to study the phenomenology of the top-philic ALP described by the Lagrangian in Eq.~\eqref{eq:top-philic} at the LHC. However, as already mentioned before, while most of the EFT data entering Eq.~\eqref{eq:top-philic} are equal to zero at  tree level, they are not when loop corrections are taken into account. Therefore, we first analyse the leading contributions induced by said loop corrections, beginning with the ALP-fermion couplings.

\subsubsection{Couplings to fermions}
\label{subsec:fermions}
Loop corrections from $c_t$ induce non vanishing contributions to the axial combinations of the ALP couplings to fermions, namely nine of the thirteen reparametisation invariants, which consistently with Eq.~\eqref{eq:ctdef} we define as
\be
c_f \equiv c_{f_R}^{ii}-c_{f_L}^{ii}\, . \label{eq:cfdef}
\ee 
On the one hand, these interactions feature a one-loop suppression $\sim y_t^2/16 \pi^2$. On the other, they are logarithmically sensitive to the cut-off scale of the EFT, which shall be indicated generically by $\Lambda$, encoding the renormalisation group running from $\Lambda$ to the electroweak (EW) scale, in particular the top mass $m_t$. $\Lambda$ is associated with the mass scale of the UV states, {\it i.e.}, $\Lambda \simeq g_\ast f_a$, with $g_\ast$ the typical interaction strength in the UV model, and we therefore take $\Lambda\lesssim 4\pi f_a$.
Taking into account the aforementioned logarithmically enhanced contributions, one finds \,\cite{Bonilla:2021ufe,Bauer:2020jbp}:\footnote{\label{footnote:1TeV}Whenever $\Lambda \lesssim 1 \,{\rm TeV}$ additional corrections $\sim m_t/\Lambda$ may become numerically important beyond the leading logarithm considered here.}
\be
\label{eq:cthird}
c_t(m_t) = c_t(\Lambda) \left( 1 - 9 \frac{y_t^2}{16\pi^2} \log \frac{\Lambda}{m_t} \right), \quad
c_b(m_t) \equiv c_b^{33} - c_Q^{33} = 5 c_t(\Lambda) \frac{y_t^2}{16 \pi^2} \log \frac{\Lambda}{m_t}\,,
\ee
while for the other fermions $f=u,d,c,s,e,\mu,\tau$ one has
\be
\label{eq:clight}
c_f(m_t) \equiv c_{f_R}^{ii} - c_{f_L}^{ii} = - 12\, c_t(\Lambda) \frac{y_t^2}{16 \pi^2} T_3^f \log \frac{\Lambda}{m_t}\,,
\ee
where $T_3^f$ represents the isospin component of $f_L$.

As we can see, starting from the ALP coupling to the top quark only, the model generates similar derivative couplings to all of the light fermions with a one-loop suppression. We anticipate that these derivative couplings may actually be traded for pseudo-Yukawa ALP interactions with the SM fermions plus anomalous terms, as discussed in Sec.\,\ref{sec:nonder}.

\subsubsection{Coupling to gluons and photons}
\label{subsec:gluphot}

Let us now  discuss the ALP coupling to gauge bosons, starting from the effective coupling to gluons that is generated at one loop via the tree-level ALP-top coupling in Eq.~\eqref{eq:top-philic}.
From the direct computation of the top-quark loop one obtains the following amplitude, $\mathcal{A}^{\mu \nu}$, for a process connecting the ALP to two gluons, for generic (off-shell) momenta:
\be
\label{eq:aggampl}
i \mathcal{A}^{\mu \nu} \left( a(k) \TO g(p) g(q) \right) = i \frac{\alpha_{S}}{\pi} \frac{c_t}{ f_a} \delta^{ab} p_\alpha q_\beta \epsilon^{\mu \nu \alpha \beta} 
\left[ 1 + 2m_t^2 C_0(p,q;m_t^2)\right],
\ee
where $k = p + q$. The $C_0$ function in Eq.~\eqref{eq:aggampl} is the standard scalar one-loop three-point integral \cite{tHooft:1978jhc, Passarino:1978jh} with all internal masses set equal to $m_{t}$  (see also Eq.~\eqref{eq:C1}). 

Despite being loop-induced, the amplitude in Eq.~\eqref{eq:aggampl} controls several ALP production and decay processes at the LHC, such as the decay of the ALP into gluons, direct ALP production via gluon fusion, and the production of an ALP in association with a jet. We will consider all of the aforementioned processes in this work. 

Let us discuss some relevant limits of this amplitude, starting from the case in which all of the energy scales involved in the $a \TO gg $ process are small compared to the top-quark mass. 
In this limit we can formally expand the $C_0$ function in the $m_t \TO \infty$ limit  
 \be
 C_0(p,q;m_t^2) \TO -1/2m_t^2\qquad {\rm for}~~ m_t \TO \infty \, .
 \ee
 From above,  we see that in Eq.~\eqref{eq:aggampl} an exact cancellation takes place at the leading order in the $1/m_t^2$ expansion. This is expected since the initial top-philic Lagrangian in Eq.~\eqref{eq:top-philic} is invariant under the ALP-shift symmetry $a \TO a + \rm c$, with the exception of the  $m_a^2$ term, which however does not enter in the top-quark loop under consideration. Indeed, if at LO in the  $1/m_t^2$ expansion the amplitude in Eq.~\eqref{eq:aggampl} were non-vanishing in the $m_t \TO \infty$ limit, it would indicate the presence of a loop-induced shift-symmetry breaking operator of the form $(a/f_a) G \tilde G$.
Instead, this operator is not generated at low energy scales, denoted here by $Q^2$, and the first non-zero contribution to this amplitude has to be necessarily suppressed by powers of $Q^2/m_t^2$. In other words, the $a \TO g g$ amplitude shows a decoupling behavior in the limit of a heavy top quark.

Let us consider for concreteness the amplitude entering the resonant production of the ALP via gluon fusion, or equivalently the ALP decay into gluons. Setting $k^2=m_a^2$, so that also $Q^2 = m_a^2$, and $p^2=q^2=0$, the amplitude in Eq.~\eqref{eq:aggampl} may be interpreted as inducing the $(a/f_a) G \tilde G$ operator in Eq.~\eqref{eq:full_Lag_a}, but with an effective coupling  
\be
\label{eq:cggtop}
c_{GG}^{{\rm eff}, \,t}(p^2=q^2=0) = \frac{1}{2} B_1\left( \frac{4 m_t^2}{m_a^2}\right) c_t = -\frac{m_a^2}{24 m_t^2} c_t + \mathcal{O}\left(\frac{m_a^4}{ m_t^4}\right),
\ee
where the $m_a^2/m_t^2$ suppression is a remnant of the original ALP shift symmetry of our model. 
In deriving Eq.~\eqref{eq:cggtop} we have implicitly used  the relation
\be
1 + 2 m_t^2 C_0(p^2=0, q^2=0; m_t^2) = B_1(4m_t^2/k^2)\,,
\ee
for the on-shell configurations where, following the notation in Ref.~\cite{Bauer:2017ris},
\be
\label{eq:B1explicit}
B_1(\tau) \equiv 1 - \tau f(\tau)^2, \quad f(\tau) \equiv \begin{cases}
    {\rm arcsin}(1/\sqrt{\tau}) \quad \quad \tau \ge 1\,, \\
    \frac{\pi}{2} + \frac{i}{2}{\log} \frac{1 + \sqrt{1-\tau}}{1 - \sqrt{1-\tau}} \quad \,\tau < 1\,.
\end{cases}
\ee

\begin{figure}
\begin{minipage}[u]{0.48\linewidth}
    
    \centering
    \includegraphics[width=\linewidth]{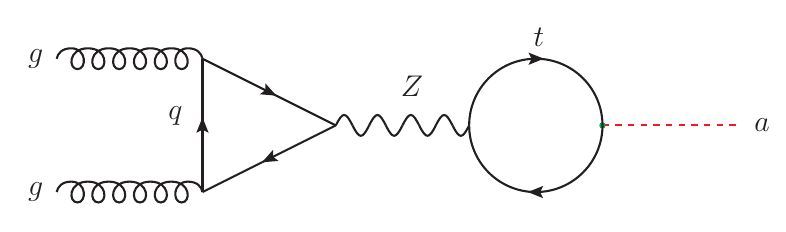}
    \caption{The one-particle-reducible diagram contributing to the $gg\TO a$ amplitude at two-loop level.}
    \label{fig:Diagramsep}
    \end{minipage}
\hfill
\begin{minipage}[u]{0.48\linewidth}
        \centering
\includegraphics[width=0.95\linewidth]{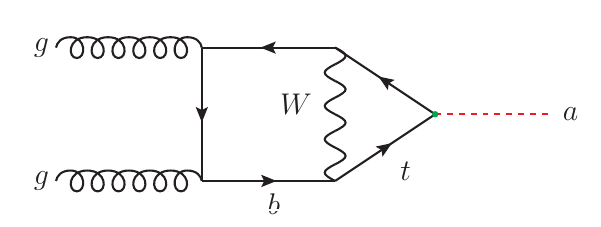}
    \caption{Example one-particle-irreducible diagram contributing to the $gg\TO a$ amplitude at two-loop level.}
    \label{fig:Diagramuniq}
\end{minipage}
\end{figure}

The large suppression in Eq.~\eqref{eq:cggtop} for relatively light ALPs raises the question of whether higher-loop contributions involving lighter fermions can actually become the dominant ones for $m_a \ll m_t$.
While a full two-loop computation of $\mathcal{A}^{\mu \nu}(a\TO gg)$ is beyond the scope of this work, we can estimate the size of this contribution. To this end, we consider the one-loop diagrams involving light fermions that couple to the ALP with the effective (one-loop) couplings in Eq.~\eqref{eq:cthird} and Eq.~\eqref{eq:clight}. These contributions correspond to diagrams such as those in Fig.~\ref{fig:Diagramsep} and Fig.~\ref{fig:Diagramuniq}. Thus, the
effective interaction of Eq.~\eqref{eq:cggtop} generalises to
\be
\label{eq:Cgg2loop}
c_{GG}^{\rm eff}(p^2=q^2=0) = \frac{1}{2} \sum_{f = \,{\rm quarks}} B_1 \left( \frac{4 m_f^2}{m_a^2}\right) c_f\,,
\ee
where $c_f$ are the effective couplings in Eq.~\eqref{eq:cthird} and Eq.~\eqref{eq:clight}. As we will see in Sec.~\ref{sec:basics}, this estimated two-loop contribution from light quarks can in fact overcome the one-loop contribution from the top-quark loop for relatively light ALPs.
Notice that the contribution from up and down quark flavours that are much lighter than the ALP will cancel among each other due to the opposite sign of $ T_3^f$ in the effective couplings (see Eq.~\eqref{eq:clight}). This implies that the main contribution for a GeV-scale ALP actually originates from the induced coupling to the bottom quark.
Equation \eqref{eq:B1explicit}  also implies that the individual contributions to ALP production via gluon fusion of quarks much lighter than the ALP, such as the $u,d$ and $s$ in our work, do not vanish and are independent of the value of $m_{a}$. Indeed,  for  $m_{f}/m_{a}\TO 0$ we have that  $B_1 ( 4 m_f^2/m_a^2)\TO 1$, in particular 
\be
\label{eq:charmcares}
B_1 \left( \frac{4 m_f^2}{m_a^2}\right)=1- \frac{ m_f^2}{m_a^2}\left[\pi+i \log\left( \frac{4 m_f^2}{m_a^2}\right)\right]^{2} +\mathcal{O}\left(\frac{ m_f^4}{m_a^4}\right).
\ee

For the range we will consider in our work,  $10~{\rm GeV}\lesssim m_{a}\lesssim 200~{\rm GeV}$, Eq.~\eqref{eq:charmcares} shows that contributions from up, down and strange quarks are negligible, while the charm mass is relevant only in the proximity of $m_{a}\sim 10~{\rm GeV} $. Also, in the range $2m_{b}\ll m_{a} \ll 2m_{t}$, combining  Eqs.~\eqref{eq:cthird}, \eqref{eq:Cgg2loop} and \eqref{eq:charmcares} one obtains
\begin{align}
\label{eq:Cgg2loopintermediate}
c_{GG}^{\rm eff}(p^2=q^2=0){\Big|}_{2m_{b}\ll m_{a} \ll 2m_{t}} &\simeq -\frac{m_a^2}{24 m_t^2} c_t(m_{t}) +\frac{1}{2}  c_b(m_{t}) \nonumber \\
 &\simeq -\frac{m_a^2}{24 m_t^2} c_t(\Lambda)+\frac{5}{2} c_t(\Lambda) \frac{y_t^2}{16 \pi^2} \log \frac{\Lambda}{m_t}\,.
\end{align}
First we notice, that at small $m_{a}$ values the bottom-quark contribution,
with $c_{b}$ obtained from one-loop induced $c_{t}$ effects, is  dominant w.r.t.~the top-quark one. For  larger $m_{a}$, the top contribution takes over. 
Second, in the mass-window of phenomenological interest in our paper, $10~{\rm GeV}\lesssim m_{a}\lesssim 200~{\rm GeV}$, $c_{GG}^{\rm eff}$ never surpasses 0.05, contrary to the $O(1)$ expectations from typical ALP models. Finally, there is a cancellation between the top and bottom contributions, arising at one- and two-loop level, respectively. We will discuss this in Sec.~\ref{sec:BRs} and in Appendix \ref{app:dip}, noting that the exclusion limits that will be derived in this work are independent of a precise computation of this effect.

\medskip

Our discussion so far has focused on low-energy amplitudes below the mass of the top quark, but in our work we also consider the opposite limit ($Q^2 \gg m_t^2$), {\it e.g.}, when computing the cross section of $pp\TO a+{\rm jet}$ production in Sec.~\ref{sec:xsALP} to set bounds on $c_t$ in Sec.~\ref{sec:DM-CS}. In this high-energy regime, there is no reason to expect the same suppression $\propto m_a^2/m_t^2$.
This behaviour can be explicitly checked in some kinematic limits. Let us, for instance, consider the case of an on-shell ALP with negligible mass, $k^2 = m_a^2 \approx 0$, and take one gluon to be on-shell, $p^2=0$, while the second gluon is off-shell by an amount $q^2=-Q^2$, with $Q^2 >0$. One then finds that the form factor in Eq.~\eqref{eq:aggampl} behaves as
\be
\label{eq:aGGoffshell}
1 + 2 m_t^2 C_0(p^2=0,q^2=-Q^2; m_t^2) = 1 - \frac{m_t^2}{Q^2} \,\,{\log}^2 
\left[ 
\frac{1 + (1 + 4 m_t^2/Q^2)^{-1/2}}{1 - (1 + 4 m_t^2/Q^2)^{-1/2} }
\right],
\ee
and it asymptotes to $1$ for large $Q^2$, corresponding to an effective coupling\footnote{The apparently accidental simplicity of Eq.~\eqref{eq:tadaaa!} will be more clear in Sec.~\ref{sec:xsALP}, where the computation of the associated production of an ALP and a light jet in the non-derivative basis is discussed.} 
\be
\label{eq:tadaaa!}
c_{GG}^{{\rm eff}, \, t}(Q^2 \gg m_t^2) = c_t/2\, .
\ee
As we can see, no suppression with $m_a^2$ is found in this case, so that the main contribution at high energies will be indeed provided by the top quark at one loop, unlike the previously discussed case where the dominant contribution comes from two-loop diagrams involving bottom quarks. Nevertheless, when all fermions are considered, since $Q^2 \gg m_t^2$ implies $Q^2 \gg m_f^2$, we have 
\be
\label{eq:tadaaa!2}
c_{GG}^{{\rm eff}}(Q^2 \gg m_t^2) = \sum_{f = \,{\rm quarks}} \frac{c_f}{2}=\frac{c_t+c_{b}}{2} \simeq\frac{ c_{t}(\Lambda)}{2}\, ,
\ee
where in the r.h.s.~of the equation we have taken into account only the leading contributions to $c_{t}(\Lambda)$ by using Eq.~\eqref{eq:cthird}.

\medskip

Finally, let us notice that the ALP coupling to photons shares the same qualitative features of the coupling to gluons. In particular, the amplitude for $a \TO \gamma \gamma$ also decouples $\propto Q^2/m_t^2$ for low $Q^2$.
When considering for concreteness the ALP decay into two on-shell photons, $p^2=q^2=0$, we may describe this process analogously to Eq.~\eqref{eq:Cgg2loop} by considering an effective coupling,
\be
\label{eq:Cgamma2loop}
c_{\gamma \gamma}^{\rm eff}(p^2=q^2=0) = \sum_{f} B_1 \left( \frac{4 m_f^2}{m_a^2}\right) N_c^f Q_f^2 c_f,
\ee
with $c_{\gamma \gamma}^{\rm eff}$ related to the operator
\be
\label{eq:aFF}
 c_{\gamma \gamma} \frac{\alpha}{4\pi}\frac{a}{f_a} F \tilde F  \subset \mathcal{L}_a.
\ee
In Eq.~\eqref{eq:Cgamma2loop} we have included the (formally two-loop) contribution from light quarks via their effective couplings $c_f$, as these will be dominant for $m_a \ll m_t$. The sum then runs over all SM fermions $f$, and $N_c^f$ represents the number of colours. In the limit of vanishing ALP mass,
$m_a^2 \TO 0$, this effective interaction would vanish, in agreement with the ALP shift symmetry.

\medskip

Analogously to the case of the gluons in Eq.~\eqref{eq:aGGoffshell}, in the high-energy regime the effective $c_{\gamma \gamma}$ coupling induced by the top quark will no longer feature the $m_a^2/m_t^2$ suppression, and the leading-order contribution to this amplitude will be given by the top at one loop. However, unlike the case of gluons, this kinematic regime is not relevant to our analysis. 

\medskip

\subsubsection{Coupling to electroweak gauge bosons}
\label{sec:weak_gauge}

We now turn to discussing the ALP couplings to EW gauge bosons, focusing in particular on the amplitudes $a \TO Z \gamma, ZZ,  W^+ W^-$. In the following, we will only consider the case of on-shell SM states, since similarly to the $a \TO \gamma \gamma$ case the high-energy regime is not relevant for our phenomenological analysis.

By referring to the following parameterisation below the EW symmetry breaking scale, 
\be
\label{eq:ewcouplings}
\mathcal{L} \supset -\frac{1}{4} g_{a \gamma Z} \, a \, F_{\mu \nu} \tilde Z^{\mu \nu} -\frac{1}{4} g_{a Z Z} \,a \, Z_{\mu \nu} \tilde Z^{\mu \nu} -\frac{1}{2} g_{a W W} \, a \, W_{\mu \nu}^+ \tilde W^{- \,\mu \nu},
\ee
one obtains from direct computation involving the top interaction in Eq.~\eqref{eq:top-philic} at one-loop\,\cite{Gunion:1991cw,Quevillon:2019zrd,Arias-Aragon:2022iwl,Bonilla:2021ufe}:
\be
\label{eq:ALPgaugeshift}
g_{a Z \gamma}^{\rm eff} = \frac{\alpha}{\pi \s c_{\rm w}} \frac{c_t}{f_a} A^t(k^2), \quad g_{a Z Z}^{\rm eff} = \frac{\alpha}{\pi \s^2 c_{\rm w}^2} \frac{c_t}{f_a} B^t(k^2), \quad g_{a W W}^{\rm eff} = \frac{\alpha}{\pi \s^2} \frac{c_t}{f_a} C^t(k^2),
\ee
where, assuming the same convention used for labelling external momenta in Eq.~\eqref{eq:aggampl}, $A^t$, $B^t$ and $C^t$ are form factors depending on the incoming ALP momentum, $k^2$, and the SM states are taken on-shell. 

We may again consider the limit of these effective couplings as the top quark becomes infinitely heavy w.r.t.~the other scales.  One finds that all of the form factors $A^t,B^t,C^t$ do not decouple for $k^2, m^2_{W,Z} \ll m_t^2$, and instead approach a non-zero, constant value. This can be interpreted as the top quark loop effectively generating unsuppressed contact interactions between the ALP and the EW gauge bosons.
In the limit of $m_t^2 \gg k^2$, and also $m_t^2 \gg m^{2}_{W,Z}$ for simplicity,
\be
A^t \approx Q_t T_t^3 N_c,\, 
\quad B^t \approx \left(\frac{(T^t_3)^2}{12} - \s^4 Q_t^2\right) N_c\,, \quad C^t \approx \frac{3}{16} N_c\,, \label{eq:nomtoQ}
\ee
raising the question whether this is consistent with the ALP shift symmetry. Indeed, although the operators in Eq.~\eqref{eq:ewcouplings} appear to break the shift symmetry, unlike the case of the $a\gamma\gamma$ and $agg$ vertices, in the limit $m_{t} \TO \infty$ no $m_{a}/m_{t}$ suppression arises in the form  factors of Eq.~\eqref{eq:nomtoQ}. In fact, the operators in Eq.~\eqref{eq:ewcouplings} are not truly shift-symmetry breaking. According to Eq.~\eqref{eq:EFTdata}, one of the rephasing invariant quantities is for instance $c_{WW} - \frac{1}{2}\, \text{Tr}\, \left(3 c_Q + c_L\right)$.  Therefore, a non-zero $c_{WW}$ can always be traded for $c_Q$ and $c_L$, which are associated to purely derivative ALP interactions.

This argument applies to  $g_{a W W}^{\rm eff}$ but can be also repeated for the case of  $g_{a Z Z}^{\rm eff} $ and $g_{a Z \gamma}^{\rm eff} $. 
In conclusion, via a rephasing, the terms Eq.~\eqref{eq:ewcouplings} can be rewritten in terms of purely derivative ALP interactions and therefore are shift symmetric. Notice that the same argument does not apply to $c_{\gamma \gamma}^{\rm eff}$ in Eq.~\eqref{eq:Cgamma2loop}, consistent with its $(m_{a}/m_{f})^{2}$ suppression. Indeed, at  tree level using the Lagrangian in Eq.~\eqref{eq:full_Lag_a}, the $a\gamma\gamma$ coupling is proportional to $c_{WW}+c_{BB}$, which is one of the rephasing invariants of Eq.~\eqref{eq:EFTdata}.

\subsection{Non-derivative basis and relation to a top-philic pseudoscalar}
\label{sec:nonder}
In order to simulate processes involving the ALP at the LHC, it is more convenient to use a different operator basis than the one presented in Eq.~\eqref{eq:full_Lag_a}. In particular, the one with non-derivative interactions between the ALP and the fermions, which lead to dimension-four, Yukawa-like interactions after electroweak-symmetry breaking. 
From a computational point of view, especially when loops are involved and have to be automated, having only dimension-four operators involving fermions considerably simplifies the computation.
This change of basis is achieved by performing ALP-dependent rotations of the SM fermions which, unlike those in Eq.~\eqref{eq:rephasing}, do not correspond to classical symmetries of the SM. These rotations (see {\it e.g.}~Ref.~\cite{Bonilla:2021ufe})  will generically transform shift-symmetric interactions into combinations of  apparently shift-breaking operators such as pseudo-Yukawas and contact terms with the gauge bosons.\footnote{When all contributions are taken into account the shift-symmetric nature of the ALP will of course be recovered.}

\medskip

Starting from the generic derivative couplings to fermions as in Eq.~\eqref{eq:full_Lag_a}, one can fully eliminate them in favor of pseudoscalar interactions:
\be
\label{eq:La_nonder}
\begin{split}
\mathcal{L}_a = \frac{1}{2} (\partial_\mu a)^2 - &\frac{1}{2} m_a^2 a^2 - \frac{a}{f_a} \left( 
\bar Q_L \phi \tilde Y_d d_R + \bar Q_L \tilde \phi \tilde Y_u u_R + \bar L_L \phi \tilde Y_e e_R + {\rm h.c.} \right)  \\
& 
+\tilde c_{GG} \frac{\alpha_S}{4 \pi} \frac{a}{f_a} G \tilde G + 
\tilde c_{WW} \frac{\alpha_2}{4 \pi} \frac{a}{f_a} W \tilde W + 
\tilde c_{BB} \frac{\alpha_1}{4 \pi} \frac{a}{f_a} B \tilde B\,.
\end{split}
\ee
The new effective couplings to gauge bosons are given in terms of the ones prior to the rotation as
\be
\tilde c_{GG} = c_{GG} + \frac{1}{2}{\rm Tr}\left(c_u + c_d - 2 c_Q\right)\,,
\ee
whereas for $\tilde c_{BB}$ and $\tilde c_{WW}$ we refer to Ref.\,\cite{Bauer:2020jbp} (the trace is over the flavour indices of the Wilson coefficients).
The pseudo-Yukawa matrices in Eq.~\eqref{eq:La_nonder} for the SM quarks are given by\,\cite{Bauer:2020jbp}
\be
\tilde Y_{u,d} = i (Y_{u,d} c_{u,d} - c_Q Y_{u,d})\,.
\ee
where $c_{Q}$, $c_{u,d}$ and $Y_u$ can be taken to be diagonal in the same basis according to our flavour assumptions.  Defining vector and axial combinations as
\be
c_{u,d}^V \equiv \frac{c_{u,d} + c_Q}{2}, \quad c_{u,d}^A \equiv \frac{c_{u,d} - c_Q}{2}\,, 
\ee
we can rewrite
\be
\label{eq:tildeYud}
\tilde Y_{u,d} = i \left( [Y_{u,d},c_{u,d}^V] +\{Y_{u,d},c_{u,d}^A\} \right)\,.
\ee
For $\tilde Y_u$, the part involving $c^V_u$ vanishes trivially due to the diagonal structure of the matrices. If the CKM matrix $V$ is assumed to be trivial, as done in our work, the same applies for $Y_d$. Indeed, in this basis we have $Y_d = V Y_d^{\rm diag}$. 

Before proceeding with the discussion of our simplified scenario, it is interesting to note that in general, without assuming that a trivial CKM matrix,  one finds
\be
[V Y_d^{\rm diag}, c_{d}^V]_{ii} = 0\,, \quad 
[V Y_d^{\rm diag}, c_{d}^V]_{i\neq j} = y_d^j V_{ij} \left(c_V^{jj} - c_V^{ii} \right)\,.
\ee
The off-diagonal combinations above being non-vanishing implies that more free parameters are needed to specify the theory beyond the axial combinations in Eq.~\eqref{eq:EFTdata}; these are related to differences of vector couplings between the different generations. Because of this structure, only two differences are linearly independent, {\it e.g.}, $(c_V^{11} - c_V^{33})$ and $(c_V^{11} - c_V^{22})$. This agrees with the fact that a non-diagonal CKM allows only for a combined baryon transformation rather than the three generation-specific ones\,\cite{Bonilla:2021ufe}, removing two transformations from the possible rephasings in Eq.~\eqref{eq:rephasing}. 

\medskip
Going back to our simplified scenario with a diagonal CKM matrix, where only the anti-commutator parts of Eq.~\eqref{eq:tildeYud} survive, and trading the diagonal Yukawa matrices for the mass of the corresponding quarks, we have:
\be
\label{eq:Laintnonder}
\mathcal{L}_{a, {\rm int}} = \sum_f - i c_f m_f \frac{a}{f_a} 
\bar \psi_f \gamma_5 \psi_f +\tilde c_{GG} \frac{\alpha_S}{4 \pi} \frac{a}{f_a} G \tilde G + 
\tilde c_{WW} \frac{\alpha_2}{4 \pi} \frac{a}{f_a} W \tilde W + 
\tilde c_{BB} \frac{\alpha_1}{4 \pi} \frac{a}{f_a} B \tilde B\,,
\ee
where $f=t,b,c,s,u,d,\tau,\mu,e$. This is the Lagrangian that will be used in this work. 

We note that higher dimensional operators are also present, which we have not explicitly written, yet could be relevant in specific cases.  For example, we have not included the four-point $aH\bar{\psi}\psi$ contact interactions involving the dynamical Higgs field, as they play no role in the calculations presented in this work.
Another example at order $a^2$ is the interaction~\cite{Bauer:2023czj,Galda:2023qjx}
\begin{equation}
\mathcal{L}_{a^2, {\rm int}} = \sum_f c_f^2 m_f \frac{a^2}{f_a^2}  
\bar \psi_f \psi_f \,,
\label{eq:a2term}
\end{equation}
which can contribute at tree level to processes with two ALPs or more as external states, or in loops. 
This term has no final bearing in the results presented in this work and therefore it is not included in our reference Langrangian \eqref{eq:Laintnonder}.\footnote{Technically, it  contributes to the $t \bar t$ computation at one loop presented in Sec.~\ref{subsec:ttbar}, specifically to the top-quark self-energy. However, as discussed later, it bears no physical consequence.}

The application of Eq.~\eqref{eq:Laintnonder} to our top-philic model in Eq.~\eqref{eq:top-philic} is straightforward: the couplings to fermions include the tree-level interaction with the top $c_t$, as well as the loop-induced $c_f$ couplings in Eq.~\eqref{eq:cthird} and Eq.~\eqref{eq:clight}. The coupling to gauge bosons prior to the rotation onto the non-derivative basis Eq.~\eqref{eq:La_nonder} are all vanishing, so that for instance the $\tilde c_{GG}$ coupling is simply given by
\be
\label{eq:axionmodel}
\tilde c_{GG} = \frac{1}{2} \sum_{f={\rm quarks}} c_f\,.
\ee
This precise value of $\tilde c_{GG}$ compensates the pseudo-Yukawa interactions of the ALP in Eq.~\eqref{eq:Laintnonder} such that the original ALP shift symmetry is preserved in this non-derivative basis, see also\,\cite{Bonnefoy:2022rik}.
For details on the {\aNLO} \cite{Alwall:2014hca, Frederix:2018nkq} implementation of the {\UFO} \cite{Degrande:2011ua,Darme:2023jdn} model, we refer to Appendix \,\ref{app:MGModel}.

We observe that more generally, in this non-derivative basis, we can identify two different scenarios for a pseudo-scalar particle coupling predominantly with the top quark:
\begin{itemize}
\item The top-philic ALP, defined by the relation in Eq.~\eqref{eq:axionmodel}, which descends from the original shift symmetry of the Lagrangian in Eq. \eqref{eq:top-philic}.
\item A top-philic \emph{pseudoscalar}
in which instead the ALP-gluon contact interaction is set to zero, namely $\tilde c_{GG} = 0$.
\end{itemize}
In this paper we focus on the 
top-philic ALP model and its phenomenology at the LHC. Nevertheless we will comment about the expected differences with a top-philic \emph{pseudoscalar}, leaving a detailed study of the latter to future work.

An important point to keep in mind is that the transformation from a derivative basis to a non-derivative basis mixes perturbative orders. This can be easily seen from the definition of $\tilde c_{GG}$ in Eq.~\eqref{eq:axionmodel}, which contains $c_{f}$. While $c_{t}$ appears at tree level in the top-philic Lagrangian in Eq.~\eqref{eq:top-philic}, in the non-derivative basis $c_{t}$ also enters the definition of $\tilde c_{GG}$, whose definition factorises an additional power of $\alpha_{S}$. 
Thus, depending on the perturbative order at which $c_t$ appears in a given process, contributions from $\tilde c_{GG}$ may or may not need to be included in the calculation for perturbative consistency.\footnote{In our model, the same arguments also apply for the $c_{f}$'s at one order higher in EW loops. 
}
For instance, as we will discuss in Sec.~\ref{sec:xsALP}, both in the derivative and non-derivative bases, at LO (tree level) $t\bar t a$ production receives a contribution from $c_{t}$ only. Conversely, as we will discuss in Sec.~\ref{subsec:ttbar}, for the calculation of one-loop corrections to $t \bar t$ production involving the ALP, both the $c_{t}$ and $\tilde c_{GG}$ contribution have to be consistently taken into account when the non-derivative basis is used. Similarly, the associated production of a top-philic ALP and a jet (Sec.~\ref{sec:xsALP}), since it is loop-induced, involves at LO both $c_{t}$ and $\tilde c_{GG}$. The same discussion applies to the cases of $gg\TO a$, $a\TO gg$ or $a\TO\gamma \gamma$. In the $a\TO gg$ and $gg\to a$ amplitudes, tree-level diagrams involving $\tilde c_{GG}$ must be included in the non-derivative basis. Analogously, $\tilde c_{\gamma\gamma}$ contributions to $a\TO\gamma \gamma$ and other induced EW coupling contributions to processes involving the EW gauge bosons should be included.
For the sake of clarity and simplicity, we will reserve a detailed discussion of this point for the associated production process, $pp\to a$ + jet, as we find it most instructive, understanding that the same considerations apply to other processes.

\section{Top-philic ALPs at the LHC}
\label{sec:basics}
We can now study the phenomenology of a top-philic ALP at the LHC, using the Lagrangian of Eq.~(\ref{eq:Laintnonder}).  As a first step, we calculate the branching ratios (BR) for the different decay modes of the top-philic ALP, in Sec.~\ref{sec:BRs}, and the cross sections for the dominant processes leading to the direct production of an ALP, in Sec.~\ref{sec:xsALP}. 
In Sec.~\ref{sec:existing}
we use our predictions to set bounds on $c_{t}$ via existing LHC searches that directly target new physics with resonant signatures of pseudoscalar states, where bounds on cross section times branching ratios can be simply re-interpreted.
In Sec.~\ref{sec:newprobes} we present new limits derived by analyzing the impact of the ALP in SM cross section measurements involving top quarks in the final state.

As anticipated in Sec.~\ref{sec:introduction}, we focus on the mass range between $10$ and $200$ GeV. The reason is twofold. First, for ALP masses significantly larger than the top mass ($m_{a}\gg m_t$), the loops inducing the interactions with the gluons ($c_{GG}^{\rm eff}$ in Eq.~\eqref{eq:Cgg2loop}) and photons ($c_{\gamma \gamma}^{\rm eff}$ in Eq.~\eqref{eq:Cgamma2loop}) are not suppressed. Thus, the resonant production as well as the decays into photons and gluons  dominates the phenomenology of the ALP at the LHC, as we shall see below. 
In this context,  constraints from diphoton, dijet and di-top resonance searches at LHC efficiently probe $c_{t}$ (see {\it e.g.}~Refs.~\cite{Bellazzini:2017neg,ATLAS:2020lks,Bonilla:2021ufe,Esser:2023fdo}).
Second, for $m_a \lesssim 10 \,{\rm GeV}$, limits on $c_{t}$ from flavour experiments, astrophysics and cosmology become quite stringent (see {\it e.g.}~Refs.~\cite{Bauer:2021mvw,Bonilla:2021ufe,PhysRevD.91.084011}).
Instead, as we will show in Sec.~\ref{sec:existing},  for the range of ALP masses that we consider,
the existing collider constraints are quite weak, motivating the exploration of alternative channels to close this gap in the top-philic ALP parameter space.

\subsection{Branching ratios}
\label{sec:BRs}

Besides the SM input parameters, the only two new physical free parameters for the top-philic ALP model are the ALP mass $m_{a}$ and the ratio $c_t/f_a$. The latter parameterises the $t\bar t a$ interaction and in turn, together with $m_{a}$, all other loop-induced interactions of the ALP with fermions and gauge bosons, as documented in Sec.~\ref{subsec:ALPgeneral}. 

Strictly speaking, according to Eqs.~\eqref{eq:cthird} and \eqref{eq:clight}, all of the $c_{f}$'s explicitly depend on $\Lambda$ and therefore on $f_{a}$ alone, independently of the ratio $c_t/f_a$. Thus, one may wonder if the chosen value of the scale $\Lambda$ affects our results. As we will show in Sec.~\ref{sec:basics}--\ref{sec:DM} all of our bounds on $c_t/f_a$ will exclude the region $f_{a}/c_{t}\lesssim \mathcal{O}(10^{2}) ~{\rm GeV}$. This means, assuming perturbation theory is valid ($c_{t}\lesssim 4\pi$ and $\Lambda\lesssim 4\pi f_a$), that we require $\Lambda\lesssim \mathcal{O}(10^{4}) ~{\rm GeV}$. We have checked that varying $\Lambda$ between 1 TeV and 10 or even 100 TeV, would modify the bounds presented in  Sec.~\ref{sec:basics}--\ref{sec:DM} at the level of 10\%.\footnote{ This is a reasonable assumption for the minimum value  both  for technical reasons (see also footnote \ref{footnote:1TeV}) and UV-inspired motivations.} Such an uncertainty corresponds to a level of precision that is clearly beyond the scope of this work.
For practical purposes, in the following we will always understand $\Lambda=1~{\rm TeV}$ unless stated otherwise. The only cases where this choice of scale may considerably affect our predictions is the decay of the ALP into gluons or photons and the direct production of the ALP via gluon fusion. However, we stress that these specific predictions are not relevant for the extraction of our bounds. This point will be clarified in the following sections and in Appendix~\ref{app:dip}.

\medskip 

In Fig.~\ref{fig:BRplot} we show the BR of each decay channel of the top-philic ALP as a function of $m_{a}$, in the range $10~{\rm GeV}\lesssim m_{a}\lesssim 200~{\rm GeV}$. Since all of the partial decay widths are proportional to $(c_t /f_a)^{2}$, the corresponding BR's are independent of the value of  $c_t /f_a$. Note that the fact that $m_a$ is below the top pair threshold also means that all decay channels are loop-induced.
The explicit formulae for the partial decay widths, as a function of the effective couplings discussed in Sec.~\ref{subsec:ALPgeneral}, can be found in Appendix \ref{app:decaywidths}. 
\begin{figure}
    \centering
    \includegraphics[width=0.8\textwidth]{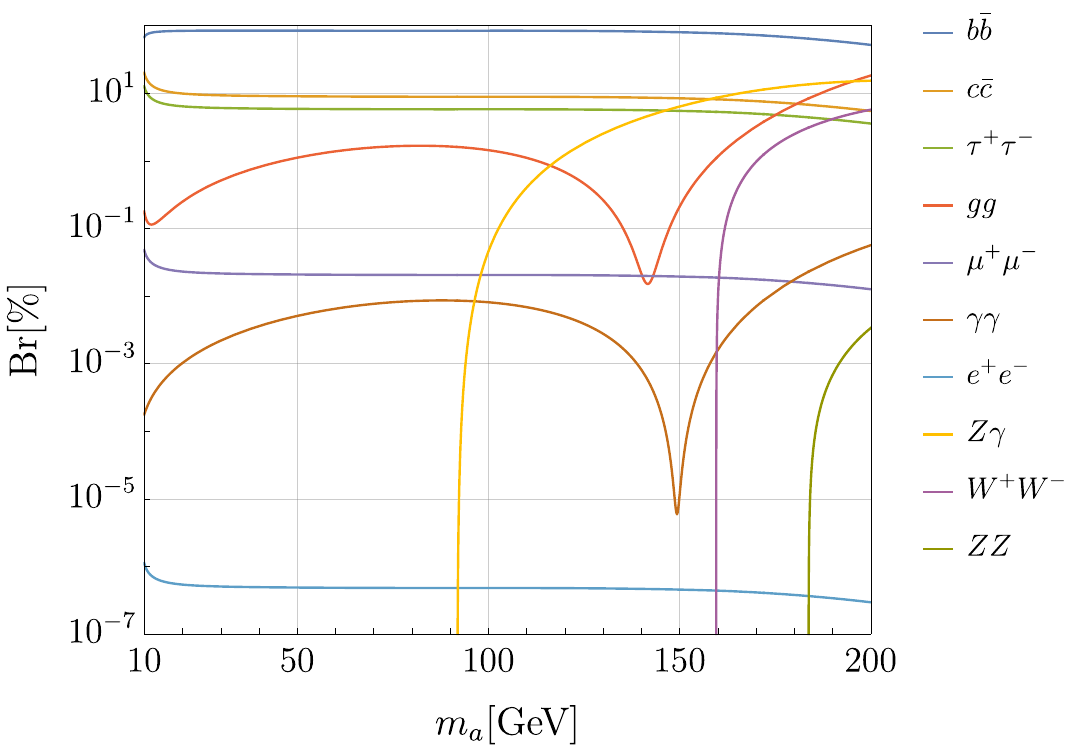}
    \caption{Branching ratios of the top-philic ALP decay into SM particles, in the mass range considered in this paper. }
    \label{fig:BRplot} 
    
\end{figure}
First, we notice that the dominant ALP decay channel in our window of interest for $m_a$ is into bottom-quark pairs. The associated BR ranges between around $50\%$ and $80\%$.
 From Eq.~\eqref{eq:Laintnonder} it is manifest that the interactions of the ALP with the fermions are proportional to the respective fermion masses. Therefore the rates of the decays into the other SM fermions are much smaller, especially for those of the first generation.

Moving to the case of $WW$, $ZZ$ and $Z\gamma$ on-shell decays, as discussed in Sec.~\ref{sec:weak_gauge} the 
corresponding loop-induced ALP couplings do not exhibit any $m_a/m_t$ suppression. However, in the approximation of on-shell $Z$ or $W$, these decays are kinematically allowed only for,   respectively, $m_{a}> 2m_{W},2m_{Z},m_{Z}$. Therefore,  at the higher end of our ALP mass window, decays into EW gauge bosons are possible and the $Z\gamma$ channel becomes particularly important, with a BR$\sim10\%$. Clearly, by consistently taking into account off-shell $Z$ and $W$ effects ($a$ decaying into four fermions or two fermions plus a photon), also in the regions $m_{a} < 2m_{W},2m_{Z},m_{Z}$ the BRs  for $WW^\ast$, $ZZ^\ast$ and $Z^\ast\gamma$ would be non-vanishing, albeit suppressed by the off-shell propagator. We will return to this point at the end of this section.

In our top-philic model, the decays into gluons and photons display a peculiar behaviour. We remind the reader that, as discussed in Sec.~\ref{subsec:gluphot}, in our approximation both decays are induced by two kinds of contributions: the one-loop top-quark contribution  and  the approximate two-loop contribution originating from a loop of light fermions where the ALP-fermion interactions, $c_{f}$, are themselves generated by another top-quark loop (see Fig.~\ref{fig:Diagramsep}). While 
in the high end of the $10~{\rm GeV}\lesssim m_{a}\lesssim 200~{\rm GeV}$ mass window the one-loop contribution is dominant, since the ratio $m_a^2/m_t^2$ is of order one, in the lower range ($m_{a}\ll m_{t}$) the two-loop contribution dominates.  We discuss in the following the case of the decay into gluons and afterwards mention the few differences w.r.t.~the analogous case of the decay into photons.

As we can see from Fig.~\ref{fig:BRplot}, the BR of the decay into gluons increases fast with $m_a$, for $m_{a}\gtrsim 150~{\rm GeV}$. In fact, it  overtakes the BR of the decay into bottom quarks at $m_a \approx 220 \,{\rm GeV}$. From this point on, the phenomenology of the ALP is dominated by its interaction with gluons and, as mentioned at the beginning of Sec.\,\ref{sec:basics},  this is one of the motivations behind our choice of considering only the region $m_a \lesssim 200 \,{\rm GeV}$ for our top-philic model.
This fact will be even clearer when we discuss the $pp\TO a$ production in Sec.~\ref{sec:xsALP}.
In the opposite part of the $m_{a}$ range, where $m_a \ll m_t$ and the $m_a^2/m_t^2$ suppression is  numerically important,  the two-loop contribution associated to light quarks becomes increasingly relevant. Since in our model we do not set the mass of the bottom quark ($m_{b}$) and of the charm quark ($m_{c}$) to zero, as motivated by Eq.~\eqref{eq:charmcares} and discussed in Appendix.~\ref{app:MGModel}, we can observe also a small change of trend in the region very close to $m_a \approx 2 m_{b}$. 

Around $m_a \approx 150 \, \rm{GeV}$ the one-loop and two-loop contributions have similar sizes but opposite signs, leading to large cancellations and a sudden drop of the gluonic  BR.
The  location of the minimum of the BR is sensitive to the exact value of the two-loop corrections, which are only estimated in our analysis, as discussed in the text around Eq.~\eqref{eq:Cgg2loop}. Therefore, in  Fig.~\ref{fig:BRplot}, the $m_{a}$ value for which the BR into gluons is minimal,  and consequently the region were the two-loop contribution is dominant,  must be considered as only indicative. We refer the reader to  Appendix~\ref{app:dip} for a more detailed inspection of this cancellation effect. However, we emphasise that the limits on $c_t/f_a$ that we will derive in our phenomenological study will not rely on the smallness of the BR into gluons in this region and on the precise value of $m_{a}$ that minimises it. 
To summarise, if an analysis were to be sensitive to the decay $a\TO gg$, 
an exact calculation of two-loop effects, including the contributions from the diagram in Fig.~\ref{fig:Diagramuniq},  would be crucial. This is not the case for the analyses presented in our work. This argument also applies to the case of the $a\TO \gamma\gamma$ decay and the direct $gg\TO a$ production discussed in Sec.~\ref{sec:xsALP}. 

\medskip

The decay into photons follows a pattern very similar to the one discussed for the case of the decay into gluons.
However, in this case, the loop-induced contributions involve electric charges and hence the minimum of the BR occurs at a numerically different value of $m_a$ w.r.t.~the decay into gluons. The same argument regarding the  location of the minimum and the irrelevance for our analysis applies to this decay channel. In particular, the relevant information for our analysis is that the BR of this channel is always (much) smaller than 0.1\%.
We also observe that, unlike the case of the gluons, there is no small increase in the region close to $m_a \approx 2 m_{b}$; rather, there is an even stronger decrease, due to the additional dependence on the electric charge of the light fermion in the loop. 

We now comment on the impact of the decay channels $a\rightarrow WW^*$ and  $a\rightarrow ZZ^*$, with the subsequent decay of the gauge bosons into fermions. This case cannot be derived in general from Eq.~\eqref{eq:ALPgaugeshift}, since in the derivation of those formulae both of the vector bosons were assumed to be on-shell; for $m_a<2 m_V$, at least one $V$ has to be off-shell.
Moreover, the {\tt ALPTopNLO\_3flav} {\UFO} model presented in Appendix \ref{app:MGModel} cannot be used to perform these calculations, because the required ingredients for generating processes involving loops of EW gauge bosons have not been implemented.
On the other hand, the impact of such decay modes can be estimated by comparing the case of the ALP with that of the SM Higgs boson, which has been studied in Ref.~\cite{Denner:2011mq} for various values of $m_H$. For the $ZZ^*$ case, the Higgs branching ratio for one virtual $Z^*$ ($m_H< 2 m_Z\simeq180$ GeV) never surpasses the value of the branching ratio at $m_H=200\gev$, where both $Z$ bosons are on-shell. We expect a similar behaviour for the ALP, such that in the entire  $m_a$ range considered the branching ratio will never exceed $0.01\%$, the value of BR($a\rightarrow ZZ$) for $m_a=200$ GeV. Furthermore, it is expected to decrease rather quickly for lower masses. 

The $WW^*$ case is slightly more delicate, as in the case of the Higgs in the mass window between $160$-$180\gev$ the branching ratio of $H\rightarrow WW^*$ is almost constant, only decreasing as the $H\rightarrow ZZ$ channel opens up and remaining constant thereafter.
For the Higgs, this is due to the fact that the $ZZ$ branching ratio is around $20\%$, which is a very different situation from our top-philic ALP, in which it never surpasses $0.01\%$. Following this argument we expect the branching ratio of $a\rightarrow WW^*$ to remain almost constant ($\simeq5\%$) in the $160$-$180\gev$ window and to decrease for lower masses, as in the Higgs case.

\medskip

As a final remark, we remind the reader that, unlike the BRs, the lifetime of the top-philic ALP does depend on the actual value of $c_t/f_a$, as can be seen in Appendix~\ref{app:decaywidths}.
Since all of the decay rates scale with positive powers of $m_a$,
the lighter the ALP, the longer its lifetime. 
In the mass range we consider, for typical values of the couplings the ALP is short-lived. 
Indeed the inverse of the partial decay width into $b\bar{b}$, which is a good approximation for the total lifetime of the ALP given that the dominant decay mode is into bottom quarks,
is given by 
\begin{equation}
    c \tau \sim 
    \frac{1}{\Gamma[a \to b\bar b]} \simeq 2.2 \times 10^{-7} \text{cm} 
    \left(\frac{f_a/c_t}{\text{TeV}}\right)^{2}
    \left(\frac{10 \text{ GeV}}{m_a} \right) \left(1-\frac{4m_b^2}{m_a^2}\right)^{-1},
\end{equation}
such that ALP decays are always prompt in the mass range considered in this paper.
In Sec.~\ref{sec:DM} we will also consider the case in which the top-philic ALP can decay into a dark matter particle, and the previous formula is modified by the extra decay channel, leading to an even more prompt ALP decay.

\subsection{LHC production cross sections}
\label{sec:xsALP}

\begin{figure}
   \centering
    \includegraphics[width=0.9\textwidth]{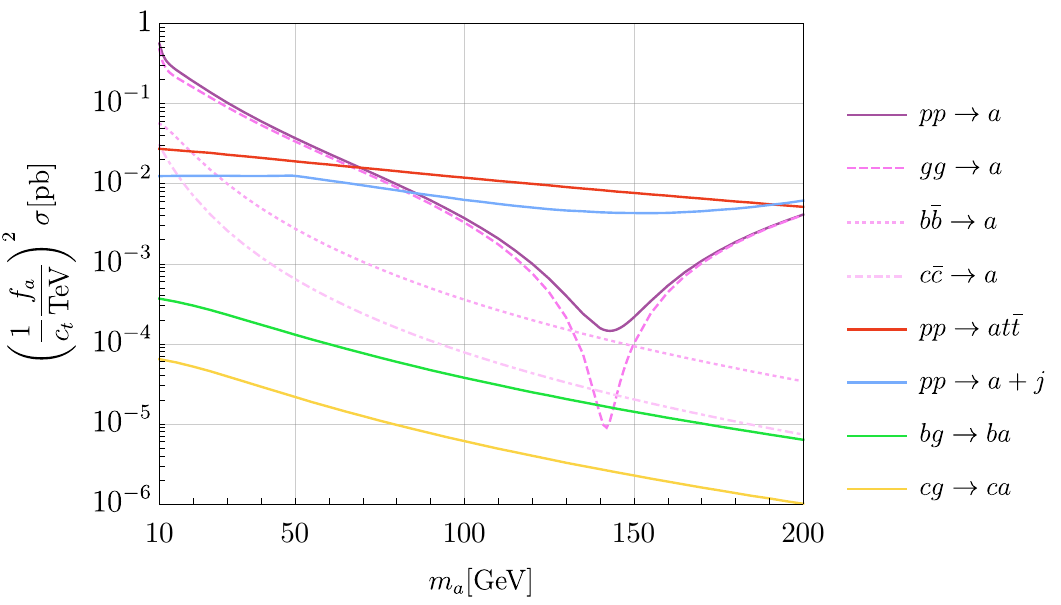}
    \caption{\label{fig:production}
    Production cross sections of the top-philic ALP. We distinguish channels with light quarks or gluons from channels with heavy quarks (charm of bottom). Individual partonic contributions to $pp\rightarrow a$ are shown by the various dashed and dotted lines. 
    }     
   \end{figure}

In this section we calculate and discuss  the cross sections for relevant  processes involving the direct production of the top-philic ALP.
The results of this section have been obtained with the help of {\aNLO} using the implementation of the {\UFO} model based on the non-derivative basis of Sec.~\ref{sec:nonder}. 
The relevant processes are:
\begin{itemize}
\item Direct production: $pp\TO a$;
\item Light-jet associated production: $pp\TO a+j$;
\item Heavy-jet associated production; $pp\TO a+j_{h}$, where we consider as heavy both the bottom and charm quarks,
\item Top-quark pair associated production: $pp\TO t\bar t a$.
\end{itemize}
In Tab.~\ref{tab:production} we list the corresponding partonic processes and the phase-space cuts that have been set for the calculation of the cross section. We also explicitly show the typical scale of the process, which we use as a factorisation and renormalisation scale in our simulations.  The quantity $m_{T}$ is the transverse mass, defined as $m_{T}\equiv \sqrt{m^{2}+p_{T}^{2}}$ and $\eta$ is the pseudorapidity.
\begin{table}[!t]
    \centering
\begin{tabular}
{|l|p{1.5in}|l|l|}

\hline
     Process & Sub-processes $(Q=c,b)$ & Scale&Cuts    \\
     \hline
      $pp\TO a$  &
     $gg\TO a$,  $Q\bar Q \TO a$ &
     $m_a$& None  \\
       
    $pp\TO a+j$&
    $gg\TO ga$,
    $gq\TO qa$,
    $q\bar q\TO ga$, $Q\bar Q\TO ga$ & 
    $m_{\rm T}(a)$&$p_{\rm T}(a)>30\,{\rm GeV},\,|\eta(a)|<4$
    \\
       
    $pp\TO a+j_{h}$&
    $gQ\TO Q a$
    &$m_{\rm T}(a)$&$p_{\rm T}(a)>30\,{\rm GeV},\,|\eta(a)|<2.5$
   \\
 
    $pp\TO t\bar t a$&  $gg  \TO t\bar t a$, $ q\bar q \TO t\bar t a$ &$m_t+\frac{m_a}{2}$ & None 
    \\
    \hline
 \end{tabular}
    \caption{Scale choice and phase-space cuts used in our calculations of ALP production processes.}
    \label{tab:production}
\end{table}
For each process listed  in Tab.~\ref{tab:production}, we plot in Fig.~\ref{fig:production} the dependence of the corresponding cross section at the 13 TeV LHC on $m_{a}$. Since all of the cross sections are proportional to 
$ (c_t / f_a)^2$ we show in Fig.~\ref{fig:production} the reference case $f_{a}/c_{t}=1~\tev$, or equivalently the cross sections normalised by $[c_{t}\,(1~\tev)/f_{a} ]^{-2}$. 

 The dependence on $m_{a}$ of the cross section for the direct production of the top-philic ALP, dominated by the $gg\TO a$ partonic channel,  follows a very similar pattern to the case of the BR of the $a\TO gg$ decay, discussed in Sec.~\ref{sec:BRs}. The main difference here is that, besides the cancellation between one-loop and two-loop contributions, there is also a dependence on the $gg$ luminosity; at small Bjorken-$x$ the gluon parton-distribution-function (PDF) grows rapidly, enhancing the $pp\TO a$ cross-section at small $m_{a}$ values. The same considerations about the uncertainties on the estimate of the photonic and gluonic BRs apply to this case. 

Since the dominant decay channel of the top-philic ALP is into bottom quarks, the direct production mainly leads to a signature that suffers from a QCD background (two $b$-jets) that is several orders of magnitudes larger, as is the case for direct Higgs production and decay via this channel. 
Moreover, a clean channel such as the ALP decaying into diphoton has a too small
cross-section times branching ratio (see Fig.~\ref{fig:BRplot}) to lead to relevant constraints on the model, as we will discuss later.

In Fig.~\ref{fig:production} we show not only the $pp\TO a$ cross-section but also the individual contribution from the $gg$, $c\bar c$ and $b \bar b$ initial states. As expected, the $gg\TO a$ is by far the dominant channel over the full $10~{\rm GeV}\lesssim m_{a}\lesssim 200~{\rm GeV}$ range, besides where the minimum of $gg\TO a$ is located. 

Imposing the cuts described in Tab.~\ref{tab:production}, the cross sections for the production of a top-philic ALP in association with either a top-quark pair or a light jet are quite similar. They are of the order $ \sigma \sim 10^{-2}\, {\rm pb} \times [c_{t}\,(1~\tev)/f_{a} ]^{2} $ and quite flat in the $10~{\rm GeV}\lesssim m_{a}\lesssim 200~{\rm GeV}$ range. Unlike the case of direct production, these predictions are used in our phenomenological studies. The former is relevant for the discussion in Sec.~\ref{sec:newprobes}, in particular for bounds from $t \bar t b \bar b$ production. The latter is crucial for the discussion in Sec.~\ref{sec:DM-CS}, where the case of a top-philic ALP predominantly decaying into DM is considered, since this process leads to the mono-jet +  $\slashed E_{T}$ signature.  We discuss in the following the calculation of the cross sections for both of these associated production processes.

At Leading-Order (LO), $t \bar t a$ production is induced by tree-level diagrams, both for the $gg  \TO t\bar t a$ and $ q\bar q \TO t\bar t a$ partonic channels. Starting from Eq.~\eqref{eq:top-philic},  diagrams with the ALP emitted directly from  the top quarks are possible at tree level.
By looking at Eqs.~\eqref{eq:Laintnonder} and \eqref{eq:axionmodel}, one may be tempted to also include the contribution from $\tilde c_{GG} = c_t/2$, leading to tree-level diagrams with the ALP stemming from gluons. However, as already anticipated at the end of Sec.~\ref{sec:nonder}, this contribution is of higher order, since there is a power of $\alpha_{S}$ in front of the $a G \tilde G$ operator. It is induced by the change  from the derivative to the non-derivative bases, which mixes perturbative orders.
This tree-level contribution in the non-derivative basis is actually a loop contribution in the derivative one and indeed should only be taken into account when NLO QCD corrections, which we do not compute in our study, are considered. 

\begin{figure}[t!]
    \centering
    \includegraphics[width=0.9\textwidth]{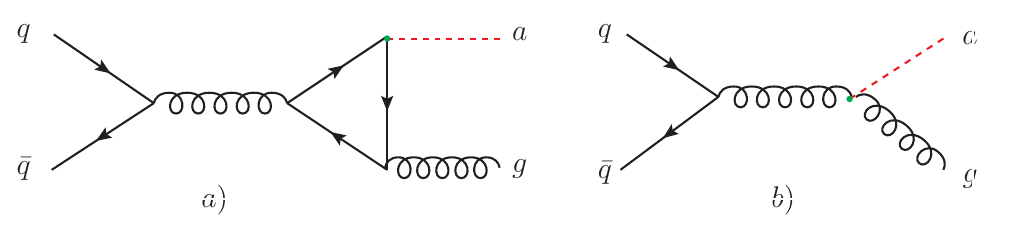}
    \caption{Representative diagrams for the process $qq\TO ag$ in the non-derivative basis.  $a)$ Diagrams common to the pseudoscalar and ALP cases. $b)$ Diagrams present only in the ALP case.}
    \label{fig:qqag}
\end{figure}

The light-jet associated production is a different story. At LO, the process is loop-induced in the derivative basis; no tree-level diagrams exist for all underlying partonic processes, listed in Tab.~\ref{tab:production}. Instead, in the non-derivative basis, both the one-loop diagrams already present in the derivative basis and additional tree-level ones are present.\footnote{Clearly, the diagrams are the same, but the Feynman rules for the $t\bar t a$ vertex are different in the two bases.} As an example, we show representative one-loop and tree-level diagrams for the $q\bar q\TO ag$ process in Fig.~\ref{fig:qqag}. Representative diagrams for the other partonic processes are shown in Appendix \ref{app:one_jet}, where analytical formulae are also  provided. In the non-derivative basis, the difference between an ALP and a top-philic pseudoscalar is precisely the contribution from these additional tree-level diagrams (See Sec.~\ref{sec:nonder}).

As in the case of $gg \TO a$, in addition to the contributions induced by $c_{t}$, the contributions from $c_{f}$ associated to $f\neq t$ cannot be neglected. These correspond to an approximate two-loop calculation that we obtain by plugging in the $c_{f}$ generated at one-loop. Concerning the change of basis from derivative to non-derivative, it is crucial to ensure that the definition of $\tilde c_{GG}$ in Eq.~\eqref{eq:axionmodel} includes $c_{f}$ for $f\neq t$ consistently with the $f$ running in the loop diagrams such as the one on the left of~Fig.~\ref{fig:qqag}. Unlike the case of direct ALP production or its decay into gluons or photons, however, predictions for the $pp\TO a+j$ production are much more reliable. Indeed, at high $p_{T}(a)$, which is the phenomenologically relevant kinematic configuration in our work, one expects a contribution from the top-quark loops of the form given in Eqs.~\eqref{eq:aGGoffshell} and \eqref{eq:tadaaa!} and a contribution from the other fermions of the form given in~\eqref{eq:tadaaa!2}. Notice that, in the non-derivative basis, this means that the purely pseudoscalar component, associated to the left-hand diagram in Fig.~\ref{fig:qqag} vanishes and all of the contribution is given by the tree-level diagram on the right. Thus, the definition of $\tilde c_{GG}$ in Eq.~\eqref{eq:axionmodel}, with only $f=t$ explains the result of Eq.~\eqref{eq:tadaaa!}. Similarly, if  the $c_{f}$ contributions are also considered, the relation Eq.~\eqref{eq:axionmodel} with  $f\ne q$ explains Eq.~\eqref{eq:tadaaa!2} and the behaviour at high $p_{T}(a)$.

\begin{figure}[!t]
    \centering
    \includegraphics[width=0.85\textwidth]{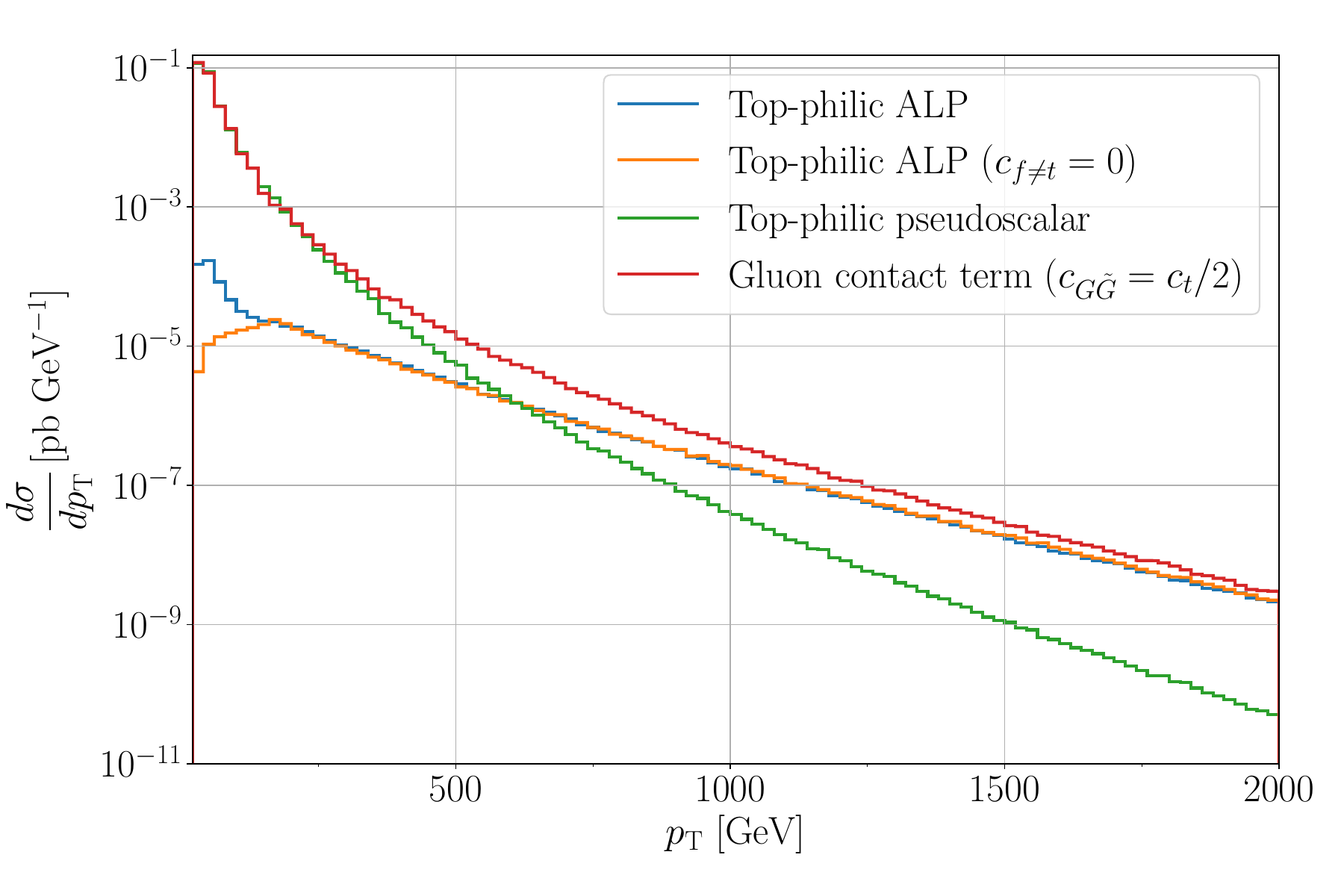}
    \caption{Transverse-momentum distribution of  $pp\TO a +j$ production for several set-ups (see text for further details): $m_a = 10~\gev$,  $p_{\rm T}(a)>30 ~\gev$, $\eta(a)<4.0$. }
    \label{fig:ptaj}
\end{figure}

In order to further investigate the difference between an ALP and a pseudoscalar, in Fig.~\ref{fig:ptaj} we plot the $p_{T}$ spectrum of the jet, or equivalently the ALP,  in $pp\TO a+j$ production for the case of $m_{a}=10~\gev$ and $f_{a}/c_{t}=1~\tev$. The blue line depicts the equivalent of what is shown in Fig.~\ref{fig:production}, which is our reference prediction for a top-philic ALP, while the orange line plots the same quantity but neglecting the contribution of $c_{f}$ with $f\ne t$ (also in the $\tilde c_{GG}$ definition). For the case $c_{f\ne t}=0$, we also plot in green the contribution of only loop-diagrams like the left-hand one in Fig.~\ref{fig:qqag} and in red the contribution of only tree-level diagrams like the right-hand one. We point out several interesting aspects of this plot below.

First of all, the sum of the green and red lines is very different from the orange line because there are (extremely) large cancellations at the level of the amplitude between the two classes of diagrams in Fig.~\ref{fig:qqag}. 
Therefore, at the level of squared amplitude the interferences between them can dominate in absolute value.  Moreover, the lines can be interpreted from different points of view. On the one hand, the green and red lines can be viewed as two different contributions to the top-philic ALP prediction in the non-derivative basis. On the other hand, the green line can be seen as the prediction of a {\it pseudoscalar}, to be compared with the one of the top-philic ALP (orange or blue lines) or with the prediction of a different kind of ALP that is dominated by $c_{GG}$ already at  tree level, with a value of $c_{GG} = c_t/2$ (red line). 

At low $p_{T}$, the prediction for the {\it pseudoscalar} and the  tree-level diagrams involving  
$\tilde c_{GG}$ are almost equal. The amplitudes have opposite sign, leading to the orange line for the top-philic ALP, with $c_{f\ne t}=0$, being several orders of magnitude smaller. In this region the difference with the case that includes contributions from $c_{f\ne t}$ (blue line) is very large, with the blue line reaching almost one order of magnitude more than the orange one. In Fig.~\ref{fig:ptaj}, the inclusive rate of the blue line is more than twice that of the orange line. This effect is induced by  $c_{f\ne t}$ which, as previously mentioned, depends significantly on the choice of $\Lambda$ and in turn on $f_{a}$. Indeed, although we do not see delicate cancellations as in the $a\TO gg$ amplitude, for small $p_{T}$ we do observe an increase of up to 500\% of our predictions when we set $\Lambda= 10 \tev$. In fact, these effects are even more significant for larger values of $m_{a}$.

However, regardless of the value of $m_{a}$ in the $10~{\rm GeV}\lesssim m_{a}\lesssim 200~{\rm GeV}$ range, the picture is completely different in the opposite regime of high $p_{T}$. There, the prediction for the top-philic ALP is almost equivalent to the contribution of the purely tree-level diagram and therefore approximately proportional to $\tilde c_{GG}^{2}\propto (c_{t}+c_{b})^{2}$. 
A change of scale $\Lambda$ from 1 to 10 TeV only has an impact of $\sim 10\%$ in this kinematic regime.\footnote{It is important to note that even exploring very large energies the correct scale for the $c_{f}$ is $m_{t}$ and the dependence on $\Lambda$ scales as in Eq.~\eqref{eq:cthird}. Indeed the relevant diagrams have a topology as the one in Fig.~\ref{fig:Diagramsep} plus a gluon emitted from the $f=q$ loop. The $af\bar f$ coupling is therefore generated at the scale $m_{a}$, and the running from $\Lambda$ stops at $m_{t}$ since we are assuming $\Lambda>m_{t}\gtrsim m_{a}$.} We have verified that this estimate already works very well from $p_{T}\gtrsim200~\gev$. We stress that this value  is the minimum $p_{T}$ considered in the analyses of Sec.~\ref{sec:DM}.\footnote{Note that while the orange and blue lines are almost one on top of the other  at  very high $p_{T}$, since the effect from $c_{b}$ is invisible on a logarithmic scale, the orange and the red are clearly different, with the latter being circa 50\% larger. Indeed, the interference of loop and tree-level diagrams is not negligible, and this behaviour is present  also in the case  of the blue line. In conclusion, the dependence of $\tilde c_{GG}^{2}$ on $\Lambda$ is indicative of the size of the dependence on $\Lambda$, but it does not give the exact result for the prediction of the cross-section for $a+j$ production, which we have instead verified by explicit computation to be of the same order.} In conclusion, while at low $p_{T}$ and therefore for the total normalisation a non-negligible dependence on $\Lambda$ is present, at high $p_{T}$ and for our study it is completely negligible.

Looking at the entire spectrum, the comparison between the green and orange lines indicates that a pseudoscalar and a top-philic ALP have completely different phenomenology. In particular, a derivative coupling with the top quark leads to much larger cross sections at high $p_{T}$. Being associated to a dimension-five operator, this coupling leads to a higher growth in energy w.r.t.~SM backgrounds and it is therefore a prime candidate for probing $f_{a}/c_{t}$. 

Finally, in Fig.~\ref{fig:production} we also plot the cross sections of the associated production of the top-philic ALP with a $c$ or $b$ jet.
In both cases, the cross section is much smaller than in the light-jet case, primarily because one of the PDFs in the initial state must be that of either a $c$ or $b$ quark, which are much smaller than those of valence quarks or gluons. We plot results for a $\eta(a)<2.5$ cut, but we have checked that the same arguments apply to the looser case of $\eta(a)<4$ used for the light jet computations. 
The main conclusion is that this process, $a+j_{h}$ is too suppressed to obtain reasonable bounds on $c_{t}$. On the other hand, this also means that the prediction for $a+j$ with $j$ light can be safely used for experimental signatures that do not distinguish between light and heavy flavour jets. In the rest of the paper, we will therefore refer to $a+j$ assuming that the contribution from heavy partons is understood.

\subsection{LHC constraints from existing BSM searches}
\label{sec:existing}

\noindent
In this subsection we will 
delineate the parameter space in the 
$m_a$ {\it vs.} $f_a/c_t$ plane that is left unconstrained by  existing LHC searches for BSM physics, for $10~\gev < m_{a}<200~\gev$. In order to do so, we exploit the results discussed in Secs.~\ref{sec:BRs} and \ref{sec:xsALP} for reinterpreting current bounds that have been set in experimental analyses. All of the bounds will refer to $c_{t}(\Lambda)$, meaning the value at the UV scale, and we will simply refer to it as $c_{t}$, understanding that the scale is set to $(\Lambda)$.

\begin{figure}
   \centering
    \includegraphics[width=0.85\textwidth]{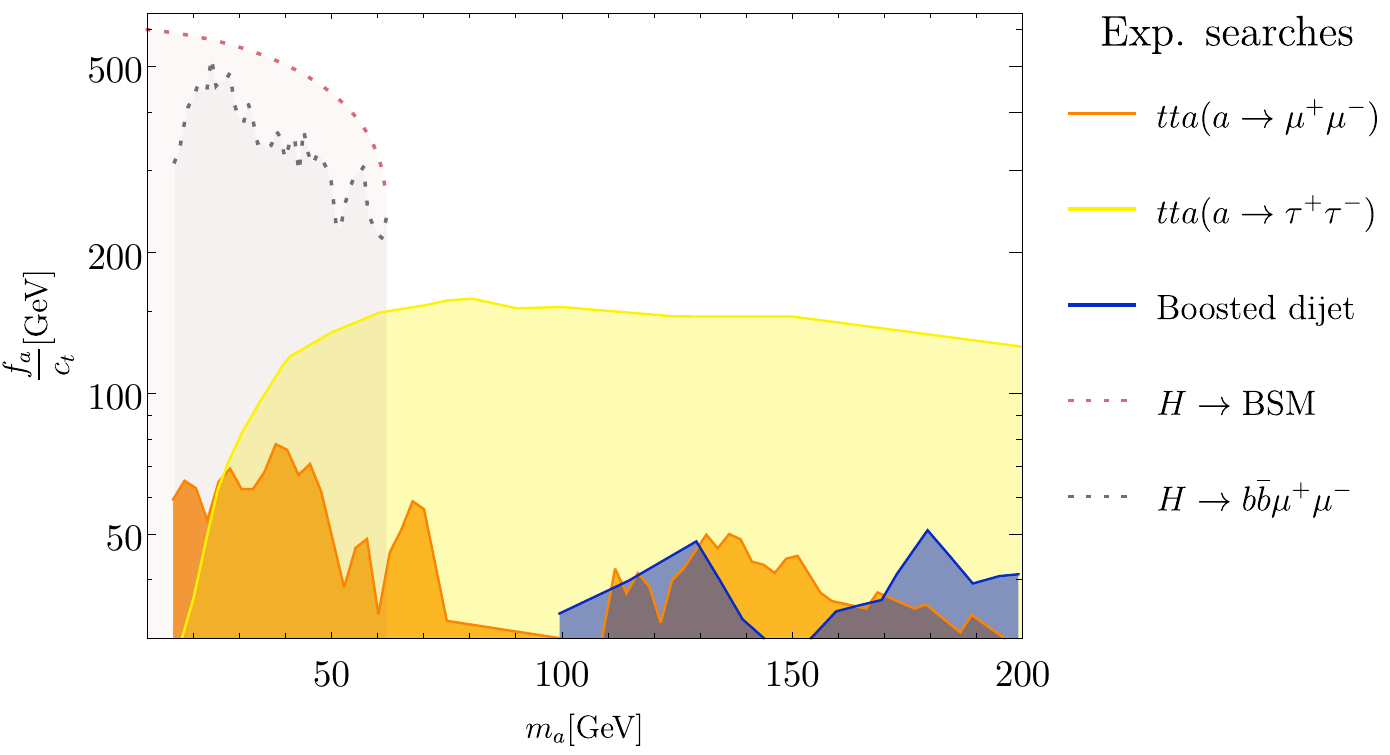}
    \caption{
   Bounds on ${f_a}/{c_t}$ obtained using LHC constraints from 
   existing searches targeting BSM states, which can be re-interpreted for our top-philic ALP model (see text for details).
     }
    \label{fig:constraints} 
        
   \end{figure}

We present a 
summary of our findings in Fig.~\ref{fig:constraints}, followed by further details of the collider searches considered and the reinterpretation in terms of a top-philic ALP. 
In Fig.~\ref{fig:constraints} we display constraints derived from experimental analyses targeting the following processes:
\begin{itemize}
\item Top-quark pair production in association with a lepton pair ($t \bar t \ell^{+} \ell^{-}$), with  $\ell^{+} \ell^{-}=\mu^+\mu^{-}$ or $\tau^+\tau^{-}$.
\item Boosted dijet production.
\item The Higgs decays channel $H\TO b \bar b \mu^{+} \mu^{-}$.
\item The Higgs decay into BSM, {\it i.e.}, the unobserved branching fraction.
\end{itemize}
Further processes have been considered but they all lead to much weaker constraints. They are discussed near the end of this section.

As we can see in Fig.~\ref{fig:constraints}, the combination of a moderate production cross section at the LHC (see Fig.~\ref{fig:production}), together with a suppressed 
branching ratio into clean final states such as diphoton  (see Fig.~\ref{fig:BRplot}),
leads to weak constraints on the decay constant of the top-philic ALP in the considered mass range.
For $m_a > m_H/2$, in the range considered,  the most stringent existing constraints are derived from searches targeting new pseudo-scalars produced in association with $t \bar t$ pairs in $t \bar t \ell^{+} \ell^{-}$. On the contrary, for $m_a < m_H/2$, the Higgs decays lead to the strongest constraints. However, there are some caveats, explained in detail later in the section, in the interpretation of the Higgs constraints in the context of a top-philic ALP. For  this reason, we have displayed the associated constraints as dashed lines in Fig.~\ref{fig:constraints}.

\paragraph{Constraints from $t \bar t \ell^+\ell^-$}
In Ref.~\cite{CMS:2022arx} the CMS collaboration searched for the associated production of 
$t \bar t$ pair with
a light pseudoscalar decaying into leptons.
As discussed in Sec.~\ref{sec:xsALP}, at LO the $pp \to a t \bar t$ production channel is independent of whether the derivative or the non-derivative version of the ALP-top coupling is used. In other words, the $pp \to a t \bar t$ prediction for a pseudoscalar coupling to the top with a Yukawa-like interaction is identical to that of a derivatively-coupled, top-philic ALP.
The experimental collaboration provides 
a 95\% CL upper limit on the cross section times branching ratio of such a topology, as a function of the pseudoscalar mass.
Hence we can directly apply their cross-section times branching ratio limits to our model, and we obtain the lines displayed in Fig.~\ref{fig:constraints} for the 
$a \to \mu^+ \mu^-$ and $a \to \tau^+ \tau^-$
decay modes.

\paragraph{Constraints from boosted dijet searches}
Given the large branching ratio of the top-philic ALP into bottom pairs, boosted dijet searches can be important probes of our model.
The ATLAS search 
\cite{ATLAS:2018hbc}
targets 
light boosted resonance in the dijet final state
in the mass range 100 to 220 GeV,
and provides an upper limit on the fiducial cross-section of such resonance.
Here we provide a crude estimate of the constraining power of this ATLAS analysis, without performing a detailed recasting.
The basic selection cut in \cite{ATLAS:2018hbc} requires a leading jet of $p_{\rm T} > 420$ GeV.
Since the (hadronic) decay of the top-philic ALP is dominated by $a\TO b \bar b$ (see Sec.~\ref{sec:xsALP}) we combine the prediction for $pp\TO a+j$, imposing the aforementioned cut on the jet, together with ${\rm BR}(a\TO b \bar b)$ in order to obtain the production cross section for 
$p p \to a +j, (a\TO b \bar b)$ in the tophilic-ALP model. Assuming one of the two jets is a  fat jet stemming from the ALP decay,  we use our prediction as an estimate for the fiducial cross section in our model and we reinterpret the limit reported by the experimental collaboration. With this procedure we obtain the 
constraint reported in blue in Fig.~\ref{fig:constraints}.

\paragraph{Constraints from Higgs decays}
The top-ALP coupling also induces a Higgs decay into a pair of ALPs or an ALP and a $Z$ boson, computed in \cite{Bauer:2017ris}. 
These exotic Higgs decays can lead to constraints on our model, both in terms of a generic unobserved branching fraction ($H \to \text{BSM}$)  \cite{ATLAS:2022vkf,CMS:2022dwd},
as well as a direct search for $H \to a a \to b \bar b \mu^+ \mu^-$ \cite{ATLAS:2021hbr}, which are displayed in Fig.~\ref{fig:constraints}. The latter combines both the benefits of a large BR for the ALP ($b \bar b$) and a clean experimental signature ($\mu^{+}\mu^{-}$). 

Regarding $H \to \text{BSM}$, if we assume that these exotic Higgs decays are the only modification of Higgs properties that are induced at this order, we can use the global signal-strength measurements to constrain $f_a/c_t$, since they scale like $1-\mathrm{BR}(H\to\mathrm{BSM})$. We use to Feldman-Cousins prescription~\cite{Feldman:1997qc} to account for the fact that unobserved branching fractions can only reduce the global signal strength, obtaining an upper limit of $\mathrm{BR}(H\to\mathrm{BSM})< 0.073$  from the combined ATLAS global signal strength measurement~\cite{ATLAS:2022vkf}.\footnote{A slightly weaker bound of 0.11 was obtained when using the measured value by CMS~\cite{CMS:2022dwd} since the ATLAS central value is more incompatible with a reduced signal strength.} We also 
verified that $H\to Za$ does not lead to significant constraints on our parameter space, nevertheless we have taken it into account when computing the $\mathrm{BR}(H \to\mathrm{BSM})$.

As already anticipated, there are some subtleties related to  the interpretation of Higgs decay in the context of the (top-philic) ALP; they are related to a dimension-six operator, therefore an effect one order higher, in the $1/f_{a}$ expansion, than the one considered in our theoretical framework discussed in Sec.~\ref{sec:topphilicALP}.

First of all, we can notice that the $H\to aa$ process arises at one loop with two insertions of $a/f_a$. 
However, in general it can also be sourced at the same $1/f_a^2$ order at tree level by the dimension-six interaction  stemming from the Lagrangian 
\begin{align}
    \mathcal{L}^{(6)}= \frac{c^{(6)}_{a H}}{f_{a}^{2}}\phi^\dagger\phi\, \partial_\mu a \partial^\mu a\equiv  \frac{c^{(6)}_{a H}}{f_{a}^{2}} \mathcal{O}^{(6)}_{a H}\,,
\end{align}
where $\phi$ is the SM Higgs doublet. The  $ \mathcal{O}^{(6)}_{a H}$ operator is  actually the sole dimension-six operator in the ALP EFT~\cite{Bauer:2017ris,Grojean:2023tsd}, but the double insertion of $c_t/f_a$ in the one-loop amplitude of $H\TO aa$, which constitutes the Born amplitude in our top-philic model (Eq.~\eqref{eq:top-philic}), mixes  with $c^{(6)}_{a H}$ via  renormalisation group (RG) evolution. This is reflected in the fact that the $c_t^2$ contribution to the $H\to aa$ partial width retains both finite and renormalisation scale ($\mu_R$) dependent pieces, the latter being associated to divergences absorbed by an $\mathcal{O}_{a H}$ counterterm. Setting $\mu_R=\Lambda$, captures the leading logarithmic contribution of $c_t^2$ to $c^{(6)}_{a H}$ when running from $\Lambda$ to $m_t$, alongside the finite piece, which then feed in to the prediction for $H\to aa$.
Therefore, neglecting the tree-level  contribution $\mathcal{O}^{(6)}_{a H}$ as we did, means that the limits that we present are obtained under the assumption that $c^{(6)}_{a H}$ vanishes at the UV matching scale $\Lambda$, \emph{i.e.}, that it is not generated by the underlying UV model. For this reason, as already mentioned, we displayed in Fig.~\ref{fig:constraints} the corresponding bounds as dashed lines, since  in principle they are model-dependent and should be interpreted with care. In  Appendix~\ref{app:UV-c6} we show that, whenever a portal exists between the complex scalar field containing the ALP degree of freedom and the SM Higgs, tree-level contributions to  $\mathcal{O}^{(6)}_{a H}$ are possible.

\paragraph{Constraints from processes with less sensitivity}
We have investigated other collider searches that could potentially constrain the top-philic ALP model. They are less constraining than those presented in Fig.~\ref{fig:constraints} and hence not shown therein. 
In particular we explored 
resonant diphoton signals \cite{Mariotti:2017vtv},
specifically in the boosted~\cite{ATLAS:2022abz} and resolved  channels~\cite{ATLAS:2021uiz,ATLAS:2023jzc}.
Diphoton resonance searches have been shown to generically constrain ALPs in this mass window~\cite{Mariotti:2017vtv,CidVidal:2018blh},
but they are not effective in our model. Indeed,  comparing  our model against typical simplified models with 
non-vanishing tree-level $c_{GG}$, the production cross section for $pp\TO a$ combined with ${\rm BR}(a\TO\gamma\gamma)$ is very suppressed. We stress again that this statement does not depend on the exact location of neither the minimum of ${\rm BR}(a\TO\gamma\gamma)$ in  Fig.~\ref{fig:BRplot} nor the one of the cross section for $pp\TO a$ in Fig.~\ref{fig:production}.
We have also verified that, for analogous motivations,  boosted resonant searches in the $b \bar b$ final state~\cite{CMS:2018pwl}
do not provide any relevant limit,
as well as resonant searches in di-$\tau$ final states~\cite{CMS:2022goy}. 
Finally we have also inspected LEP bounds and interpreted them in terms of
$Z \to \gamma a(a\TO jj)$~\cite{L3:1992kcg},
$Z\to \gamma a(a\TO\tau \tau)$~\cite{OPAL:1991acn}
and $Z \to \text{BSM}$~\cite{ALEPH:2005ab}
(see \cite{Bauer:2017ris}), all of which lead to weaker limits than those shown in Fig.~\ref{fig:constraints}.

\paragraph{Electroweak precision
constraints}

A top-philic ALP together with the heavy states typically present in any UV completion might induce modifications to electroweak precision observables at low energy, possibly leading to relevant further constraints. 
These indirect bounds depend on the details (charges, couplings and masses) of the particles in the UV completion of the model and can be computed as loop corrections involving the heavy states. Even though for a top-philic ALP, these effects are truly indirect, in the sense that since neither the ALP nor the top quark are directly observed, the link with the UV physics might be difficult to establish, such constraints are potentially relevant and should be considered. 
At low energy, effects of heavy states can be described by dimension-six operators, as obtained by the RG running, including mixing between the ALP EFT and the Standard Model EFT (SMEFT), with  Wilson coefficients matched to the UV model.   More generally, one can estimate the $\mathcal{O}(1/f_a^2)$ contributions to the evolution of SMEFT Wilson coefficients using RG methods, as proposed in Ref.~\cite{Galda:2021hbr}, and then exploit information available from  global SMEFT fits to indirectly constrain ALP couplings generated at the scale $\Lambda$. Such a method was employed in Ref.~\cite{Biekotter:2023mpd}, where the current sensitivity of precision electroweak observables, notably the $W$ boson mass and the $\rho$ parameter, to $c_t$ values has been estimated to be of order one, roughly independently of the ALP mass (with $f_a= 1$ TeV and $\Lambda=4\pi f_a$). The mixing was obtained by including only approximate two-loop effects in resummed RG evolution. Taking $\Lambda=1$ TeV in line with the rest of our calculations and $\mu\sim m_W$, this corresponds to $f_a/c_t\gtrsim500$ GeV, {\it i.e.}, competitive with the bounds presented here. While not conclusive \emph{per se}, as similarly to the constraints from mixing with $\mathcal{O}^{(6)}_{a H}$ the bounds come with the theoretical assumption that no dimension-six operators are generated in the UV, this analysis certainly motivates further studies.

\medskip
\section{New probes of the top-philic ALP at the LHC}
\label{sec:newprobes}

Considering our top-philic ALP model, the compilation of bounds reported in the previous section have been obtained via a reinterpretation into the $(m_a,\, f_a/c_t)$ space of upper limits provided by experimental collaborations for specific searches,  \emph{e.g.},  resonances produced in association with top quarks but also unobserved Higgs decays. Instead, in this section we derive new bounds by investigating the effect of the top-philic ALP on  cross-section measurements performed by the ATLAS and CMS collaborations for SM processes involving top-quark final states, in particular:
\begin{itemize}
\item Top-quark pair production in association with  a bottom-quark pair, $t \bar t b \bar b$.
\item Four-top production, $t \bar t t \bar t$.
\item Top-quark pair production, $t \bar t$.
\end{itemize} 
Since no useful upper limit that can be reinterpreted for our analysis is provided in those measurements, the extraction of the bounds in the $(m_a,\, f_a/c_t)$ space relies not only on our theoretical predictions but also on a statistical approach  for comparing the top-philic ALP with LHC datasets. 

We begin by summarising and discussing our findings in Sec.~\ref{sec:summary-indirect}. We then describe our statistical approach  (Sec.~\ref{subsec:stats}) and how we have derived the exclusion limits from $t \bar t  b \bar b$ (Sec.~\ref{subsec:ttbb}), $t \bar t t \bar t$ (Sec.~\ref{subsec:fourtop}), and $  t \bar t$ (Sec.~\ref{subsec:ttbar}). We explain the computation of the top-philic ALP predictions for these processes, which do not feature the ALP in the final state, listing the specific input data sets and discussing how our calculations have been exploited in order to obtain exclusion limits in the $(m_a,\, f_a/c_t)$ space. We also comment on the potential impact of neglected dimension-6 SMEFT effects on our bounds, when generated either in the UV or through RG evolution.

\subsection{Summary of our findings}
\label{sec:summary-indirect}

\begin{figure}[!t]
    \centering
    \includegraphics[width=0.80\textwidth]{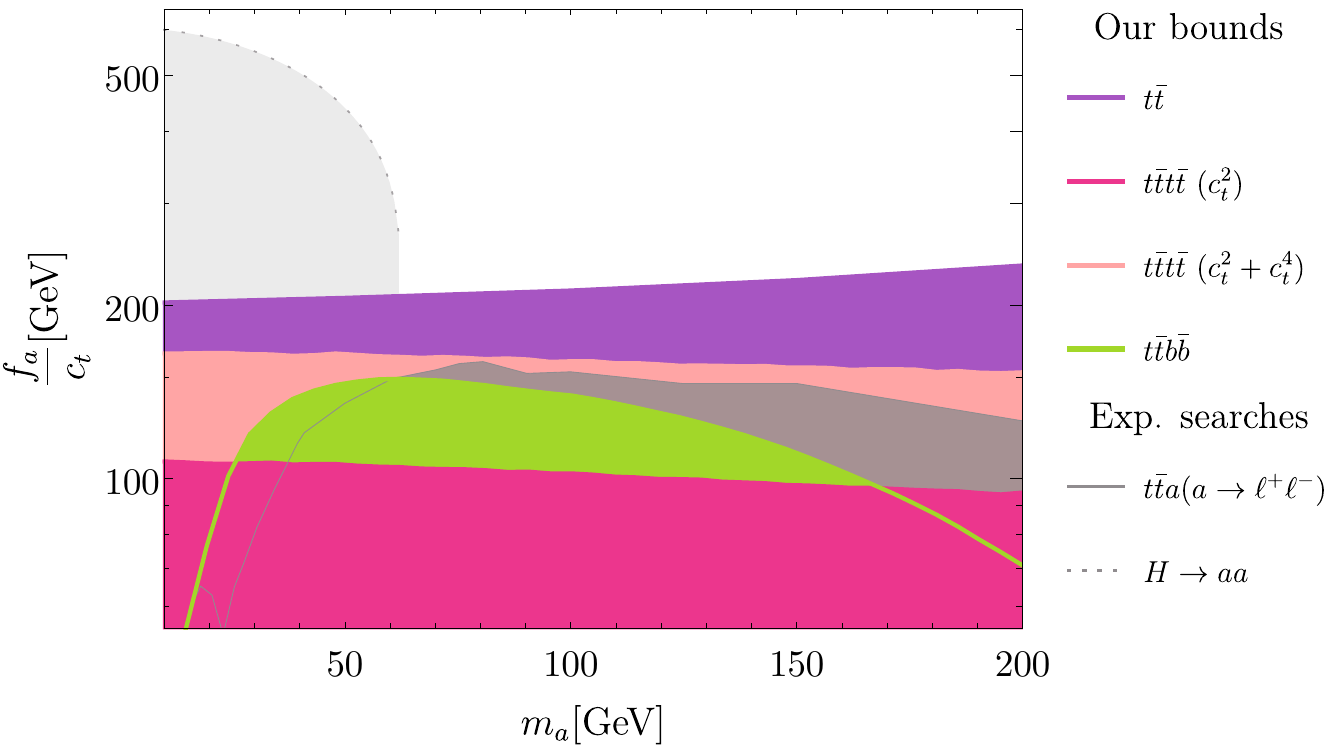}
    \caption{   
Summary plot of exclusion limits obtained via our statistical analysis for $t \bar t b \bar b$, $t \bar t t \bar t $ and $t \bar t $. The best existing bounds from $t \bar t \ell^+\ell^-$ and $H\TO aa$ already displayed in Fig.~\ref{fig:constraints} are shown here for comparison.}
   \label{fig:FullVisible}
\end{figure}

In Fig.~\ref{fig:FullVisible} we show the limits that we can set in the $(m_a,\, f_a/c_t)$ space for a top-philic ALP with $10~\gev < m_{a}<200~\gev$, via the measurement of $t \bar t  b \bar b$, $t \bar tt \bar t$ and $t \bar t$ cross sections. For all processes, the ALP is present only as a virtual particle in internal propagators, and in the case of the $t \bar t$ process it appears only via loop corrections. The first feature that should be emphasised for Fig.~\ref{fig:FullVisible} is that the bound originating from  $t \bar t$ production is much stronger than the best one  in Fig.~\ref{fig:constraints}, from $t\bar t \ell^{+ } \ell^{- }$, in the range  
 $m_{a}>m_{H}/2$ (reproduced by the solid grey area in Fig.~\ref{fig:FullVisible}). Top-philic-ALP effects from loops in $t \bar t$ production have already been investigated in Ref.~\cite{Esser:2023fdo}, but here, as documented in Sec.~\ref{subsec:ttbar}, we provide the first exact calculation of loop-induced effect of order $c_{t}^{2}/f_{a}^{2}$ for this process.  Ignoring the bounds from Higgs decays, which may be in principle affected by the dimension-six operator discussed in Sec.~\ref{sec:existing} and Appendix \ref{app:UV-c6},  the bound from $t \bar t$ production is also the strongest in the range $m_{a}<m_{H}/2$. Over the full $10~\gev < m_{a}<200~\gev$ range an almost flat exclusion limit of order
 \be
 \frac{f_{a}}{c_{t}}\gtrsim 200~\gev\, , \label{eq:bound200g}
 \ee
 can be set.
 We therefore encourage experimental collaborations to perform dedicated analysis in this direction. In Sec.~\ref{subsec:ttbar} we also suggest strategies to obtain the best information from data.
 
 Next to $t \bar t$ production, we have $t \bar t t \bar t$ production. The details about the difference between the approximations called ``$c_{t}^{2}$'' and ``$c_{t}^{2}+c_{t}^{4}$'' are described  in Sec.~\ref{subsec:fourtop}. Here we just want to say that they are both tree-level results featuring the ALP in internal propagators. As the name suggests, in addition to effects of order $c_{t}^{2}/f_{a}^{2}$, the latter approximation also takes into account effects of order  $c_{t}^{4}/f_{a}^{4}$,  which leads to more stringent bounds: $f_a/c_t\gtrsim 100~\gev$ for $c_{t}^{2}$ and $f_a/c_t\gtrsim 160~\gev$ for $c_{t}^{2}+c_{t}^{4}$.
 
Finally, we have  $t \bar t b \bar b$ production, mostly driven by resonant  $t \bar t a (a\TO b \bar b)$ production. While the exclusion limit from $t \bar t t \bar t$ production is quite flat, the one from $t \bar t a (a\TO b \bar b)$ production is not. In fact, for very small or large values in the  $10~{\rm GeV}\lesssim m_{a}\lesssim 200~{\rm GeV}$,  $t \bar t a (a\TO b \bar b)$ returns much less stringent exclusion limits than $t \bar t t \bar t$, while in the central region are only slightly less stringent.

Overall, we see that precision measurements yield comparable or better bounds on the top-philic ALP than existing resonance searches. 
That said, such constraints are still rather mild and therefore mainly apply to the case of a relatively strongly coupled ALP.  This shows how elusive this type of particle is and motivates further study into constraining top-philic ALPs, by either employing an even more global approach, or identifying new channels through which they can be constrained.
The interested reader can find many more details in the next Secs.~\ref{subsec:stats}--\ref{subsec:ttbar}. We recall here that the case of a top-philic ALP decaying predominantly into invisible DM and therefore leading different phenomenology is investigated in Sec.~\ref{sec:DM}.

Given the relatively small bound on $f_a/c_t$ from Eq.~\eqref{eq:bound200g}, one may wonder if the EFT approach is still valid, \emph{i.e.}, whether the energies probed are too high compared to the allowed scale of NP. This question has been addressed in detail in \cite{Esser:2023fdo} and we briefly recap the main argument here.  The validity of the EFT requires
$    f_a>Q(b)$
where $Q(b)$ is the typical scale of the process from which a bound $f_a/c_t<b$ is extracted.
Whenever $b>Q(b)$ we are always in the range of validity of the EFT for couplings of order one or larger. Our bounds in fact correspond to the opposite region, in which $b\lesssim Q(b)$. However, the main point to keep in mind here is that we never bound $f_a$ alone but always its ratio with $c_t$.
The validity of the EFT is then only preserved for relatively large values of $c_t\gtrsim1$, which cannot realistically exceed $\mathcal{O}(4\pi)$. Considering the above, we can summarise the validity criterion by the following condition:
\begin{equation}
\left\{\begin{split}
f_a &\ge Q(b) \\
\frac{f_a}{c_t}&\ge b \\ 
\end{split}\right.\quad\longrightarrow\quad
4\pi>c_t\ge \frac{Q(b)}{b}\,,    
\end{equation}
which is always satisfied for the analyses performed in this work.

\subsection{Statistical approach\label{subsec:stats}}

In this section we describe the common statistical approach used in deriving the bounds reported in Secs.~\ref{subsec:ttbb}--\ref{subsec:ttbar}. 
 In essence, we take as input a 
 selection of published measurements
and our theoretical predictions for the observables in the top-philic ALP model. These are combined to build a likelihood as a function of $(m_a,\, c_t/f_a)$, which is used to derive confidence intervals in the parameter space.

It is important to note that this approach allows us to combine multiple measurements from different experimental signatures
 to maximise sensitivity to the top-philic ALP.
For a dataset comprising one or more bins, that could come from multiple inclusive or differential measurements, our general strategy is as follows. We first ensure that no statistically overlapping measurements are used, since the correlations among these are rarely reported. For each bin, we construct a signal-strength, $\mu^i_{\mathrm{obs.}}$, by dividing the measured value by the corresponding SM prediction, usually quoted in the experimental paper. Correlations between bins are taken into account when they are published by the experimental collaborations. Typically, bin-by-bin covariances are reported for differential measurements but no information about inter-analysis correlations is available.
We construct a corresponding theoretical prediction, $\mu^i(m_a, c_t/f_a)$ for the top-philic ALP in each data bin by dividing our computed LO, ALP-mediated corrections by the SM prediction at the same order. This approach assumes that any SM higher-order corrections factorise from the ALP-mediated effects, since we are comparing ALP predictions relative to the SM with observed signal strengths that are typically defined  via SM model predictions involving (several) higher-order effects.
We can then construct a $\chi^2$ function
\begin{align} \label{eq:chi}
    \chi^2(m_a,\, c_t/f_a) = 
    \big(\vec{\mu}(m_a,\,c_t/f_a) -\vec{\mu}_\mathrm{obs.}\big)\cdot\mathbf{V}^{-1}\cdot\big(\vec{\mu}(m_a,\,c_t/f_a) - \vec{\mu}_\mathrm{obs.}\big),
\end{align}
where the vector dimension extends over the set of data bins and $\mathbf{V}$ is the experimental covariance matrix, accounting for statistical, systematic and theoretical uncertainties. For each ALP mass, we minimize the $\chi^2$ to find the best fit $c_t/f_a$ value, and identify  95\% confidence-level intervals via the requirement:
\begin{align}
    \Delta\chi^2(m_a,\, c_t/f_a)\equiv\chi^2(m_a,\, c_t/f_a) - \chi^2_{\mathrm{min.}}\leq 3.84\,.
\end{align}

\subsection{Top-philic ALP in $t\bar tb \bar  b$\label{subsec:ttbb}}
Since the top-philic ALP decays dominantly into $b\bar{b}$ pairs (see Sec.~\ref{sec:BRs}), the majority of ALPs produced in association with $t\bar{t}$ at the LHC will result in the $t\bar{t}b\bar{b}$ final state. To date, this process has been observed by the CMS and ATLAS experiments in a number of channels in the 13 TeV LHC dataset
\cite{ATLAS:2018fwl,CMS:2019eih,CMS:2020grm,DHondt:2023dlt}. We estimate the corresponding bounds on $f_a/c_t$ by constraining the signal contribution of a resonantly produced ALP in association with a pair of top quarks subsequently decaying into a $b\bar{b}$ pair at LO.

 Our simulation is performed at LO accuracy and in the narrow width approximation (NWA) for the top-philic ALP. For $10~\gev < m_{a}<200~\gev$, there are many other topologies through which an intermediate ALP can yield the same final state, including $t$-channel like diagrams and $s$-channel diagrams where an off-shell ALP splits into $t\bar{t}$. We checked that, being non-resonant, these contributions as well as their interference with the main signal and background are negligible compared to the on-shell $pp\to t\bar{t}a(a\to b\bar{b})$ production and we henceforth neglect them. The same applies to possible dimension-six operator contributions to the same final state that we have not considered.\footnote{ 
For example, dimension-six $b\bar{b}t\bar{t}$ or $q\bar{q}t\bar{t}$ contact interactions could contribute to this process. As with the $\mathcal{O}_{a H}$ operator discussed in Sec.~\ref{sec:existing}, we assume that such operators are not generated in the UV. However, the associated Wilson coefficients could obtain non-zero values through renormalisation group evolution~\cite{Galda:2021hbr}, albeit suppressed by a loop factor and the $b$  or light quark Yukawa coupling, respectively. The relatively small value of such coefficients, coupled with the non-resonant nature of the signature leads us to conclude that they can safely be neglected for the purposes of this exercise.
} Moreover, since all of the ALP decay modes are suppressed by at least one loop factor, we have confirmed that the ALP remains very narrow over the entire $(m_a,\,f_a/c_t)$ parameter range of interest, justifying our use of the narrow width approximation.

We subject our signal events to a minimal set of parton-level cuts associated to a final state with two $b$-jets  in the LHC experiments. We require each $b$ quark to satisfy transverse momentum and pseudorapidity cuts of $p_{\rm T}>$ 20 GeV and $|\eta|<$ 2.5, respectively, and additionally require a minimum angular distance between the $b\bar{b}$ pair of $\Delta R_{b\bar{b}} = \sqrt{\Delta\eta_{b\bar{b}}+\Delta\phi_{b\bar{b}}}>0.4$.
The $\Delta R_{b\bar{b}}$ cut  is necessary in order to avoid events that would be seen as a single $b$-jet after jet clustering. It has a significant impact on the signal efficiency for lighter ALP masses, where the ALP is produced with significant boost, leading to a collimation of its decay products. In particular, the signal efficiency is about 75\% for $m_a=100$ GeV, and becomes negligibly small for our lightest mass point of $m_a=10$ GeV. The nature of $b$-tagged final states therefore restricts the sensitivity to ALP masses to the intermediate range above a few tens of GeV. In principle, with a more advanced analysis, it would be possible to consider events where the  $b\bar b$ are collimated and lead to a single $b$-jet and gain sensitivity for lighter ALPs. More comments are given at the end of this section. 

Our signal prediction is therefore simulated in NWA as a function of $m_a$ via the inclusive $t\bar{t}a$ production rate shown in Fig.~\ref{fig:production}, the $b\bar{b}$ branching ratio shown in Fig.~\ref{fig:BRplot} and the aforementioned selection efficiency. In order to confront our ALP signal with the experimental measurements of $t\bar{t}b\bar{b}$, we divide it by a LO prediction interfaced with parton shower of 1.562 pb~\cite{Jezo:2018yaf} to define the signal strength parameter, $\mu_{t\bar{t}b\bar{b}}(m_a,\, f_a/c_t)$.\footnote{In this case, we are actually not assuming that all higher-order SM corrections factorise from the ALP-mediated effects. In this process, parton-shower effects alone induce 30\% corrections on to the fixed, LO predictions. In doing so, we are therefore conservative. }
This prediction is then confronted with a combination of four statistically independent $t\bar{t}b\bar{b}$ cross section measurements unfolded to the full phase space from the LHC~\cite{DHondt:2023dlt}: two by each of the ATLAS~\cite{ATLAS:2018fwl} and CMS~\cite{CMS:2020grm}. 
The ATLAS measurements used are in the lepton+jets and the $e\mu$ dilepton channels, where exactly four $b$-jets are required, while the CMS measurements correspond to the dilepton and lepton+jets and hadronic channels.\footnote{At this level, we deem it sufficient treat all measurements as statistically independent determinations of the $t\bar{t}b\bar{b}$ signal strength, neglecting possible kinematic acceptance differences between analyses. We nevertheless do not consider signal regions that select less than 4 $b$-jets since the exact signal acceptance for such a requirement would require detailed simulations beyond the scope of this work.} We constructed corresponding signal strengths by normalising each measurement to the nominal theory prediction used by the analysis, as summarised in Tab.~\ref{tab:ttbb_data},  where theory errors were added in quadrature to the overall systematic uncertainty.
\begin{table}[t!]
\centering
\begin{small}
\begin{tabular}{|l|l|l|l|l|}
\hline
Exp.  & Channel               & $\mu_{t\bar{t}b\bar{b}}~\pm$ stat. $\pm$ syst.  & Ref.\tabularnewline
\hline
CMS   & dilepton              & $1.36\pm 0.10\pm 0.34$            & \cite{CMS:2020grm}\tabularnewline
CMS   & lepton+jets           & $1.26\pm 0.04\pm 0.31$            & \cite{CMS:2020grm}\tabularnewline
ATLAS & dilepton ($e\mu$, 4b) & $1.75\pm 0.05\pm 0.56$             & \cite{ATLAS:2018fwl}\tabularnewline
ATLAS & lepton+jets (4b)      & $1.57\pm 0.09\pm 0.49$            & \cite{ATLAS:2018fwl}\tabularnewline
\hline
\end{tabular}   
\end{small}
\caption{\label{tab:ttbb_data}
Signal strengths for the full phase space $t\bar{t}b\bar{b}$ cross section measurements used to constrain the ($m_a,\, f_{a}/c_t$) plane of the top-philic ALP. The nominal theory prediction used to normalise the measured cross section  can be found in the corresponding experimental reference.}
\end{table}

The measurements are clearly systematically dominated, driven entirely by modelling uncertainties and $b$-tagging systematics. One can also note that there is a systematic $\sim1\sigma$ excess across the data. We obtain the ensuing bounds on $(m_a,\, f_a/c_t)$ by combining the measurements in a $\chi^2$ analysis. Since the experiments do not report correlations between the different measurements, we take the conservative approach of assuming fully correlated systematic uncertainties, especially given their primarily theoretical origin. The bounds obtained are shown in the summary plot of~Fig.~\ref{fig:FullVisible}  with a green filled area, starting at $f_a/c_t=0$ for $m_a=10$ GeV, peaking at 150 GeV for $m_a\sim 60$ GeV and slowly decreasing to $f_a/c_t=70$ GeV for $m_a=200$ GeV. As explained earlier, the low mass sensitivity is spoiled by the efficiency of the $b$-jet selection, while at high masses, the $t\bar{t}a$ production cross section and the $b\bar{b}$ branching fraction both decrease, leading to weakened bounds. 

A dedicated search for a $b\bar{b}$ resonance in this channel may improve the sensitivity to this model, as well as potentially reduce the theoretical uncertainties associated to the SM background. This would especially be useful in the boosted regime, where jet substructure techniques could be used to identify lighter ALPs that get rejected by the selection criteria of the cross section measurements. A more realistic phenomenological study beyond our simple parton-level approximation taking into account backgrounds, the proper identification of $b$-jets, and the possibility of selecting different numbers of $b$-jets is warranted, and we leave this for future work.

\subsection{Top-philic ALP in $t \bar t t \bar t$\label{subsec:fourtop}}

\begin{figure}[!ht]
    \centering
    \includegraphics[width=0.9\textwidth]{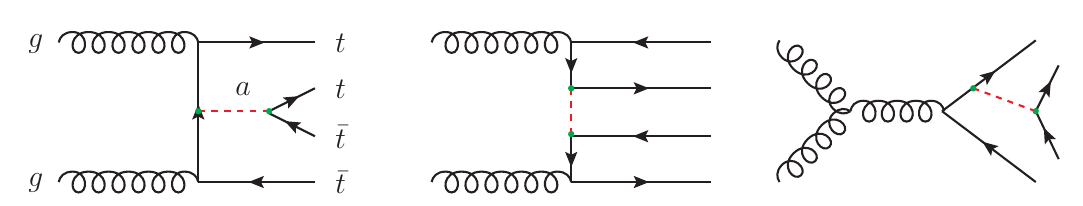}
    \caption{Representative diagrams for $t\bar t t \bar t$ production through the coupling of a virtual ALP $a$. Note that for $m_a<2 m_t$ no resonant diagram is present. \label{fig:diag4t} }
\end{figure}

In our considered mass range, top-philic ALPs contribute non-resonantly to four-top production at hadron colliders. This can happen in many different ways, through $t$- or $s$-channel-like topologies involving one or even two ALPs, via $gg$ or $q\bar{q}$ initiated processes and at various orders in $(c_t/f_a)^2$, $\alpha_S$ and $\alpha_{\rm EW}$. 

The $gg$ initial state can lead to contributions proportional to both $c_{t}^{2}$, when the ALP mediated amplitude is interfered with the purely SM one,  and  $c_{t}^{4}$ when the ALP mediated amplitude is squared. Representative diagrams  are shown in Fig.~\ref{fig:diag4t}. Instead, besides diagrams of order $c_{t}^{2}$ as in the $gg$ initial state, the one-loop induced couplings, $c_{f}$, can mediate diagrams in the $q \bar q$ initial state of order $c_{t}^{4}$ ($c_{t}^{6}$) when interfered with the SM amplitude (the order $c_{t}^{2}$ diagrams) and  $c_{t}^{8}$ when squared.  They have topologies of the form, {\it e.g.},  $q\bar q\TO a a\TO t \bar t t \bar t$.  We have calculated these contributions and found them to be negligible.

An additional complication in the context of four-top is that, already at LO for the SM~\cite{Cao:2016wib}, and even at NLO \cite{Frederix:2017wme}, more than one perturbative contribution is present in the $(\alpha_S, \alpha_{\rm EW})$ expansion. Moreover, the nominally subleading contributions are not negligible and there are large cancellations among the different perturbative contributions. For this reason, we include both contributions of purely QCD origin and the mixed QCD--EW ones when computing our top-philic ALP predictions, and studied each one separately.
Considering only tree-level contributions, all of the perturbative orders in four-top production are of the form:
\begin{equation}
\mathcal O \left(\left(\frac{c_t}{f_{a}}\right)^{2n} \alpha_{S}^{m}\alpha_{\rm EW}^{l}\right)\qquad {\rm with}~n+m+l=4 ~{\rm and}~ n\le2\,,
\label{eq:order4t}
\end{equation}
where $m,n$ and $l$ are all positive integers and the case $n=0$ corresponds to the SM.

\begin{figure}
    \centering
    \includegraphics[width=0.8\textwidth]{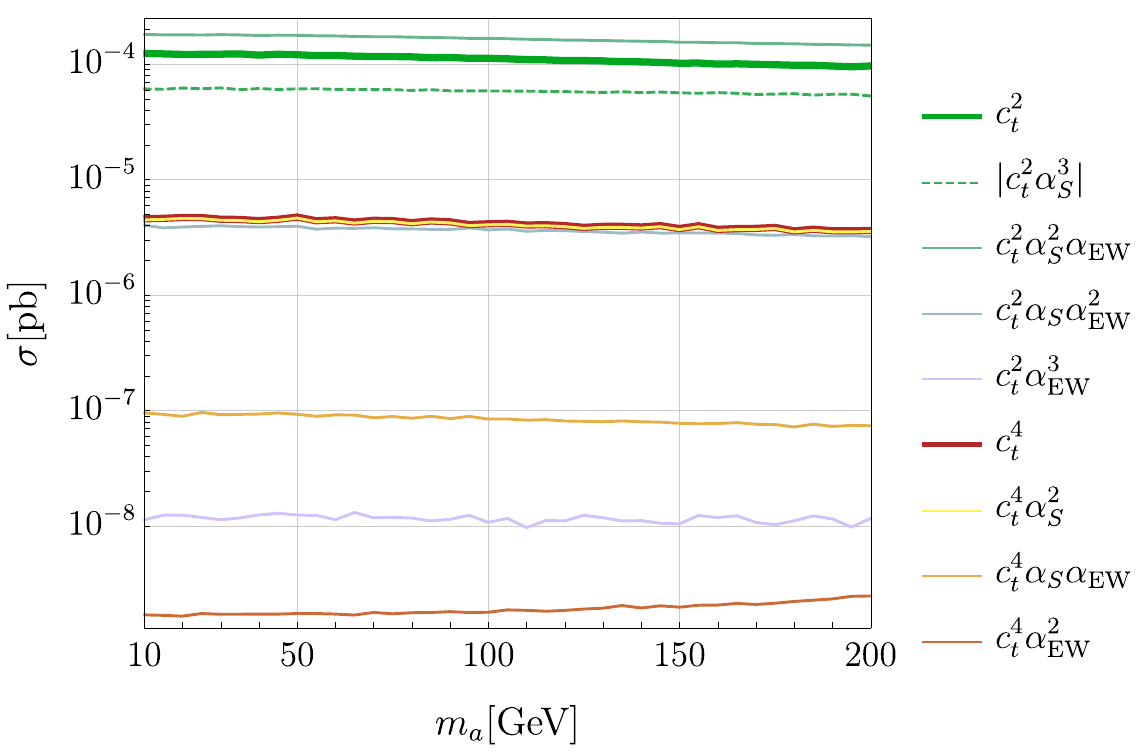}
    \caption{Breakdown of the different tree-level ALP induced contributions ($n=1,2$ in Eq.~\eqref{eq:order4t}) to four-top production at LHC 13 TeV. Orders denoted by an absolute value in the legend indicates negative contributions.
    \label{fig:4top}
    }
\end{figure}

The contributions of each order in Eq.~\eqref{eq:order4t}, excluding the SM one, are displayed in Fig.~\ref{fig:4top},  for the total cross section at 13 TeV, setting $f_a/c_t=1~{\rm TeV}$.
We explicitly denote them via their powers of $c_{t}$, $ \alpha_{S}$ and $\alpha_{\rm EW}$. When the powers of the last two are not specified, the sum over the different $m$ and $l$ combinations for a given $n$ in Eq.~\eqref{eq:order4t} is understood. 
The contribution proportional to $c_t^{2}$ is clearly dominant, featuring a cancellation between a negative, purely QCD-mediated  ($\alpha_S^3$) contribution and a positive, mixed QED-QCD mediated ($\alpha_S^2\alpha_{\rm EW}$) one.
The  contribution proportional to $c_t^{4}$ is entirely dominated by the purely QCD-mediated ($\alpha_S^2$) term. Although it may appear as very subleading w.r.t.~the one proportional to $c_t^{2}$, we will show that, given our sensitivity to $f_{a}/c_{t}$, this contribution is also relevant.  

The information provided in Fig.~\ref{fig:4top} can be exploited to set bounds on $f_{a}/c_{t}$ by comparing the theoretical predictions to experimental measurements. Indeed, the LHC experiments have recently reported the observation of four-top production with significances of 5.6 and 6.1 $\sigma$ from the ATLAS~\cite{ATLAS:2023ajo} and CMS~\cite{CMS:2023ftu} experiments, respectively, in the same-sign dilepton and multi-lepton (more than three leptons) channels. As in the case of $t\bar{t}b\bar{b}$, we use a set of statistically independent measurements of the four-top signal strengths to constrain the top-philic ALP parameter space. In addition to the two aforementioned analyses, we also incorporate measurements in the complementary single lepton and opposite-sign dilepton channels accompanied with multiple jets from the CMS~\cite{CMS:2023zdh} and ATLAS~\cite{ATLAS:2021kqb} collaborations. Three of the analyses report a signal strength measurement using the complete NLO prediction of 12.0$^{+2.2}_{-2.5}$ fb as the reference cross section \cite{Frederix:2017wme}, while the CMS multilepton observable only quotes a measured cross section of 17.7$^{+3.7}_{-3.5}$(stat.)$^{+2.3}_{-1.9}$(syst.) fb. The SM prediction has recently been improved to include QCD threshold resummation at next to leading logarithmic accuracy, yielding a value of 13.4$^{+1}_{-1.8}$ fb, representing a 12\% increase in the inclusive cross section and reduced theoretical uncertainties~\cite{vanBeekveld:2022hty}. We make use of the updated prediction to construct the signal strength for the CMS multilepton measurement, adding the theory error in quadrature to the systematic uncertainty. For the other measurements, we conservatively rescale the signal strengths to the new reference cross section while keeping the uncertainties unchanged.

The input signal strengths used in our analysis are summarised in Table~\ref{tab:4top_data}.
\begin{table}
\centering
\begin{small}
\begin{tabular}{|l|l|l|l|}
\hline
Exp.  & Channel & $\mu_{t\bar{t}t\bar{t}}~\pm$ stat. $\pm$ syst. & Ref.\tabularnewline
\hline
ATLAS & SSDL+ML & $1.70 \pm 0.40^{+0.7}_{-0.4}$   & \cite{ATLAS:2023ajo}\tabularnewline
ATLAS & OSDL+1L & $2.00 \pm 0.70^{+1.5}_{-1.0}$   & \cite{ATLAS:2021kqb}\tabularnewline
CMS   & SSDL+ML & $1.32 \pm 0.27^{+0.2}_{-0.23}$  & \cite{CMS:2023ftu}\tabularnewline
CMS   & OSDL+1L & $2.20 \pm 0.50 \pm 0.50$        & \cite{CMS:2023zdh}\tabularnewline
\hline
\end{tabular}
\end{small}
\caption{\label{tab:4top_data}
Signal strengths for the inclusive $t\bar{t}t\bar{t}$ cross section measurements in the same-sign dilepton + multilepton (SSDL+ML) and the opposite-sign dilepton + single lepton (OSDL+1L) channels used to constrain the ($m_a$, $c_t$) plane of the top-philic ALP. The recent theoretical prediction of 13.4$^{+1}_{-1.8}$ fb~\cite{vanBeekveld:2022hty} is used as a reference value throughout, such that the values of $\mu_{t\bar{t}t\bar{t}}$  are about 10\% lower than those reported by the experimental publications.
}
\end{table}
As in the previous section, we combine the four measurements into a $\chi^2$ analysis to extract bounds on $(m_a,\,f_a/c_t)$. We assume uncorrelated errors in this case since most measurements have a significant statistical component and the sources of systematic uncertainty are not theoretically dominated.
As with $t\bar{t}b\bar{b}$, there is a systematic albeit not so significant excess across the four-top measurements such that we find the SM hypothesis of $\mu_{t\bar{t}t\bar{t}}=1$ is just barely allowed.
We observe that-higher order effects in $c_t/f_a$ have a non-negligible impact on the bounds obtained. Truncating the ALP prediction at order $(c_t/f_a)^2$, we find bounds of $f_a/c_t\gtrsim100 ~\gev$ for $m_a=10 ~\gev$, with a mild dependence on $m_a$, such that the bound weakens by about 10\% at $m_a=200$ GeV (dark pink area labelled as ``$c_{t}^{2}$'' in Fig.~\ref{fig:FullVisible}). When we include the higher-order $(c_t/f_a)^4$ contributions, the $m_a=10~\gev$ bound strengthens to $f_a/c_t\gtrsim160~ \gev$ with a slightly milder mass dependence (light pink area labelled as ``$c_{t}^{2}+c_{t}^{4}$'' in Fig.~\ref{fig:FullVisible}). 

The fact that higher order effects in $1/f_a$ make a difference means that potential effects from higher-dimension operators ($\rm D>5$) could be relevant. Indeed, in deriving this bound, we have only considered the contribution to the four-top cross section from off-shell ALPs, neglecting possible model-dependent contributions from higher-dimension operators generated from integrating out heavy states in the UV. This is somewhat analogous to the discussion in Sec.~\ref{sec:existing} of the bounds from Higgs decays, where we have assumed that the dimension-six operator, $\mathcal{O}^{(6)}_{a H}$ is not generated at the matching scale. Instead, for the four-top process (see also a similar discussion for $t\bar{t}b\bar{b}$ in Sec.~\ref{subsec:ttbb}),  dimension-six SMEFT operators could contribute at $\mathcal{O}(1/f_a^2)$, such as four-top contact, $q\bar{q}t\bar{t}$, or the chromomagnetic dipole operators (see, {\it e.g.}, Ref.~\cite{Aoude:2022deh} and references therein). We are therefore analogously assuming that such operators are not generated, or at least relatively suppressed, at the matching scale. Assuming this is the case, we do not expect RG mixing between $c_t^2$ and dimension-six operators to play a significant role here, since our four-top bounds are derived from tree-level $c_t^2$ effects, while the RG running arises at one loop.

\subsection{Top-philic ALP in $t \bar t$\label{subsec:ttbar}}

New states coupled to the top quark can also modify production rates via one-loop corrections, as has already been investigated in Refs.~\cite{Esser:2023fdo,Bruggisser:2023npd} for the case of an ALP  in  top-quark pair production. In this section we  present the first exact calculation of one-loop induced $c_{t}$ effects on the $t \bar t$ production cross section.  We provide numerical results for  various differential distributions and we use them to constrain the coupling of the top-philic ALP. Many technical details regarding this calculation are left for a future publication\,\cite{Maltoni:2024wyh}, where we will also provide analytical formulae.

\begin{figure}[!t]
    \centering
    \includegraphics[width=0.8\textwidth]{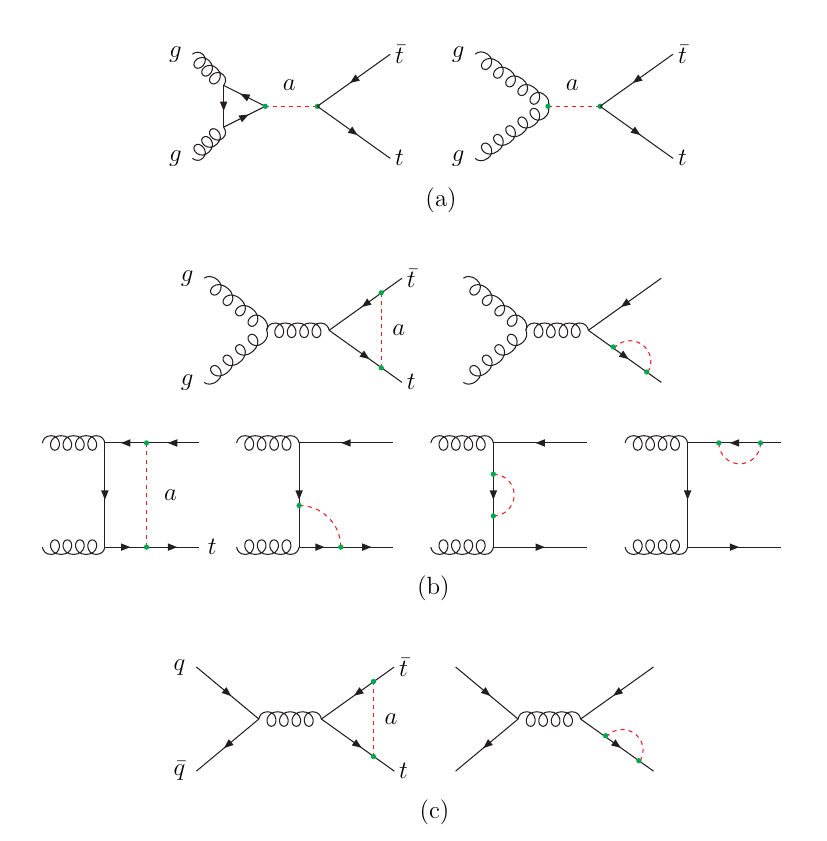}
    \caption{Representative diagrams for the ALP virtual corrections to $pp\TO t\bar t$ process, as obtained from the non-derivative Lagrangian of Eq.~\eqref{eq:Laintnonder}. We highlight the presence of the right diagram in $(a)$ which is absent in the  case of a pseudo-scalar particle $A$.}
    \label{fig:feynmandiagram_mtt}
\end{figure}

Expanding the inclusive cross section of top-quark pair production in powers of $c_t/f_a$, one obtains
\begin{equation}
\label{eq:ttbarNP}
\sigma =\sigma_{\rm SM}+\sigma_{\rm NP, \, virt}+\sigma_{\rm NP,\, real}\,.
\end{equation}
where the various quantities are defined in the following.
The term $\sigma_{\rm SM}$ is the SM prediction, factorising no powers of $c_t/f_a$. At this stage we can consider the SM prediction as the LO one, induced by  QCD tree-level diagrams from the $gg\TO t\bar t$ and $q\bar q\TO t\bar t$ processes. The first new physics effects (NP) from $c_{t}$ are given by either one-loop corrections or ALP real emissions. 
The term $\sigma_{\rm virt,\,NP}$ is given by the interference of the tree-level amplitudes with the one-loop Feynman diagrams shown in Fig.~\ref{fig:feynmandiagram_mtt}. Since we perform the calculation in the non-derivative ALP basis (Eq.~\eqref{eq:Laintnonder}), as discussed at length in Sec.~\ref{sec:xsALP} for the case of $pp\TO a+j$, the top-right diagram involving the $aG\tilde{G}$ contact term also has to be considered.

The term $\sigma_{\rm NP,\, real}$ denotes contributions coming from the real-radiation process $p p \TO t \bar t a$. If the ALP is unresolved, {\it e.g.}~because the ALP transverse momentum ($p_{\rm T}(a)$) is too small or the ALP is invisible, the process is indistinguishable from top-quark pair production. Both $\sigma_{\rm NP,\, virt}$ and $\sigma_{\rm NP,\, real}$ are of order $(c_{t}/f_{a})^{2}$ and therefore must, in principle, be considered together. 

\begin{figure}[!t]
    \centering
    \includegraphics[width=0.49\textwidth]{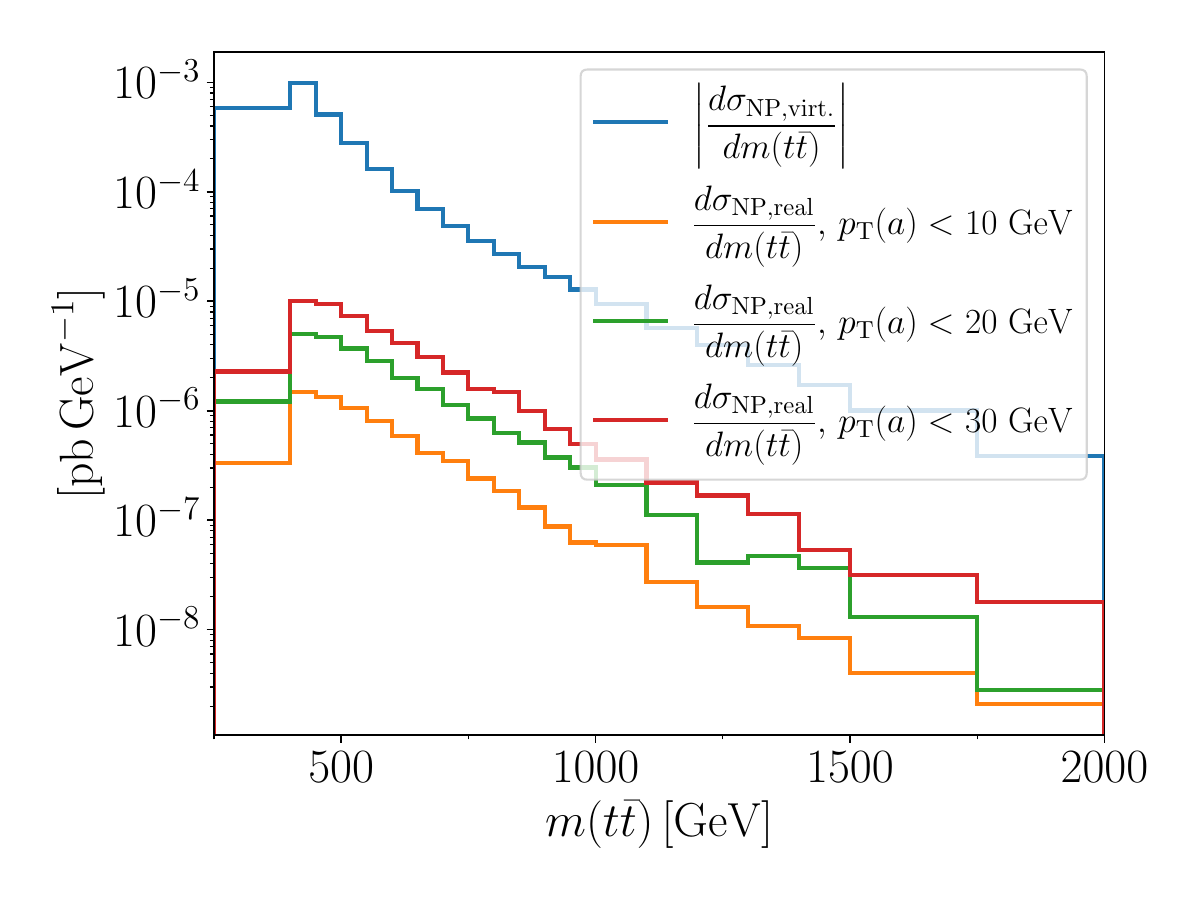}
    \includegraphics[width=0.49\textwidth]{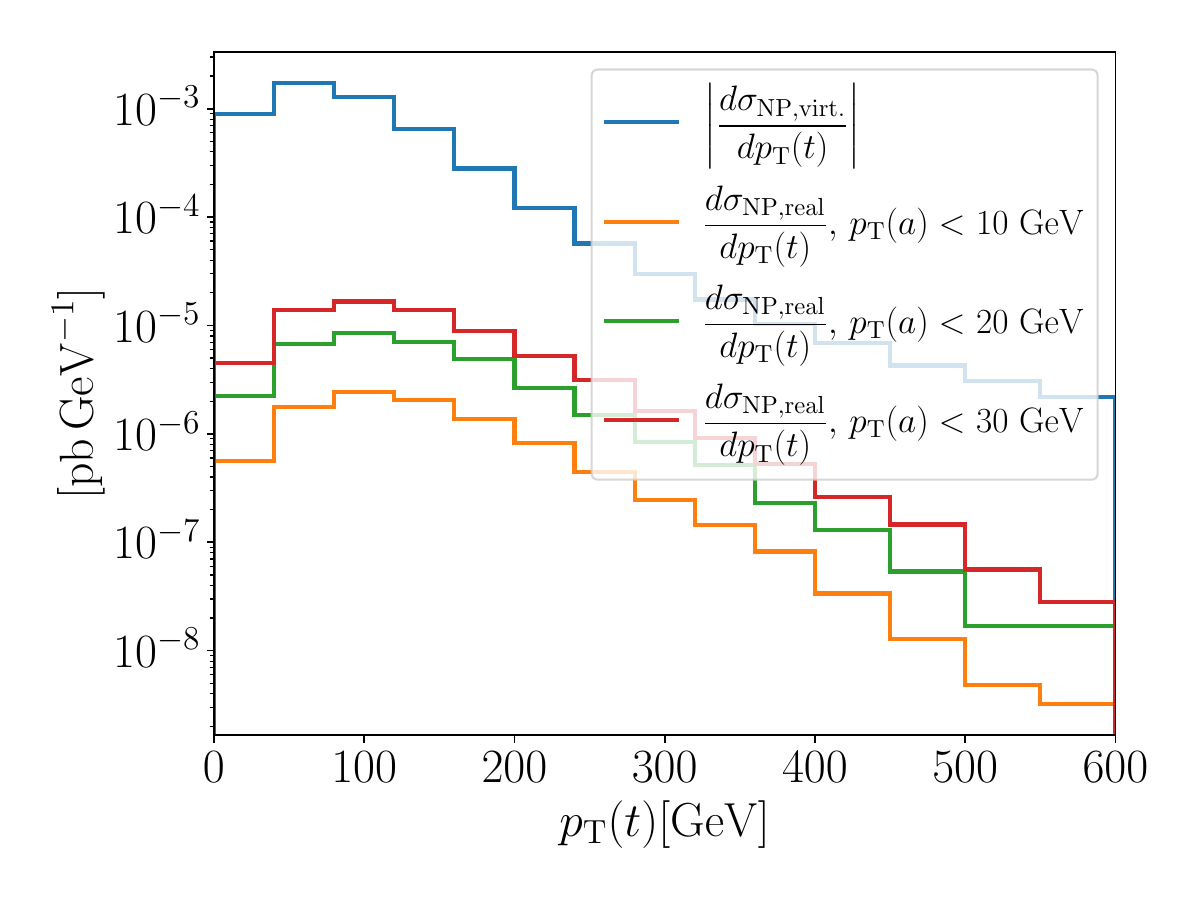}
    \caption{NP corrections for the $m(t\bar{t})$ (left) and $p_{\rm T}(t)$ (right) distributions: $m_a=10\, {\rm GeV}$, $c_t/f_a=1\, {\rm TeV}^{-1}$. Blue: only $|\sigma_{\rm NP,\, virt}|$. Other colours:  $\sigma_{\rm NP,\, real}$ for different $p_{T}(a)$ cuts. The virtual dominates the real by more than two orders of magnitude. 
    \label{fig:realvsvirt}}
\end{figure}

Before discussing some details on the $\sigma_{\rm NP,\, virt}$ calculation, we want to show that we can safely ignore the contribution from $\sigma_{\rm NP,\, real}$,
since its effect is subdominant w.r.t.~the one of $\sigma_{\rm NP,\, virt}$.  This can be seen in  Fig.~\ref{fig:realvsvirt}, which compares the spectrum in $m(t\bar{t})$ (left) and $p_{\rm T}(t)$ (right) from $|\sigma_{\rm NP, \, virt}|$ (blue) to that of $\sigma_{\rm NP, \, real}$ with an upper cut on $p_{\rm T}(a)$ of respectively 10 (orange), 20 (green) and 30 (red) GeV. Even for the weakest cut of 30 GeV, which most probably includes part of the phase space where the ALP would be experimentally resolved, the real emission contribution is consistently 1--2 orders of magnitude smaller than the virtual one. This comparison has been performed for $m_{a}=10~\gev$ and clearly for larger values of $m_{a}$ the gap between virtual and real would be even larger. 

The quantity $\sigma_{\rm NP,\, virt}$  has previously been estimated by considering only the $s$-channel diagram in Fig.~\ref{fig:feynmandiagram_mtt}$(a)$ using an effective $aG\tilde G$ interaction dressed by an approximated form-factor \cite{Esser:2023fdo}.  In Ref.~\cite{Bruggisser:2023npd} the $s$-channel diagram has been calculated exactly, but the remaining loop diagrams from Fig.~\ref{fig:feynmandiagram_mtt} have been only estimated. 
In our calculation all of the one-loop diagrams of order $(c_t/f_a)^2$, as those in  Fig.~\ref{fig:feynmandiagram_mtt}, are exactly taken into account. These diagrams in general contain UV divergencies, which have to be renormalised via the counterterms for the top-quark mass and its wave function.  We perform renormalisation in the on-shell scheme, so no renormalisation scale dependence is present for what concerns $m_{t}$.  We note in passing, that the higher-order term of Eq.~\eqref{eq:a2term}   generates a one-loop correction to the self-energy of the top quark, see {\it e.g.}~Ref.~\cite{Galda:2023qjx}. However, the corresponding diagram is a tadpole and therefore does not depend on the $p^2$ of the top quark. It amounts to a universal constant that can be reabsorbed into the definition of the top mass on-shell, bearing no physical consequences. More details on the full computation will be given in Ref.~\cite{Maltoni:2024wyh}.
In conclusion, since we exactly calculate all of the one-loop contributions, further effects w.r.t.~Refs.~\cite{Esser:2023fdo, Bruggisser:2023npd} are taken into account, such as new kinematic dependencies and also contributions from the $q\bar{q}$-initiated channels (Fig.~\ref{fig:feynmandiagram_mtt}$(c)$).

\begin{figure}[!t]
    \centering
    \includegraphics[width=0.49\textwidth]{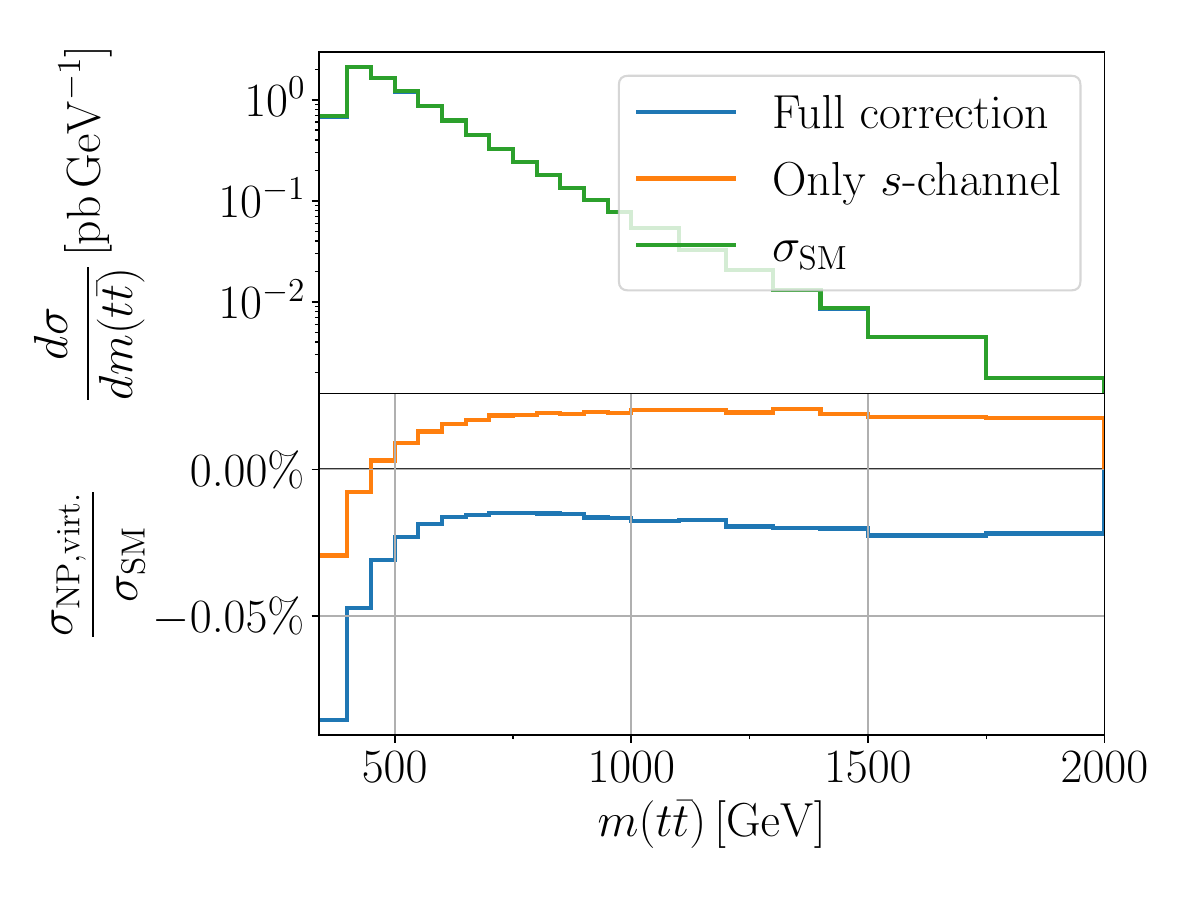}\hfill
    \includegraphics[width=0.49\textwidth]{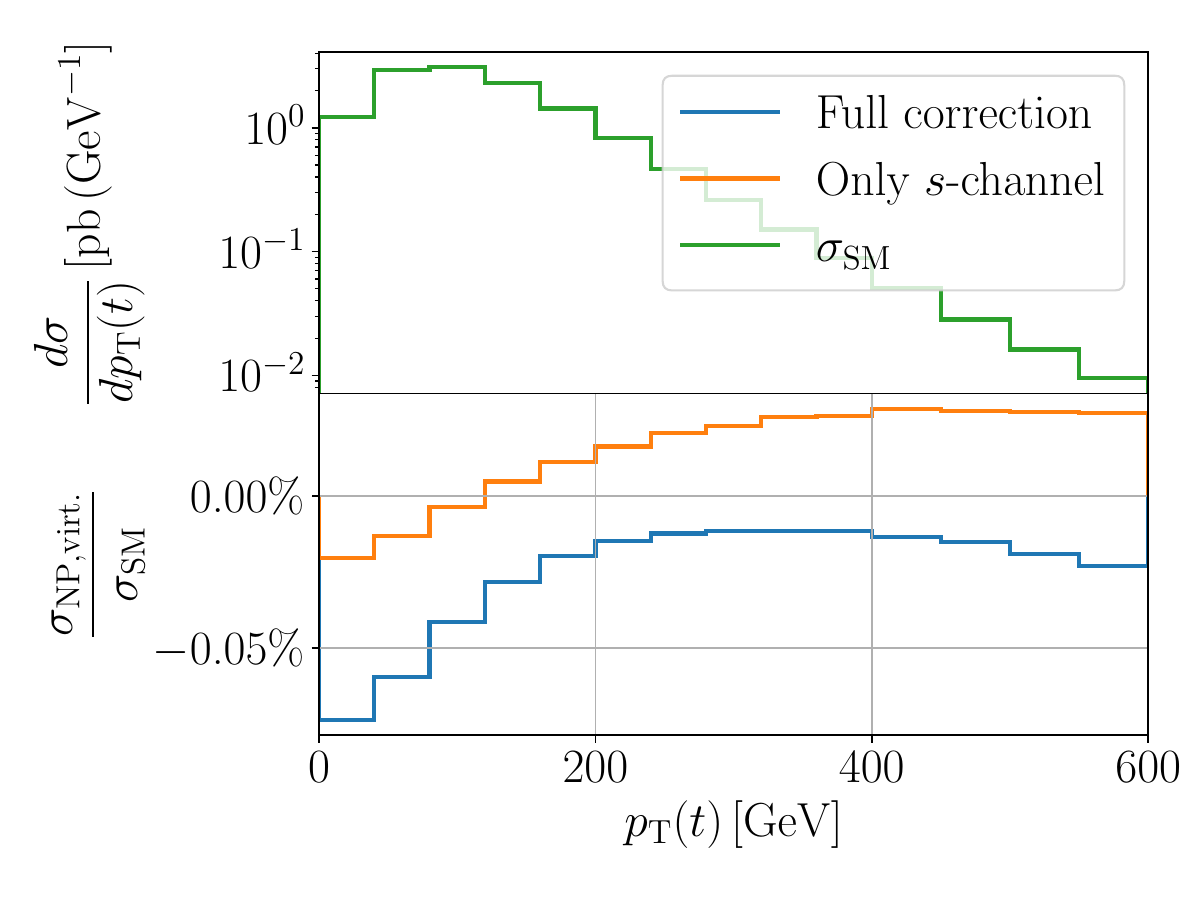}
    \caption{Relative impact of $|\sigma_{\rm NP,\, virt}|$ (bottom part of the plots) and $\sigma_{\rm SM}$ (upper part) for  the $m(t\bar{t})$ (left) and $p_{\rm T}(t)$ (right) distributions: $m_a=10 \,{\rm GeV}$, $c_t/f_{a}=1 ~{\rm TeV}^{-1}$. Lower panel:  only the $s$-channel diagrams (orange), exact $\mathcal{O}\left((c_t/f_a)^2\right)$ calculation (blue).   }
    \label{fig:ttdist_diffchannel}
\end{figure}

 The relative effects of the one-loop corrections to the $m(t\bar{t})$ and $p_{\rm T}(t)$  spectra are shown in Fig.~\ref{fig:ttdist_diffchannel}.
For $m_{a}=10~\gev$, we show our exact calculation and also, as a  comparison, the contribution from only the $s$-channel diagrams displayed in the top of Fig.~\ref{fig:feynmandiagram_mtt}.\footnote{The latter cannot be directly compared with Ref.~\cite{Esser:2023fdo} since therein the contribution was considered only under the approximation $m(t\bar{t})\gg 2m_t$.}
Two main differences emerge between the exact calculation and the contribution from only $s$-channel diagrams. First, the corrections from the exact calculation in the bulk of the distributions ($m(t\bar{t})\simeq 2m_t$ and $ p_{\rm T}(t) \simeq 0$) are much larger in absolute value than in the case of the $s$-channel only contribution. Second, the two approaches lead to corrections of different signs for $m(t\bar{t})\gg 2m_t$. Since going beyond the $s$-channel approximation leads to rather different predictions for the top-philic ALP corrections to top pair production at hadron colliders, we expect that top-quark data will lead to different constraints on the parameter space than previously estimated.

An additional aspect that we would like to stress is that
we do not include effects from the squared one-loop diagram in Fig.~\ref{fig:feynmandiagram_mtt}$(a)$. These contributions are  of $\mathcal{O}\left((c_t/f_a)^4\right)$ and are clearly positive by definition. However, they  only account for a subset of the complete corrections at that order, which also include two-loop ALP-mediated diagrams interfering with the SM amplitude. Since we have demonstrated the importance of such virtual corrections at one loop, we argue that the full two-loop calculation should be performed in order to reliably derive constraints at this order. 

\begin{figure}[t!]
    \centering
    \includegraphics[width=0.49\textwidth]{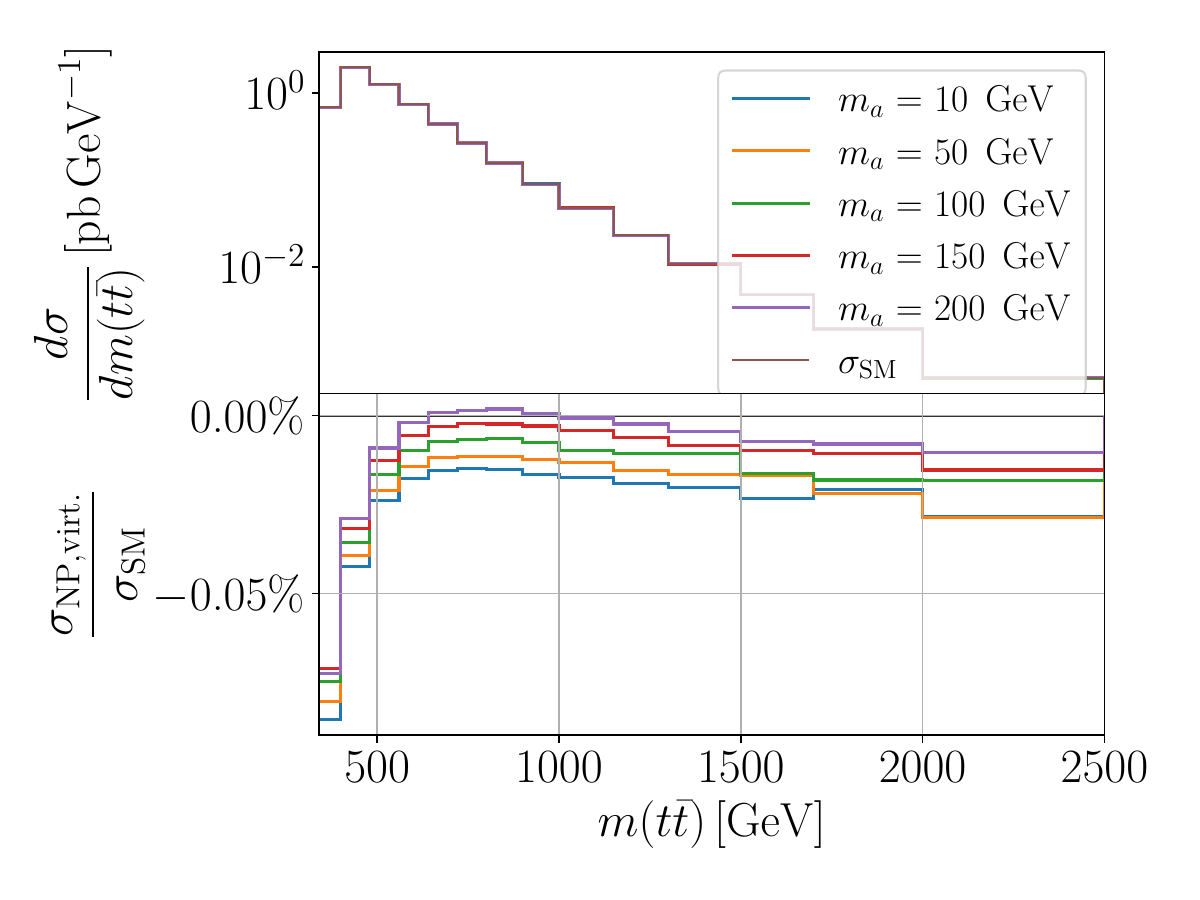}
    \includegraphics[width=0.49\textwidth]{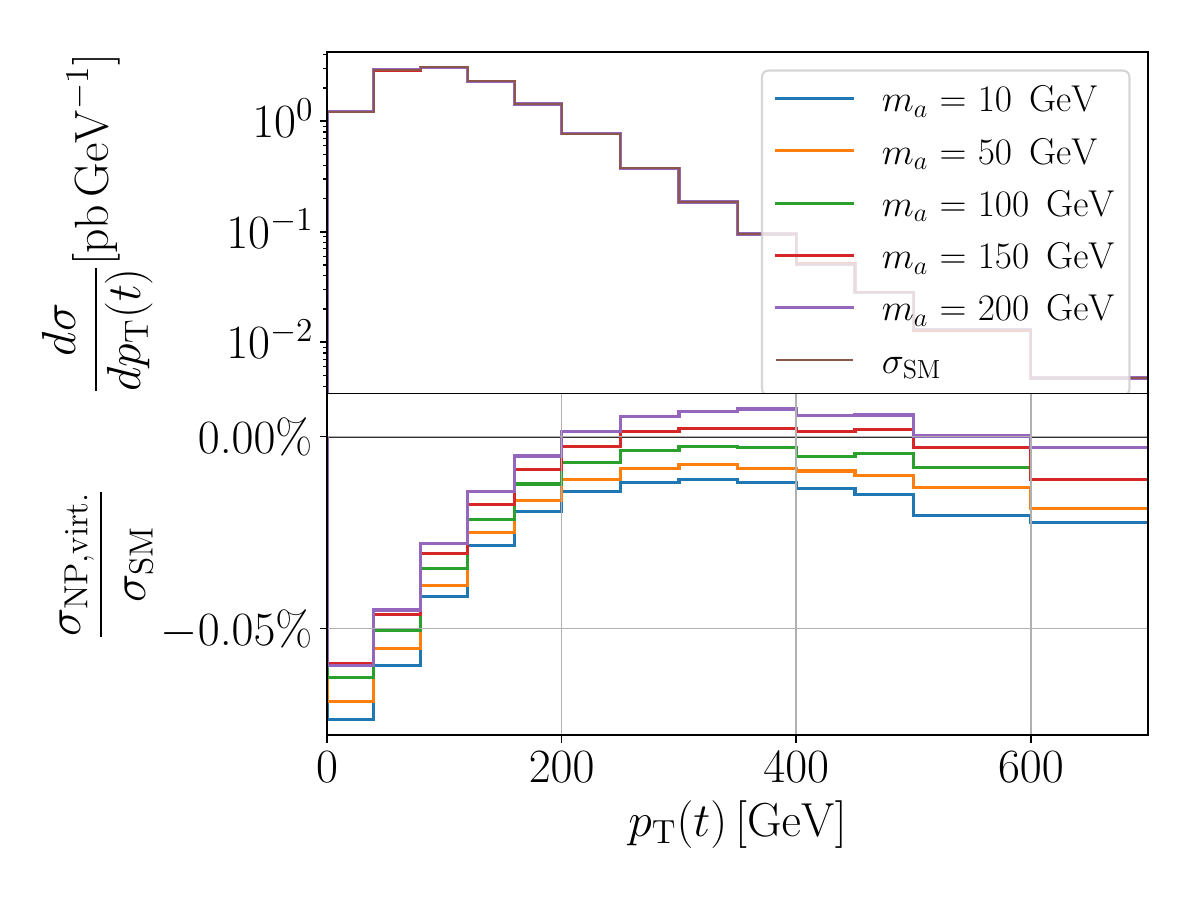}
    \caption{Differential $m(t\bar{t})$ (left) and $p_{\rm T}(t)$ (right) distributions similar to Fig.~\ref{fig:ttdist_diffchannel} but for different values of $m_a$ at  $c_t/f_a=1$ TeV$^{-1}$. Only the exact calculation is considered here unlike with Fig.~\ref{fig:ttdist_diffchannel}. }
    \label{fig:tt_differentmass}
\end{figure}

Our prediction for different values of $m_{a}$ is shown in Fig.~\ref{fig:tt_differentmass}.
As can be seen, the relative impact of $\sigma_{\rm NP,\, virt}$ is very mildly dependent on the value of $m_{a}$ in the considered mass range. Thus, it is not surprising that the corresponding exclusion limits on $f_a/c_t$ are also very mildly dependent on $m_{a}$ as we have already seen in Fig.~\ref{fig:FullVisible}. We describe in the following in more details how they have been obtained.

\medskip

We use our new predictions to derive exclusion limits on $f_a/c_t$, via the statistical approach described in Sec.~\ref{subsec:stats}. We combine the information from a selection of top-quark measurements at the LHC listed in Tabs.~\ref{tab:mt} and~\ref{tab:pt}, some of which are taken from the \verb|fitmaker| database~\cite{Ellis:2020unq}.
 They consist of  unfolded $m(t\bar{t})$ and $p_{\rm T}(t)$ differential distributions in  published by the ATLAS and CMS collaborations, based on measurements from LHC Runs 1 and 2.
 \begin{table}[!t]
  \small
  \centering
    \begin{tabular}{|c|c|c|c|l|}
   
\hline
      $\sqrt{s}$   & Collab. & Channel & bins & Ref.\\
      \hline
      8 TeV & ATLAS & Dilepton&6&\cite{1607.07281} \\ 
      8 TeV & ATLAS & $\ell$+jets&7& \cite{1511.04716}\\ 
      8 TeV & CMS & Dilepton&6& \cite{1505.04480}(a)\\
      8 TeV & CMS & $\ell$+jets &7&\cite{1505.04480}(b)\\ 
      13 TeV & ATLAS & $\ell$+jets &9&\cite{1908.07305} \\ 
      13 TeV & CMS & Dilepton &7& \cite{1811.06625}\\ 
      13 TeV & CMS & $\ell$+jets &10& \cite{1803.08856}\\ 
      13 TeV & CMS & $\ell$+jets &15&\cite{2108.02803}  \\
         \hline
    \end{tabular}
    \caption{Experimental $m(t\bar{t})$ differential distributions used in  the global fit.}
    \label{tab:mt}
     
\end{table}
\begin{table}[!t]\centering
  
   \small
    \centering
    \begin{tabular}{|c|c|c|c|l|}
   
\hline
      $\sqrt{s}$    & Collab. & channel & bins&Ref.\\
      \hline
    8 TeV&ATLAS&$\ell$+jets&8&\cite{1511.04716}\\
     
    8 TeV&CMS&Dilepton&5&\cite{1505.04480}(a)\\
    
    8 TeV&CMS&$\ell$+jets&8&\cite{1505.04480}(b)\\
     
    13 TeV&ATLAS&$\ell$+jets&8&\cite{1908.07305}\\
    
    13 TeV&CMS&Dilepton&6&\cite{1811.06625}\\
   
    13 TeV&CMS&$\ell$+jets&17&\cite{2108.02803}\\
         \hline
    \end{tabular}
    \caption{Experimental $p_{\rm T}(t)$ differential distributions used in the global fit.}
    \label{tab:pt}
 
\end{table}
 We computed predictions for each differential distribution 
  in {\aNLO}, at a fixed renormalisation and factorisation scale $\mu_{\rm R}=\mu_{\rm F}=m_t$ for QCD, and using the {\tt NN23NLO} PDF sets~\cite{Ball:2013hta}. Our predictions for $\mu(m_a,\,c_t/f_a)$ entering Eq.~\eqref{eq:chi} are the following for each bin 
 \be
 \label{eq:mufortt}
 \mu(m_a,\,c_t/f_a)=\frac{\sigma_{\rm LO}+\sigma_{\rm NP,\, virt}}{\sigma_{\rm LO}}\,.
 \ee 
 As also mentioned in Sec.~\ref{subsec:stats}, this approach assumes that radiative corrections in the SM, which are very large (see {\it e.g.}~Refs.~\cite{Czakon:2017wor, Czakon:2019txp}), also factorise the NP contribution from ALP loops.

We derive bounds on $f_a/c_t$ for  different $m_{a}$ values, in the range $10~{\rm GeV}\lesssim m_{a}\lesssim 200~{\rm GeV}$, using Eq.~\eqref{eq:mufortt} for the theory predictions to be compared with data from Refs.~\cite{1607.07281}, \cite{1511.04716}, \cite{1505.04480}(a)(b), and \cite{1803.08856}, for the $m(t\bar{t})$ distribution, and from Refs.~\cite{1908.07305} and \cite{1811.06625} for the $p_{\rm T}(t)$ distributions. The requirement of combining statistically independent measurements, \emph{i.e.} those that either involve different top-quark decay modes, collider energies or experimental collaborations, would have allowed for several different combinations of input distributions. Our selection was identified as the one generally leading to the most stringent bounds.

 \begin{table}[!t]\small
    \centering
    \begin{tabular}{|c|c|c|c|c|c|}
    
    \hline
        $m_a$ [GeV]& 10 & 50 & 100 &150 &200\\
       
        \hline
        &&&&&\\[-3.ex]
        $\dfrac{f_a}{c_t} $ [GeV] & 201&206&212&221&234
       \\[1.5ex]

        \hline
    \end{tabular}
    \caption{ Upper limits on $f_a/c_t$ for representative values of $m_{a}$ from $t\bar t$ measurements  at the LHC. }
    \label{tab:constraints}
\end{table}
We report representative results in Tab.~\ref{tab:constraints} and, as already mentioned, the limits on $f_a/c_t$  are nearly independent of the ALP mass, which reflects how insensitive the virtual ALP corrections are to this quantity. In Fig.~\ref{fig:FullVisible}, the bounds from $t \bar t$ production correspond to  the interpolation of the data  reported in Tab.~\ref{tab:constraints}. As discussed in Sec.~\ref{sec:summary-indirect}, top-quark pair production data leads to the strongest constraints on $c_{t}$ in our model. Given the shape of the corrections induced by $\sigma_{\rm NP}/\sigma_{\rm SM}$, see Fig.~\ref{fig:tt_differentmass}, we expect that a finer binning close to the threshold would lead to stronger constraints. We encourage experimental collaborations to follow this strategy to further improve the bounds provided in this paper.  

As with the other top-quark processes considered so far, we comment on possible contributions from dimension-six operators. Top-quark pair production is sensitive to a number of $q\bar{q}t\bar{t}$ contact interactions, as well as the chromomagnetic top-quark operator of the SMEFT. We emphasise again that our bounds are derived assuming no such operators are generated in the UV, although we note that one may not expect a UV model generating a top-philic scenario to lead to significant modifications of interactions involving the light quarks. In any case, their impact on the $t\bar{t}$ process and the associated bounds on such operators from LHC data are known (see, {\it e.g.},~\cite{Ellis:2020unq,Ethier:2021bye} and references therein). However, even if we assume that they are not generated at the UV scale $\Lambda$, since our $t\bar{t}$ bounds are derived using one-loop order predictions in the ALP-EFT, RG mixing effects between $c_t$ and the dimension-six SMEFT operators are potentially relevant.  Looking at the RG equations for the operators in question, we see that the terms in the RG equations for $q\bar{q}t\bar{t}$ operators that depend only on $c_t$ always come with a power of the light quark Yukawa, such that their impact on $t\bar{t}$ production can be neglected. The chromomagnetic top operator is only sourced by a combination of $c_t$ and $c_{GG}$, so we can neglect it for the purposes of our study. It has also been shown that four-top operators can affect $t\bar{t}$ rates at  one-loop level. If such operators were generated at tree level in the UV, $t\bar{t}$ data may lead to relevant constraints. In our case, these operators are only generated by running, such that we can again neglect this effect as subleading in the perturbative expansion. Overall we conclude that the constraints on $c_t$ from $t\bar{t}$ data that we have derived  are accurate up to the one-loop and $1/f_a$ order that we consider in our work. 

\section{The top-philic ALP as a portal to Dark Matter}\label{sec:DM}

In this section we consider the possibility that the top-philic ALP acts as a mediator between the SM and the dark-matter (DM) sector. 
Similar scenarios, 
for ALP mediators with other combinations of couplings with SM turned on, have previously been considered in, {\it e.g.}, Refs.~\cite{CidVidal:2018blh,Acanfora:2023gzr,Fitzpatrick:2023xks,Dror:2023fyd,Ghosh:2023tyz,Armando:2023zwz}
(see also Refs.~\cite{Arina:2016cqj,Arcadi:2017wqi} for the case of (pseudo)scalars). 
From the phenomenological point of view, it
is an interesting possibility. On the one hand, since the ALP can feature an invisible decay, it opens  up signatures with missing transverse momentum at the LHC. On the other hand,  indirect limits obtained from the virtual exchange of a top-philic ALP, such as those from $t\bar t$ and $t \bar t t \bar t$ cross section measurements discussed in Sec.~\ref{sec:newprobes}, remain applicable. Thus, both cases provide constraints that can be combined together.

We focus on the regime where the interaction is large enough to lead to a freeze-out production of DM in the early Universe. For definiteness, we consider a DM sector including a Majorana fermion $\chi$ with mass $m_{\text{DM}}$ and coupling to the top-philic ALP with strength $c_{\text{DM}}$, 
\begin{equation}
\label{eq:DMlag}
    \mathcal{L}_{\chi} \supset 
    i  \bar \chi \slashed{\partial} \chi 
    - m_{\text{DM}} \bar \chi \chi - i c_{\text{DM}}\frac{m_{\text{DM}}}{f_a} a \bar \chi \gamma^5 \chi\,.
\end{equation}
If $2 m_{\text{DM}} < m_a$ the ALP can decay invisibly into DM with a partial width
\begin{equation}
    \Gamma(a \to \bar \chi \chi) =
c_\text{DM}^2 \frac{m_a m_{\text{DM}}^2}{2\pi f_a^2} \sqrt{1 - 4 m_{\text{DM}}^2/m_a^2}\,.
\end{equation}
In the following we will set $c_{\text{DM}}=c_t(\Lambda)$ 
and $f_a = 1$ TeV. 

Depending on the value of the dark-matter mass  $m_{\text{DM}}$, there are two possible scenarios for what concerns the ALP decay modes:
\begin{itemize}
    \item[A)] $m_a < 2 m_{\text{DM}}$: the ALP decay into dark matter is kinematically closed. The collider phenomenology of the top-philic ALP is in this case equivalent to the analysis of the previous sections.

    \item[B)] $m_a > 2 m_{\text{DM}}$: the ALP decay into dark matter is kinematically open. Given that the other competing decay channel is into $b \bar b$ pair, whose loop induced coupling is of order $\sim c_t/16\pi^2$, the decay into dark matter always dominates in this regime. Thus, one can use the approximation $\text{BR}(a \to  \bar \chi \chi) \simeq 1$ and the resulting phenomenology at collider is the one of an invisible top-philic ALP.
\end{itemize}
Note that the limits derived in Sec.~\ref{sec:summary-indirect} from $t \bar t$ and $t t \bar t \bar t$ are valid in both scenarios, and exclude roughly $f_a/c_t \lesssim 200$ GeV in the ALP mass range we consider.

In the next section we will first explore the parameter space of the model, focusing on the dark matter relic density constraint. As we will see, the dark matter production mechanism in the early Universe is different depending on the mass ratio $m_{\text{DM}}/m_a$. After that, we will study the collider phenomenology of an invisible top-philic ALP.

\subsection{The relic density constraint}
\label{sec:relicdensity}

\begin{figure}[t]
    \centering
    \includegraphics[width=0.55\textwidth]{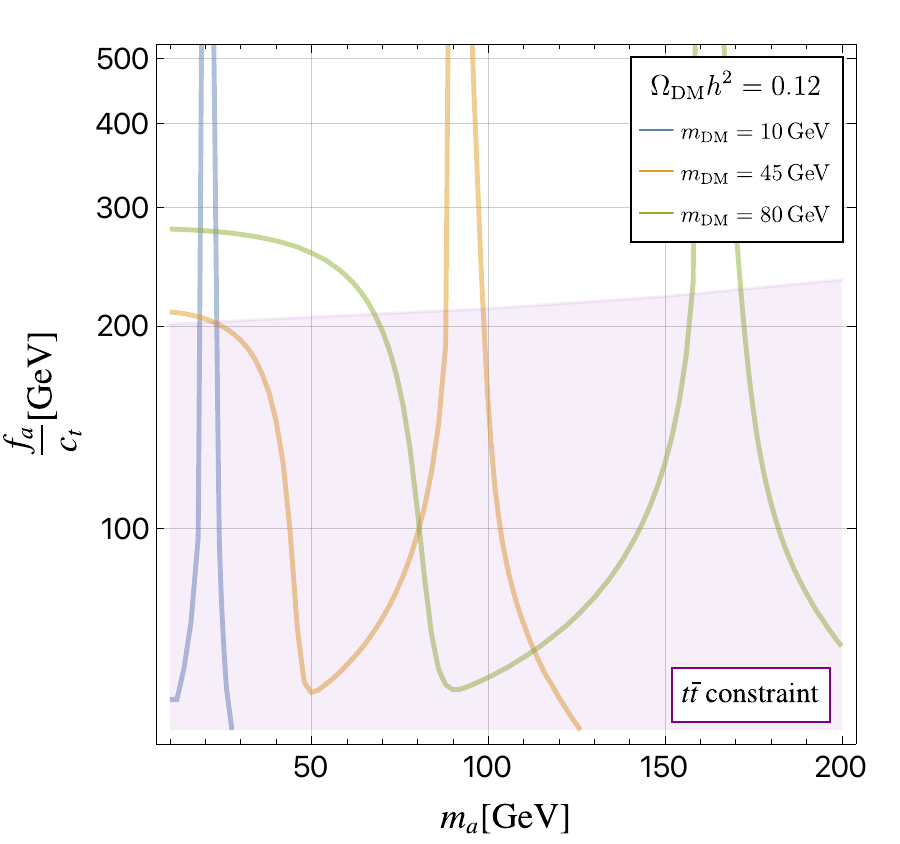}
    \caption{
    Relic density lines with $\Omega h^2 = 0.12$, 
    assuming $c_{\text{DM}} = c_t (\Lambda)$.
    Different lines represent different DM masses,
    respectively $m_{\text{DM}}= 10, 45, 80$ GeV in (blue, yellow, green).
The shaded purple region is excluded by top-philic ALP modifications of $t\bar t$ cross section at the LHC, as shown in Sec.~\ref{subsec:ttbar}. 
}.
    \label{fig:DM_ALP}
\end{figure}

In order to study the dark matter abundance from freeze-out one can implement the model to any of the public tools for computing relic densities, such as {\tt{MadDM}} \cite{Arina:2021gfn} or {\tt{micrOMEGAs}}
\cite{Belanger:2001fz}. The results to be presented in the following have been obtained from the latter by implementing 
the Lagrangian with the RG-induced couplings as presented in Sec.~\ref{sec:top-philic}.

There are two types of processes that can drive freeze-out in the early universe, whose relative importance depends on the mass ratio between the dark matter and the ALP.
If $ m_a>2 m_{\text{DM}}$ (scenario B introduced above), then DM annihilation
occurs through processes such as $\bar \chi \chi \to \text{SM--} \text{SM}$ mediated by the ALP.
In this case all SM--ALP interactions are in principle involved. 
Notice that for freeze-out, the typical energy of the annihilation process is of order $2 m_{\text{DM}}$. 
In this energy range, according to what has been discussed in Sec.~\ref{sec:BRs}, we expect that the relevant DM annihilation channel into SM pairs is mediated by the ALP decaying into a pair of bottom quarks.
In our  implementation we hence neglect the radiatively induced couplings of the ALP into gauge bosons. 
From the previous discussion about the loop functions and our analysis of the branching ratios, we expect that considering the annihilation into bottoms will suffice to give a good estimate of the 
$\bar \chi \chi \to \text{SM-} \text{SM}$ annihilation process.\footnote{
We postpone to future work a more detailed computation of the 
$\bar \chi \chi \to \text{SM-} \text{SM}$
annihilation including full loop effects.}

On the other hand, if $m_a < m_{\text{DM}} $ the DM annihilates mainly into a pair of ALPs with a $t$-channel diagram, $\bar \chi \chi \to a a$, and the process is fully determined by $c_{\text{DM}}$. In this case the ALP is assumed to be in thermal equilibrium with the SM bath. In the intermediate mass range $ m_{\text{DM}}<m_a<2 m_{\text{DM}}$, both processes can be important to determine the DM abundance. In the last two cases 
the ALP can only decay into SM states (scenario A introduced above). 

Within our approximations 
and the implementation in 
{\tt{micrOMEGAs}},
we now explore the conditions under which we can obtain the correct dark matter relic abundance in the top-philic ALP mediator model. Lines of correct relic density in the plane
$f_a/c_t$ {\it vs.} $m_a$ are displayed in Fig.~\ref{fig:DM_ALP}, assuming $c_{\text{DM}}=c_t (\Lambda)$,
and for fixed representative values of the dark matter mass,
respectively $(10,45,80)$
GeV.
We note that
generically, since the ALP coupling to DM increases with $m_{\text{DM}}$ (see Eq.~\eqref{eq:DMlag}),  one can obtain the correct relic abundance with smaller $c_t/f_a$ when the DM is heavier.

For $2 m_{\text{DM}} < m_a$, and the ALP acts as an $s$-channel mediator, 
the $b\bar{b}$ annihilation is suppressed when increasing the ALP mass, and hence the relic density lines bends towards smaller values of $f_a/c_t$ when increasing $m_a$.
For $2 m_{\text{DM}} \sim m_a$ there is a resonant enhancement in the annihilation, and the correct relic abundance can be obtained with a small interaction strength (large $f_a/c_t$).
For $m_{\text{DM}} > m_a$ the dark matter annihilates into pairs of ALPs $\bar \chi \chi \to a a$. In this case the relic density lines are hence independent of the ALP mass for $m_a \ll m_{\text{DM}}$.
The transition between the two regimes of $\chi \chi \to \text{SM--} \text{SM}$ annihilation or $\bar \chi \chi \to a a$ annihilation, passing through the resonance, is clearly visible in the shape of the lines in Fig.~\ref{fig:DM_ALP}. 
From this figure, it is clear that, unless the mass ratio is such that the annihilation is resonantly enhanced, large values of $c_t/f_a$ are typically needed to obtain the correct relic abundance (as also noticed in {\it e.g.}~Refs.~\cite{Acanfora:2023gzr,Fitzpatrick:2023xks} for similar ALP-mediated models). 
This is a generic conclusion, which is independent of our specific choice for $c_{\text{DM}}$ that fixes
$c_{\text{DM}} = c_t(\Lambda)$.

In Fig.~\ref{fig:DM_ALP}, we also include the LHC limits we derive from the previous section by studying the ALP effects in $t \bar t$,
as a purple shaded area,
which are independent of the value of the dark matter mass and coupling.
We see that these bounds already rule out significant parts of the lines where the correct relic density can be obtained.
Referring to the two previously introduced scenarios, for ALP masses lower than the resonant peak we are in scenario A, where the ALP decays into SM. In this case the other constraints of Sec. \ref{sec:summary-indirect},
specifically $t \bar t  b \bar b$ and $ t \bar t \ell^+ \ell^-$, also apply.
For ALP mass larger than the resonant peak we are instead in scenario B, {\it i.e.} an invisible ALP,
of which the specific collider signatures will be discussed in the next subsection.

As a final aside, 
note that in the regime $2 m_{\text{DM}} < m_a$ we also expect stringent constraints from indirect detection, given the existing results from the Planck satellite data on thermal dark matter annihilating predominantly to $b \bar b$ \cite{Fermi-LAT:2015att,Fermi-LAT:2016uux}. Intriguingly, this is also the typical model and range of masses that could accommodate the galactic center excess (see {\it e.g.}~Refs.~\cite{Calore:2014nla,PerezdelosHeros:2020qyt}
and references therein). 
Furthermore, also in the other mass regime, $m_a < m_{\rm DM}$, there might be indirect detection constrains from DM annihilation into SM pairs, even if these channels are subleading in setting the relic abundance (see {\it e.g.}~Ref.~\cite{Armando:2023zwz}).
A detailed investigation of direct and indirect detection constraints on this parameter space is left for future work.

\subsection{Invisible top-philic ALP at the LHC}
\label{sec:DM-CS}

In this section we focus on
scenario B), {\it i.e.}, the case where $m_a \gtrsim 2 m_{\text{DM}}$, and the ALP decays dominantly into dark matter.
From Fig.~\ref{fig:DM_ALP} we observe that in the $m_a > 2 m_{\text{DM}}$ region a very large $c_t/f_a$ coupling is needed in order to get the correct relic density. This is due to the fact that the relevant coupling for DM annihilation (the coupling to bottom quarks) is effectively one-loop suppressed (see  Eq.~\ref{eq:cthird}).
The only cases which are viable with moderate values of $c_t/f_a$ are those close to the resonance. 
We first study the LHC reach on the invisible top-philic ALP, and then afterwards discuss the implication for the dark matter relic density.

\begin{figure}[!t]
    \centering
    \includegraphics[width=0.95\textwidth]{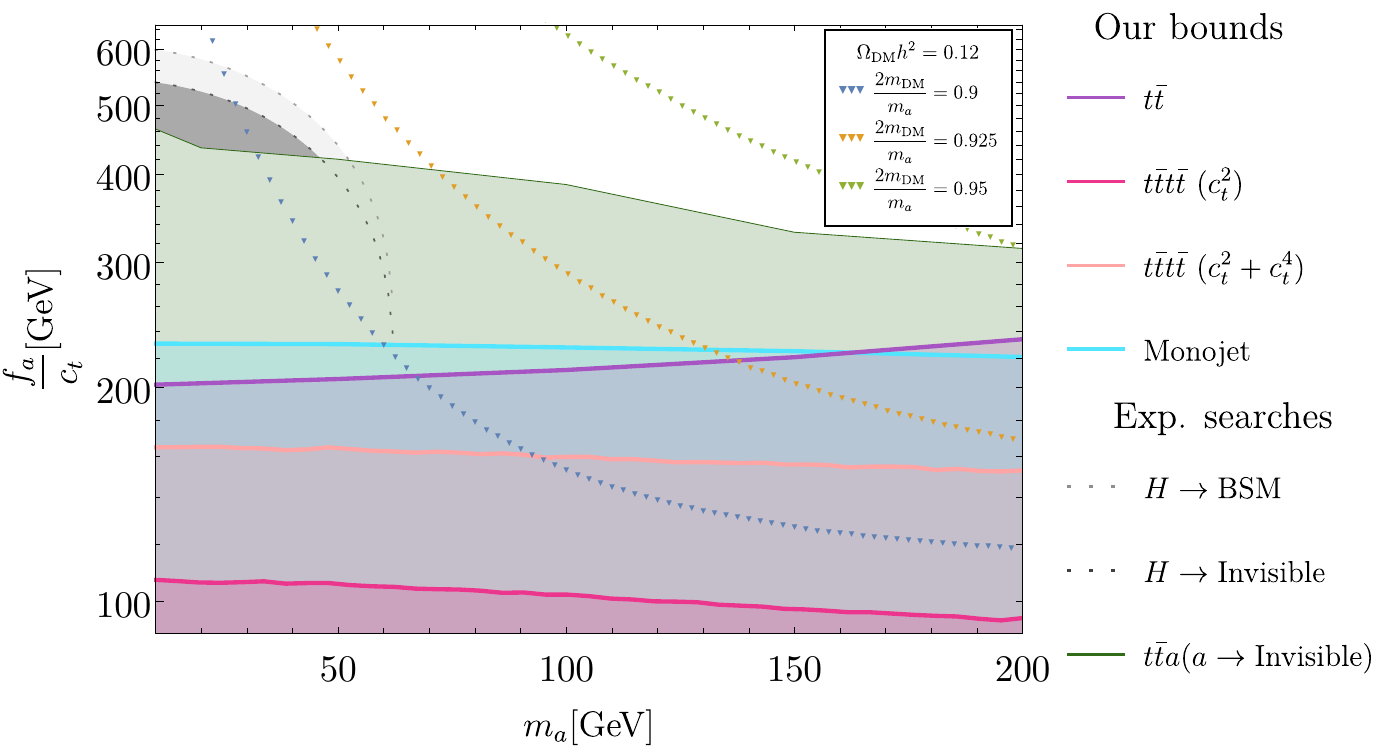}
    \caption{
Combination of exclusions from current experimental searches and our proposed probes for a top-philic ALP decaying invisibly with a $100\%$ BR.
The limit derived from the ATLAS search \cite{ATLAS:2021hza} of $t \bar t a, (a\TO {\rm invisible})$ is shown in olive green.
The solid lines show the limits we have obtained from $t \bar t t \bar t$, $t \bar t$, and also from a re-interpretation of mono-jet limits (see text for details).
Regions with dashed contours indicate the limits from Higgs decay which rely on some assumptions on dimension-6 operators, as explained in the text.
Finally, for reference, we show three lines (dotted lines) where the simplified DM model in Eq.~\eqref{eq:DMlag} can lead to the correct relic abundance
through freeze out (thanks to a resonant enhancement in the DM annihilation cross section).
    \label{fig:consraints_invisible}
}
\end{figure}

In Fig.~\ref{fig:consraints_invisible} we summarise the LHC constraints in the top-philic ALP parameter space, assuming $\text{BR}(a \to \text{invisible}) =1$.
First,  we recall that constraints obtained in the previous section from virtual corrections to $t \bar t$ and $t \bar t t \bar t$  observables are applicable in this scenario, as they only depend on the ALP--top interaction and its mass, not on its width. However, in this case, the direct production of the ALP can now be observed in $t \bar t + \text{MET}$  and the mono-jet signatures.
The constraint from $t \bar t + \text{MET}$ is taken directly from an ATLAS analysis~\cite{ATLAS:2021hza}, where a simplified model with pseudoscalar DM mediator coupled predominantly to the top  quark is considered (a similar CMS analysis was published in Ref.~\cite{CMS:2021eha})
in our mass range of interest (see \cite{Esser:2023fdo} for a complete recasting of this search extended also to lower ALP masses). The model and constraints derived in \cite{ATLAS:2021hza} can be directly mapped to our top-philic ALP by identifying $g_t \equiv v \frac{m_t}{f_a}$, with $v=246$ GeV.
Indeed, for what concerns the tree-level production in association with $t \bar t$,
a top-philic ALP or pseudoscalar are equivalent (see discussion in Sec.~\ref{sec:nonder}).
The line from mono-jet constraints, instead, is obtained by simulating $p p \to a j$ with the {\aNLO} model and comparing it with the signal regions in Ref.~\cite{ATLAS:2021kxv}
as explained in the Appendix \ref{app:mono-jet}.

In Fig.~\ref{fig:consraints_invisible}
we also report the limits from $H \to a a$ decay, now with $a$ decaying invisibly.
The experimental bound comprises both the channel in which the Higgs decays completely invisibly ($H\TO aa\TO \bar\chi\chi\bar\chi\chi$) and the one where the decay is partially invisible ($H\TO Za\TO Z\bar\chi\chi$). Using the most recent and best experimental measurement~\cite{ATLAS:2023tkt} we obtain a bound similar to that obtained for the Higgs decay into BSM particles in Sec.~\ref{sec:existing}.
These bounds should be considered featuring the same caveats, discussed around Fig.~\ref{fig:constraints}, regarding the underlying assumptions on the dimension-six interaction generated by the UV theory.
For completeness we also checked the possibility of strengthening the constraints exploiting the one-loop-induced couplings to weak gauge bosons (see Sec. \ref{sec:weak_gauge}), arising {\it e.g.}~
from mono-$Z$ via  $pp \to Z a$ and $a$ leading to $\slashed{E}_{T}$.
For this purpose we estimated the limits 
on $c_t/f_a$
from the bounds derived in Refs.~\cite{Brivio:2017ije,Bonilla:2022pxu}
on $ c_{\tilde W}$,
and found that even projected sensitivities with $300 \text{~fb}^{-1}$ were below the $c_t/f_a$ range displayed in Fig.~\ref{fig:consraints_invisible}.

From Fig. \ref{fig:consraints_invisible}
we see that the ATLAS search targeting the missing energy signature of the top-philic ALP in association with $t \bar t$ production is the one that best constrains the decay constant (green region in the plot).
Nevertheless, the limit from $t \bar t$ is 
only moderately weaker, and it is
comparable to the mono-jet reach, emphasising the
effectiveness of looking for this particle through its indirect modifications in top final states.

The results of our analysis have the following implications for the dark matter relic abundance. The existing collider limits exclude the relic density line in the scenario $m_a > 2 m_{\text{DM}}$ unless we lie on the resonantly enhanced region where $m_a \gtrsim 2 m_{\text{DM}}$.
In order to provide an indication of the LHC reach of this special case, in the LHC exclusion limit of Fig. \ref{fig:consraints_invisible} we show for illustration the lines where the model predicts the correct relic abundance by assuming $2m_{\text{DM}}/m_{a} = \{0.9, 0.925,0.95\}$ with $c_{\text{DM}} = c_t (\Lambda)$.
We conclude that resonantly enhanced dark matter annihilation through top-philic ALP is still viable, but significantly constrained by the collider limits. In particular, following our estimate of the relic abundance, a tuning finer than  $O(10\%)$ in the dark matter mass is necessary.

In general, our analysis for the top-philic ALP acting as a portal to DM shows that the LHC limits severely bound the fraction of the parameter space leading to the correct dark matter relic density through freeze-out.

\section{Conclusions}
\label{sec:conclusions}
In this paper we have discussed and analysed the LHC phenomenology of an ALP in the mass range $m_a  \subset [10,200]$ GeV, coupling only derivatively to the top quark ({\it i.e.},  ``top-philic") at high scales. We discussed how the presence of such a coupling in the UV implies  the generation of ALP couplings to all of the SM fermions at low scales, as well as  to gauge bosons, leading to a very rich phenomenology
for the top-philic ALP (see {\it e.g.}~its many decay channels in the branching-ratio plot, Fig.~\ref{fig:BRplot}).
We have focussed our work on the $[10,200]$ GeV mass window, where a top-philic ALP remains
remarkably elusive to direct LHC searches and the exploration of new strategies to probe this particle are strongly motivated. 

We have investigated in detail the properties a top-philic ALP and its production modes at the LHC, in particular its production in association with one extra jet in different kinematical regions. In order to be able to calculate loop effects, we have created a dedicated {\UFO} model and performed our calculation with the help of {\aNLO}. In doing so, we have inspected and elucidated technical aspects related to the change of basis in the ALP EFT, derivative {\it vs.}~non-derivative, and discussed the connection between an ALP and a generic pseudoscalar state. 

We have then inspected the reach of existing LHC searches which target low-mass resonances in association with $t \bar t$, or with a jet, as well as possible constraints from Higgs decays.
The elusive nature of the top-philic ALP in the $[10,200]$ GeV mass range is due to the fact that the main decay channel is into hadronic final states ({\it e.g.}, $b\bar b$ final state) and the gluon fusion production cross section is suppressed.

We then proposed new strategies to probe a top-philic ALP and to constrain its decay rate, in increasing complexity.
Given its branching ratio pattern, we first studied the impact of a top-philic ALP into the $t \bar tb\bar b$ SM cross-section measurements, which could be affected by $t \bar t a, (a\to b\bar b)$. In this case, an experimental study of the further constraints that could be derived from a differential $t \bar tb\bar b$ cross section measurement by taking into account the resonant nature of the ALP decay $tt (a \to bb)$ would be welcome. 
Second, we investigated the contribution of a virtual top-philic ALP into four-top final states.
Third, we studied the (loop-induced) corrections to $t \bar t$ final state at the LHC.
In summary, we have found that the reach of direct LHC searches is comparable (or weaker) than that based on 
the virtual contributions to top final states.

We have then considered a less minimal model, where in addition to the top-philic ALP we have added to the SM also a dark matter particle, for which the ALP acts as a portal to the SM.
This DM model serves as a benchmark scenario to analyse the case in which the top-philic ALP decays invisibly at the LHC.
We have first explored the parameter space of the model, identifying the regions leading to the correct dark matter abundance via freeze-out. 
Afterwards, we have confronted this parameter space with LHC limits, by employing our previous analysis based on virtual effects of the top-philic ALP, as well as by re-interpreting existing LHC searches for MET signatures.
Even though our DM phenomenology analysis should be considered  preliminary, we can conclude that the region of parameter space leading to the correct relic abundance via freeze-out is already significantly constrained by the LHC.  We leave for future work possible studies of the freeze-in regime (see {\it e.g.}~Refs.~\cite{Bharucha:2022lty,Dror:2023fyd}).

In summary, our study shows that top-philic ALPs are quite elusive in the $[10,200]$ GeV mass range and that precision measurements at the LHC using top-quark final states can help significantly in improving the limits.  We find that the visible and invisible scenarios can be constrained at the same level and that various channels can be used with similar sensitivities. An interesting direction that we leave for future work is to extend the mass window to fully cover the range up to the $2 m_t$ threshold,  where the ALP decays into (weak) gauge bosons become relevant and could  provide a powerful handle to constrain the model further.

\section*{Acknowledgements}

We thank Ilaria Brivio and Jesus Bonilla for many insightful discussions. DP and ST thank the Galileo Galilei Institute for Theoretical Physics for the hospitality during the completion of this work. FM is grateful to  Alessandro Vicini for help with {\sc \small POWHEG}. SB thanks Marco Gorghetto, Hyungjin Kim and Andrea Mitridate for useful discussions. SB and AM are supported in part by the Strategic Research Program High-Energy Physics of the Research Council of the Vrije Universiteit Brussel and by the iBOF “Unlocking the Dark Universe with Gravitational Wave Observations: from Quantum Optics to Quantum Gravity” of the Vlaamse Interuniversitaire Raad. SB, FM and AM are supported in part by the “Excellence of Science - EOS” - be.h project n.30820817. SB is in part supported by FWO-Vlaanderen through grant numbers 12B2323N. This work is supported by the Deutsche Forschungsgemeinschaft under Germany’s Excellence Strategy - EXC 2121 Quantum Universe - 390833306. KM is supported by an Ernest Rutherford Fellowship from the Science and Technologies Facilities Council (STFC), grant No. ST/X004155/1, and was previously supported by STFC grant No. ST/T000759/1.
ST is supported by a FRIA (Fonds pour la forma-
tion `a la Recherche dans l’Industrie et dans l’Agriculture)
Grant of the Belgian Fund for Research, F.R.S.-FNRS
(Fonds de la Recherche Scientifique-FNRS)
\begin{appendices}
\section{UV completion}
\label{app:UVmodel}

In this appendix we present a simple UV model that can reproduce, at low energies, the top-philic ALP Lagrangian discussed in the main body of the paper. In addition, we derive the expected size of the coefficient of the dimension-six operator $\mathcal{O}_{a H}^{(6)}$ considering a renormalisable portal interaction between the Higgs and the Peccei--Quinn-like scalar.

\subsection{A toy UV model for the top-philic ALP}

We present a simple UV model leading to the EFT where the ALP couples only to the top quark at the leading order.
The SM is extended by including two vector-like (in terms of SM symmetries) fermions $T$ and $\Psi$, whose mass is much above the EW scale. These fermions share the same quantum numbers as the right-handed top quark (top partners). Differently from the top quark they are however chirally charged under a new $U(1)$ symmetry, which is spontaneously broken by the vacuum expectation value of a complex scalar field $\Phi$. The corresponding (pseudo) Nambu-Goldstone boson is the ALP  $a$.

The charges under this new $U(1)$ symmetry are 
\be
Q(\Phi) = 1, \quad Q(T_R) = -1, \quad Q(\Psi_R) = 1\,, 
\ee
and all other charges are zero. This implies that the $U(1)$ symmetry is anomaly free as far as the SM gauge group is concerned.
Compatibly with this charge assignment, we can write the following Yukawa interactions and mixings:
\be 
\label{eq:yukmix}
\mathcal{L}_{\rm UV} = y \Phi \bar{T}_L T_R + \delta \bar{T}_L t_R + y^\prime \Phi^\ast \bar{\Psi}_L \Psi_R + \delta^\prime \bar{\Psi}_L t_R + \,{\rm h.c.}\, ,
\ee
where $\delta$ and $\delta'$ have a dimension of a mass. In the following, we set $\delta^\prime = 0$ for simplicity as it does not affect our conclusions.  Thus, the fermion $\Psi$ is only responsible for anomaly cancellation.
The model is designed such that the ALP couples to the SM via the mixing between the $t_R$ and its partner $T_R$.

In the $U(1)$ breaking vacuum we have: 
\be
\label{eq:Phivev}
\Phi = \frac{1}{\sqrt{2}}(f_a + \rho) e^{i a/f_a}\,,
\ee
and we can therefore make the following chiral rotations on the top partners,
\be
\label{eq:chiralrot}
T_R \TO e^{-i a/f_a} T_R\,, \quad \Psi_R \TO e^{i a/f_a} \Psi_R\,.
\ee
This makes the ALP interactions manifestly shift symmetric,
\be
\mathcal{L}_{\rm UV} = -\frac{1}{f_a} \partial_\mu a \, \bar{T}_R \gamma^\mu T_R + \frac{1}{f_a} \partial_\mu a \, \bar{\Psi}_R \gamma^\mu \Psi_R -
m_T \bar T T - m_\Psi \bar \Psi \Psi + \left( \delta \bar T_L t_R + \,{\rm h.c.}\right),
\ee
where we have neglected the radial mode $\phi$.
Notice that no anomalous terms are generated by the rotation Eq.~\eqref{eq:chiralrot}.
In addition, we can read the mass terms for $T$ and $\Psi$ as $m_T = y f_a/\sqrt{2}$ and $m_\Psi = y^\prime f_a/\sqrt{2}$, respectively.

The mixing term in Eq.~\eqref{eq:yukmix} implies that only one combination, $c_\theta t_R + s_\theta T_R$, remains massless above EW symmetry breaking, where the angle $\theta$ depends on $\delta$ and $m_T$. As mentioned before, this mixing is the source of the ALP couplings to the SM.
In fact, integrating out $T$ at  tree level one finds (neglecting gauge interactions):\footnote{The same result can be obtained by simply diagonalising the fermion mass matrix with a rotation of the right-handed components only.}
\

\be
\frac{\delta \mathcal{L}_{\rm UV}}{\delta \bar T_L} = \frac{\delta \mathcal{L}_{\rm UV}}{\delta \bar T_R} = 0 \quad \TO \quad 
T_R = \frac{\delta}{m_T} t_R\,, \quad T_L = - \frac{\delta}{m_T^2 f_a} \partial_\mu a \gamma^\mu t_R\,.
\ee
Substituting these relations back in the Lagrangian, we obtain the top-philic interaction in Eq.~\eqref{eq:top-philic}:
\be
\mathcal{L}_{a, {\rm int}} = -\frac{\delta^2}{m_T^2} \frac{1}{f_a} (\partial_\mu a)
\,\bar{t}_R \gamma^\mu t_R\,,
\ee
where we recognize $c_t = -\delta^2/m_T^2$.

\subsection{On the dimension-six operator $\mathcal{O}_{a H}^{(6)}$}
\label{app:UV-c6}
Let us now turn to discuss the generation of the dimension-six operator involving the Higgs doublet $\phi$ and two ALPs, namely
\be
\mathcal{O}_{a H}^{(6)} = c_{a H}^{(6)} \frac{1}{f_a^2} \phi^\dagger \phi \, (\partial_\mu a \partial^\mu a)\,. 
\ee 
As we shall see below, this operator can be generated at  tree level whenever a portal exists between the complex scalar field containing the ALP degree of freedom and the SM Higgs. To show this, we refer to a simple Lagrangian given by 
\begin{equation}
    \mathcal{L} \supset |\partial_\mu \Phi|^2 - V(|\Phi|^2) + \kappa |\Phi|^2 \phi^\dagger \phi\,,
\end{equation}
where $\kappa$ is the portal interaction. The potential for $\Phi$ shall be taken to be 
\begin{equation}
 V(|\Phi|^2) = - \mu^2 |\Phi|^2 + \lambda |\Phi|^4\,.
\end{equation}
Below the symmetry breaking of the $U(1)$, the complex field $\Phi$ gets a vev, as in Eq.~\eqref{eq:Phivev}. In terms of the radial mode and the ALP mode, one has (see {\it e.g.}~Ref.~\cite{Weinberg:2013kea})
\begin{equation}
    \mathcal{L} \supset \frac{1}{2} (\partial_\mu \rho)^2 + \frac{1}{2} (\partial_\mu a)^2 + \frac{\rho}{f_a} (\partial_\mu a)^2 - \frac{1}{2} m_\rho^2 \rho^2 + \frac{\kappa}{2} (f_a^2 + 2 \rho f_a + \rho^2) \phi^\dagger \phi + \mathcal{O}(\rho^3)\,,
\end{equation}
where $m_\rho^2 = \lambda f_a^2$ and we neglect the self interactions of the radial mode.
We can now integrate out $\rho$ at tree level to derive the value of the coefficient $c_{a H}^{(6)}$ for this simple model. We find:
\begin{equation}
    c_{a H}^{(6)} = \frac{\kappa}{\lambda}\,.
\end{equation}
Dimension-eight operators are generated by the same tree-level exchange such as $(\partial_\mu a \, \partial^\mu a)^2$, as well as additional contributions to the Higgs potential.

We conclude that the dimension-six operator contributing  to the $h \TO a a$ decay can be generated at  tree level provided that a portal exists between the complex scalar and the Higgs.

\section{ALP decays}\label{app:decaywidths}

In this section we provide the analytical expressions for the ALP decay channels that enter in the evaluation of its branching ratios.

The decay into gluons can be evaluated starting from the effective interaction $c_{GG}^{\rm eff}$in Eq.~\eqref{eq:Cgg2loop}. One has \cite{Bauer:2020jbp}:
\be
\Gamma(a \TO gg) = \frac{\alpha_{S}^2 m_a^3}{8\pi^3 f^2}|c_{GG}^{\rm eff}|^2\,.
\ee
Similarly, the decay width into photons is controlled by $c_{\gamma \gamma}^{\rm eff}$ in Eq.~\eqref{eq:Cgamma2loop} and reads:
\be
\Gamma(a \TO \gamma \gamma) = \frac{\alpha^2 m_a^3}{64\pi^3 f_a^2} |c_{\gamma \gamma}^{\rm eff}|^2\,.
\ee
The decay into fermions is controlled by the effective couplings $c_f$ in Eq.~\eqref{eq:cthird} and Eq.~\eqref{eq:clight}:
\be
\Gamma(a \TO f \bar f) = \frac{m_a m_f^2}{8 \pi f_a^2} |c_f|^2 N_c^f \sqrt{1 - \frac{4 m_f^2}{m_a^2}}\,.
\ee
The decay into EW gauge bosons can be extacted from the couplings in Eq.~\eqref{eq:ewcouplings}. 
One has\,\cite{Bonilla:2021ufe}:
\begin{eqnarray}
&&\Gamma(a \TO Z \gamma) = \frac{m_a^3 |g_{a \gamma Z}^{\rm eff}|^2}{128 \pi}
\left( 1 - \frac{M_Z^2}{m_a^2}\right)^3\,, \\
&&\Gamma(a \TO Z Z) = \frac{m_a^3|g_{a Z Z}^{\rm eff}|^2}{64 \pi}
\left( 1 - \frac{4 M_Z^2}{m_a^2}\right)^{3/2}\,,\\
&&\Gamma(a \TO W W) = \frac{m_a^3|g_{a W W}^{\rm eff}|^2}{64 \pi}
\left( 1 - \frac{4 M_W^2}{m_a^2}\right)^{3/2}\,.
\end{eqnarray}

\section{Madgraph implementation and UFO model details}\label{app:MGModel}

In this Appendix we provide details of the {\UFO} model implementation that we have  used to perform the calculations in Sec.~\ref{sec:xsALP} and Sec.~\ref{sec:newprobes}. This model, which we call {\tt ALPTopNLO\_3flav}, has been produced via the help of {\sc \small NLOCT} \cite{Degrande_2015} and {\sc \small FeynRules} \cite{Alloul:2013bka}. It takes into account the SM interactions and the interactions of the top-philic ALP with both quarks and gluons, also for loop diagrams such as those shown in Fig.~\ref{fig:qqag} or in Appendix \ref{app:one_jet}. Three quark flavours ($c,b$ and $t$) are treated as massive, as discussed in the paper.

The primary motivation for the creation of the {\tt ALPTopNLO\_3flav} {\UFO} model was the possibility of automating the calculation of the ALP--gluon--gluon  ($agg$) and ALP--gluon--gluon--gluon  ($aggg$) interaction, which in the case of a top-philic ALP are induced only by  quark loops. 
For the sake of simplicity, we implement the non-derivative basis, where only dimension-four operators are involved in the generation of the $agg(g)$ vertex at one loop making the use of {\sc \small NLOCT} straightforward. 

ALP EW interactions have not been taken into account in this model, meaning for instance, that it cannot be used to calculate the ALP decay into EW gauge bosons. On the other hand, for tree-level interactions, it is sufficient to use {\sc \small FeynRules}  for the creation of a dedicated {\UFO} model and both the derivative and non-derivative bases can be employed.  

\begin{figure}[t!]
    \centering
    \includegraphics[width=0.9\textwidth]{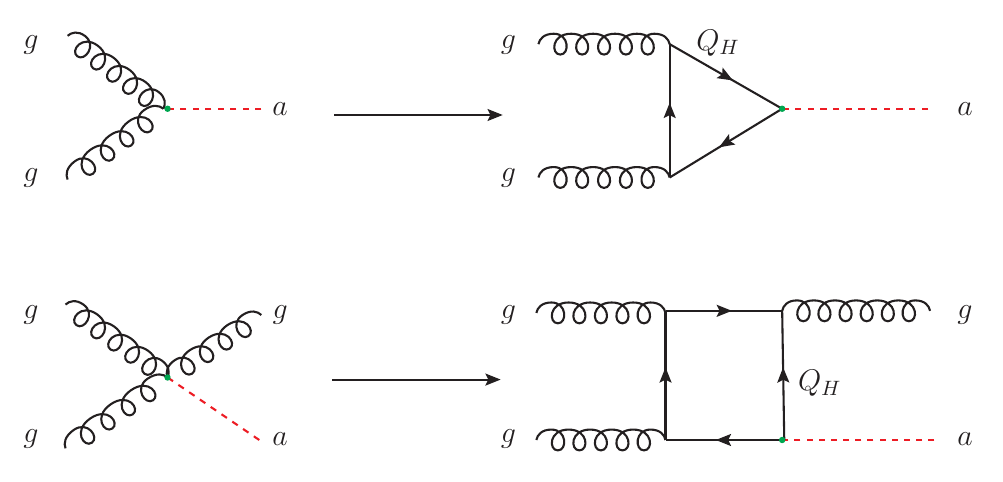}
    \caption{Pictorial representation of the prescription to translate dimension-five ALP--gluon contact term into an equivalent loop with an heavy-quark running in it, that couples to the ALP with a dimension-four operator. }
    \label{fig:prescription}
\end{figure}

The implementation of $agg(g)$ loop-induced interactions is more involved. As discussed in detail in Secs.~\ref{sec:nonder} and \ref{sec:xsALP}, the $agg(g)$ loop in the derivative basis is equivalent to the same loop in the non-derivative basis plus a tree-level contribution (see {\it e.g.}~Fig.~\ref{fig:qqag} and related discussion) both at the same perturbative orders in $\alpha_{S}$ and $c_{t}/f_{a}$. However, at the moment, setting up a calculation in {\aNLO} with amplitudes receiving contributions from both tree-level and one-loop diagrams at the same perturbative order is not straightforward. 
In order to bypass this limitation, we introduce an additional, auxiliary heavy fermion $Q_{H}$ with mass $m_{Q_{H}}$ whose role is to mimick the $agg(g)$ tree-level interaction parameterised by the $\tilde c_{GG}$ parameter (see Eqs.~\eqref{eq:Laintnonder} and \eqref{eq:axionmodel}). This allows us to reproduce the $agg(g)$ vertex stemming from the $aG\tilde{G}$ operator in the limit $m_{Q_{H}}\TO\infty$. It has to be noted that the introduction of $Q_{H}$ is just a technical trick and it bears no relation to any of the fermions in the UV model discussed in Appendix~\ref{app:UVmodel}.  In practice, we perform the following substitution for  Eq.~\eqref{eq:Laintnonder},
\begin{equation}\label{eq:prescription}
 \tilde c_{GG} \frac{\alpha_{S}}{4\pi}\frac{a}{f_a}G\tilde{G}\Longrightarrow i c_{Q_{H}} \frac{m_{Q_{H}}}{f_a} a\bar Q_{H}\gamma_5 Q_{H} \, ,
\end{equation}
with
\begin{equation}\label{eq:prescription2}
c_{Q_{H}}=\frac{1}{2}\sum_q c_q=\frac{c_{t}-c_{b}}{2}= \frac{c_t(\Lambda)}{2}\left(1-4\frac{y_t^2}{16 \pi^2} \log \frac{\Lambda}{m_t},\right).
\end{equation}
This substitution is depicted in Fig~\ref{fig:prescription}.

We have checked via the analytic formulae of Sec.~\ref{app:one_jet} that, in the limit $m_{Q_{H}}\TO \infty $, we recover exactly the $agg(g)$ vertex induced by the $aG\tilde G$ operator. In the numerical simulations, where a value for $m_{Q_{H}}$ has to be chosen, we set $m_{Q_{H}}=100\tev$. We have verified that this value is suitable for our purposes and does not induce any appreciable corrections to the limit of a contact interaction.

\medskip

The model also leaves open the possibility of setting the kinematic mass of the charm and the bottom equal to zero without altering their coupling with the ALP. Rather than being useful for the evaluation of the loop-induced $agg(g)$ vertex, this feature is intended for computing tree-level processes with the charm and the bottom as external states. To this end, as also done in some cases for models with different flavour schemes, we introduce a kinematic mass $m_{q}$ and an interaction parameter $\tilde m_{q}$ for the bottom, charm, top, and the heavy auxiliary quark, $Q_{H}$. In conclusion, the part of the Lagrangian describing the interactions of the ALP is    
\begin{equation}\label{eq:model}
\mathcal{L}_{\rm int}=-i\sum_{q=c,b,t} c_q a \frac{\tilde m_q}{f_a}\bar q \gamma_5 q + i c_{Q_{H}} a \frac{\tilde m_{Q_{H}}}{f_a}\bar Q_{H} \gamma_5 Q_{H}\,,
\end{equation}
with $c_q$ as in Eqs.~\eqref{eq:cthird}--\eqref{eq:clight} and $c_{Q_{H}}$ as in Eq.~\eqref{eq:prescription}. 

\medskip

We conclude this section by listing the free parameters  present  in the  {\tt ALPTopNLO\_3flav} {\UFO} model in addition to the SM ones that can modified in the associated {\tt param\_card.dat}. All values, apart from $c_{t}$ which is dimensionless, are given in GeV.

\begin{itemize}
\item {\tt fa}: the value of $f_a$. Default value: $10^3$.
\item {\tt Lambda}: the value of $\Lambda$, the UV scale. Default value: $10^3$.
\item {\tt Muscale}: the value of the scale $\mu_{\rm IR}$ setting the lower bound of the RG running from $\Lambda$. Default value: $m_t$, as in all equations of the paper.
\item {\tt ct } the tree-level coupling of the  top quark with the ALP. Default value: 1.
\item  {\tt MHq}: the value of the kinematical mass $m_{Q_{H}}$. Default value: $10^5$.
\item  {\tt AxMHq}: the value of the mass $\tilde m_{Q_{H}}$ entering Eq.~\ref{eq:model}. Default value:  $10^5$.
\item  {\tt AxMt}: the value of the mass $\tilde m_{t}$ entering Eq.~\ref{eq:model}. Default value: $m_{t}$.
\item  {\tt AxMb}: the value of the mass $\tilde m_{b}$ entering Eq.~\ref{eq:model}. Default value: $m_{b}$. 
\item  {\tt AxMc}: the value of the mass $\tilde m_{c}$ entering Eq.~\ref{eq:model}. Default value: $m_{c}$.  
\end{itemize}

Since we distinguish kinematical mass ($m_{q}$) and the interaction masses entering Eq.~\ref{eq:model}, the following features are available:
 \begin{itemize}
     \item For tree-level diagrams, it is possible to turn off the interaction of a given quark $q$ without having it massless: $m_{q}\ne 0$ and $\tilde m_{q}= 0$. Note that if the $agg(g)$ vertex is calculated, the interaction of the given quark $f$ will still be present in the $\tilde c_{GG}$ definition, since Eq.\eqref{eq:prescription2} assumes that up-type and down-type contributions cancel each other.  
     \item It is possible to set a quark $q$ as massless without turning off its interaction with the ALP: $m_{q}= 0$ and $\tilde m_{q}\ne 0$. Obviously, this works only if the scale involved in the process are larger than the real value of $m_{q}$.
     \item It is possible to completely turn off the $agg(g)$ tree-level interaction ($\tilde c_{GG}$) obtaining  a model equivalent to $a$ being a pseudoscalar and not an ALP: $\tilde m_{Q_{H}}=0$.
     \item It is possible to turn off RG running effects and restore the top-philic ALP model at  tree level: $\Lambda=\mu_{\rm IR}$
 \end{itemize}

\section{Analytical formulae and verification of the Madgraph implementation}

\label{app:one_jet}
In this Appendix we report the analytical formulae for (some of) the $2\TO 2$ amplitudes entering the $pp\TO a + j$ calculation in Sec.~\ref{sec:xsALP} and the $pp\TO t \bar t$ calculation in Sec.~\ref{subsec:ttbar}. These amplitudes have been used for testing the {\tt ALPTopNLO\_3flav} {\UFO} model presented in Appendix \ref{app:MGModel}. We will also  emphasise the difference between an ALP and a pseudoscalar, meaning $\tilde c_{GG}=0$ in the non-derivative basis. 

In this Appendix we assume that, on top of the ALP interaction with fermions, only QCD is present. As we are not interested in the correlations among the $c_{f}$ we treat them all as independent. For the case of the pseudoscalar, denoted here as $A$, the interacting Lagrangian is  
\be
\mathcal{L}_{A, \, \rm int.}=\sum_q - i c_q \frac{m_q}{f_a} A\bar\psi_q \gamma_5 \psi_q\,\,, 
\ee
while in the case of an ALP $a$ we have
\begin{eqnarray}
\label{eq:ALPforanal}
\mathcal{L}_{\rm a, \, \rm int. }&=&\sum_q \frac{c_q}{2} \frac{(\partial_\mu a)}{f_a}\bar\psi_q\gamma_\mu\gamma_5\psi
\nonumber\\
&=&\sum_q - i c_q \frac{m_q}{f_a} a\bar\psi_q \gamma_5 \psi_q+\sum_q c_q\frac{\alpha_{S}}{8\pi}\frac{a}{f_a}G\tilde{G} +\mathcal{O}(1/f_{a}^{2})\,,
\end{eqnarray}
where the second term in the r.h.s.~of Eq.~\eqref{eq:ALPforanal} is what in the main text we referred as the contribution proportional to $\tilde c_{GG}$. 

In the following, we will always assume the momenta as incoming ($p_{1}$ and $p_{2}$ for the initial state and $(-p_{3})$ and $(-p_{4})$ for the final state) and therefore the  Mandelstam variables are defined as
\begin{align}
\hat{s}&=(p_1+p_2)\,,\\
\hat{t}&=(p_1+p_3)\,,\\
\hat{u}&=(p_1+p_4)\,.
\end{align}

The $C_0$ loop scalar integral that will appear in the following formulae is defined as  
\begin{align}
\label{eq:C1}
C_0(p_i,p_j; m_q^{2}) \equiv \frac{1}{i \pi^2} \int
         \frac{d^4 \! q}
              {[q^2-m_q^2][(q+p_i)^2-m_q^2][(q+p_{ij})^2-m_q^2]}\,.
\end{align}
where $p_{ij}^2=(p_i+p_j)^2$. It is therefore the usual $C_{0}$ function for the three internal masses all set equal to $m_q$ and incoming momenta $p_{i}$, $p_{j}$ and $(-p_{i}-p_{j})$. For the comparisons with numerical results, we have used the {\sc \small LoopTools} package \cite{Hahn:1998yk}.

\subsection{$a/A$ + jet final states}
\subsubsection*{The $gq \TO aq$   and $gq \TO Aq  $ process}\label{app:gqaq}
\begin{figure}[t!]
    \centering
    \includegraphics[width=0.9\textwidth]{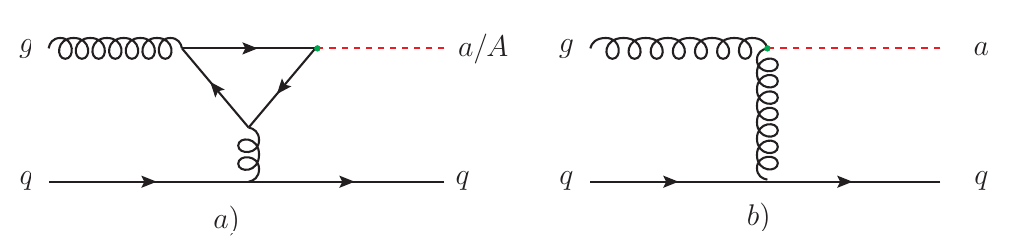}
    \caption{Representative diagrams for the process $gq\TO aq$ (or $gq\TO Aq$).  $\mathbf a)$ Diagrams common to the pseudoscalar and ALP cases. $\mathbf b)$ Diagrams present only in the ALP case. }
    \label{fig:qgqa}
\end{figure}
Starting with the case of a pseudoscalar,  we consider the $g(p_1)q(p_2) \TO A(-p_3)q(-p_4)$ process .  The diagrams associated to this process are those involving a loop of fermions, such as the left one of Fig.~\ref{fig:qgqa}. The corresponding
squared matrix element, after summing(averaging) over the final-state(initial-state) spin and  colour states, is 
\begin{equation}
\label{eq:pseudoanal1}
\overline{\sum}|\mathcal{M}_{g q\TO A q}|^2
 = - \frac{ \alpha_{S}^3 }{2 N_c \pi f_a^2}\;
\left\lvert\sum_q c_q m_q ^2 C_0(p_1,-p_4;m_q^2)\right\rvert^2 \; \frac{ \hat{s}^2+\hat{t}^2}{\hat{u}}.
\end{equation}
This result is consistent with the formulae in Ref.~\cite{Field:2003yy}.

The corresponding case where instead of a pseudoscalar $A$ we consider an ALP $a$ involves also the diagram on the right of Fig.~\ref{fig:qgqa}.  The formula for the squared matrix element reads
\begin{equation}
\label{eq:pseudoALP1}
\overline{\sum}|\mathcal{M}_{gq \TO a q}|^2
 = -\frac{\alpha_{S}^3}{2 N_c \pi f_a^2}\;
\left\lvert \sum_q c_q \left[\frac{1}{2}+m^2_q C_0(p_1,-p_4;m_q^2)\right]\right\rvert^2 \; \frac{ \hat{s}^2+\hat{t}^2}{\hat{u}}\,.
\end{equation}

Since for $ - \hat t  \gg m_{q}^{2}$ 
\be
\label{eq:DLNP}
C_0(p_{1},-p_{4}; m_q^2) =  \frac{1}{2\hat t} \,\,{\log}^2 
\left[ 
\frac{1 + (1 - 4 m_q^2/\hat t\,)^{-1/2}}{1 - (1 - 4 m_q^2/\hat t\,)^{-1/2} }
\right]
\simeq
 \frac{1}{2\hat t}\log^{2}\left(\frac{-\hat t}{m_{q}^{2}}\right)\,,
\ee
from the comparison of Eqs.~\eqref{eq:pseudoanal1} and \eqref{eq:pseudoALP1} it is clear that while at large transverse momentum the amplitude for a pseudoscalar is strongly suppressed, in the case of an ALP that is not true, in line with what is observed in Fig.~\ref{fig:ptaj}.

It is also interesting to note that if one considers an ALP that is not only top-philic but that already has a $c_{GG}$ interaction term at tree-level, Eq.~\eqref{eq:DLNP} gives additional information. The diagram on the right of Fig.~\ref{fig:qgqa} would lead to the LO prediction and while the diagram on the left of Fig.~\ref{fig:qgqa} can be considered a loop correction, which depends on $c_{t}$ and exhibits at large $\log\hat t$, an IR effect that cannot be captured via RG running.

\begin{figure}[t!]
    \centering
    \includegraphics[width=0.9\textwidth]{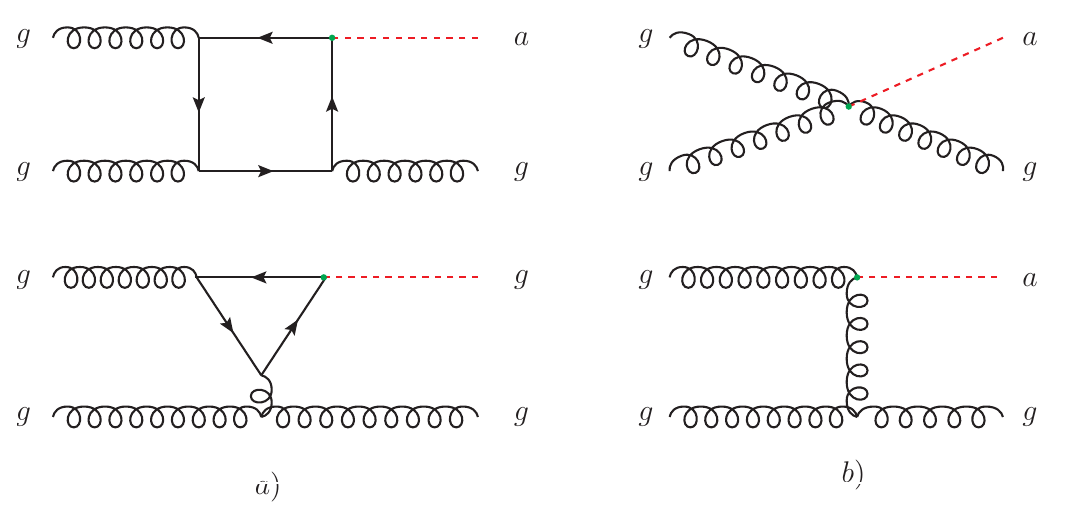}
    \caption{Representative diagrams for the process $gg\TO ag$ (or $gg\TO Ag$).  $a)$ Diagrams common to the pseudoscalar and ALP cases. $b)$ Diagrams present only in the ALP case.}
    \label{fig:ggag}
\end{figure}

\subsubsection*{The $q\bar{q }\TO ag $ and $q\bar{q }\TO Ag $   processes}\label{app:qqag}

The second case that we consider consists of the partonic processes $q(p_1)\bar{q}(p_2)  \TO A(-p_3)g(-p_4)$, for the pseudoscalar, and $q(p_1)\bar{q}(p_2) \TO a(-p_3)g(-p_4)$, for the ALP.
The relevant diagrams for this partonic process are already shown in the Fig.~\ref{fig:qqag} in the main text and we do not repeat them here.

In the case of a pseudoscalar we obtain
\begin{equation}
\overline{\sum}|\mathcal{M}_{q {\overline q}\TO Ag }|^2
 = \frac{(N_c^2-1) \alpha_{S}^3 }{2 N_c^2 \pi f_a^2}\;
\left\lvert\sum_q c_q m_q ^2 C_0(p_1,p_2;m_q^2)\right\rvert^2 \; \frac{ \hat{t}^2+\hat{u}^2}{\hat{s}}\,,
\end{equation}
which is consistent with the formulae in Ref.~\cite{Field:2003yy}, while in the case of an ALP we obtain
\begin{equation}
\overline{\sum}|\mathcal{M}_{q {\overline q}\TO ag }|^2
 = \frac{(N_c^2-1) \alpha_{S}^3}{2 N_c^2 \pi f_a^2}\;
\left\lvert \sum_q c_q\left[\frac{1}{2}+ m^2_q C_0(p_1,p_2;m_q^2)\right]\right\rvert^2 \; \frac{ \hat{t}^2+\hat{u}^2}{\hat{s}}\,.
\end{equation}

The same considerations given for the case of $gq \TO aq$   and $gq \TO Aq  $ also apply here, with the substitution of $- \hat t$ with $\hat s$.

\subsubsection*{The $gg \TO ag$ and $gg \TO Ag$   process}\label{app:ggag}
The analytical formula for the $gg\TO Ag$ amplitude is quite involved, due to the presence of the box and triangle diagrams, such as the one on the left of Fig.~\ref{fig:ggag}, which we do not reproduce here. The numerical results obtained via our model  for the pseudoscalar case have been tested  (point by point in phase space) against the  POWHEG \cite{Nason:2004rx,Frixione:2007vw,Alioli:2010xd} implementation of Ref.~\cite{Bagnaschi:2011tu}, where the real radiation component of the NLO QCD corrections precisely contains the analogous process for a pseudoscalar Higgs.

The contribution to the total amplitude coming from the contact term squared (diagrams $b)$ in Fig. \ref{fig:ggag}) is
\begin{equation}
\overline{\sum}|\mathcal{M}^{\rm contact}_{gg\TO ag }|^2= \frac{1}{\pi}\frac{\alpha_s^3}{f_a^2} \frac{N_c}{N_c^2-1}\left \lvert\sum_q \frac{c_q}{2} \right\rvert^2\frac{m_a^8+s^4+t^4+u^4}{stu}\, .
\end{equation}
This result was checked point-by-point in phase space directly with our numerical implementation.

\subsection{The $gg \TO a \TO t\bar t$ and $gg \TO A \TO t\bar t$ processes}
\begin{figure}[!t]
    \centering
    \includegraphics[width=0.9\textwidth]{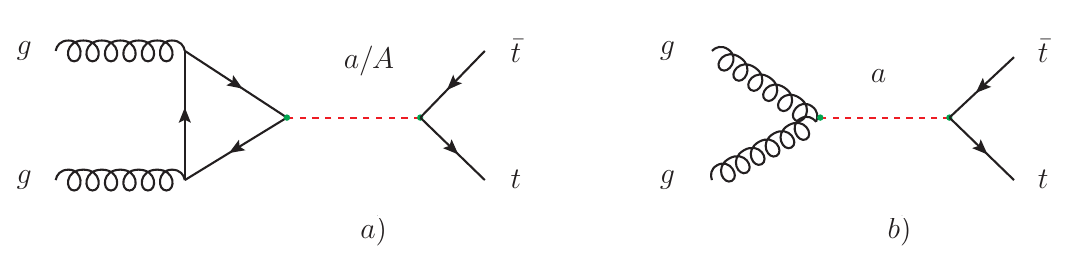}
    \caption{Diagrams for the process $gg\TO a\TO t\bar t$ (or $gg\TO A\TO t\bar t$).  $a)$ Diagrams common to the pseudoscalar and ALP cases. $b)$ Diagrams present only in the ALP case.}
    \label{fig:ggatt}
\end{figure}

The partonic process $gg\TO t \bar t$ receives contributions both from the SM and, staring at one-loop, from the pseudoscalar $A$ or the ALP $a$, as discussed for the latter in Sec.~\ref{subsec:ttbar}. 
Here we consider only the $s$-channel mediated diagram of Fig.~\ref{fig:ggatt}. 

Let us start with the pseudoscalar $A$. 
The NP contribution includes 
not only the one-loop squared amplitude $\left\lvert\mathcal{M}_{gg\TO A \TO t\bar t}\right\rvert^2$ but also its interference with the SM contribution from QCD ($\mathcal{M}_{gg\TO t \bar t}$). The partonic cross section for the process $gg \TO t\bar t$ reads \cite{Dicus:1994bm}:
\begin{eqnarray}
\frac{d\sigma_{\rm NP}}{d\cos\theta} & = & \frac{1}{32\pi}\frac{1}{s}\beta\,\overline{\sum}\left[\left\lvert\mathcal{M}_{gg\TO A \TO t\bar t}\right\rvert^2 + 2\Re\left(\mathcal{M}_{gg\TO t \bar t} \mathcal{M}^{*}_{gg\TO A \TO t\bar t}\right) \right]\label{eq:ggaxttpseudo}\,,
\end{eqnarray}
where
\begin{equation}
\beta=\sqrt{1-\frac{4m_t^2}{s}}\,,
\end{equation}
and
\begin{align}
\overline{\sum} \left\lvert\mathcal{M}_{gg\TO A \TO t\bar t}\right\rvert^2=\frac{3\alpha_{S}^2}{32\pi^2}s^3 c_t^2 \frac{m_t^2}{f_a^2} \left\lvert\sum_q \frac{c_q }{f_a}~ \frac{m_q^{2} C_0(p_1,p_2;m_q^2)}{s-m_A^2+im_A\Gamma_A(s)} \right\rvert^2\,,
\label{eq:squaredA}
\end{align}
\be
\overline{\sum} \, 2\Re\left(\mathcal{M}_{gg\TO t \bar t}\, \mathcal{M}^{*}_{gg\TO A \TO t\bar t}\right)= \frac{c_t m_t}{f_a}\left(\frac{\alpha_{S}^2  m_t s}{1-\beta^2\cos^2\theta}\right)
{\Re}\left[\sum_q \frac{c_q }{f_a}~ \frac{m_q^{2} C_0(p_1,p_2;m_q^2)}{s-m_A^2+im_A\Gamma_A(s)}\right].
\label{eq:intA}
\ee

In the case of an ALP, the analogue of Eq.~\eqref{eq:ggaxttpseudo} is simply
\begin{eqnarray}
 \frac{d\sigma_{\rm NP}}{d\cos\theta} & = & \frac{1}{32\pi}\frac{1}{s}\beta\,\overline{\sum}\left[\left\lvert\mathcal{M}_{gg\TO a \TO t\bar t}\right\rvert^2 + 2\Re\left(\mathcal{M}_{gg\TO t \bar t} \mathcal{M}^{*}_{gg\TO a \TO t\bar t}\right) \right]
\label{eq:ggaxttALP}\,,
\end{eqnarray}
with
\begin{align}
\overline{\sum} \left\lvert\mathcal{M}_{gg\TO a \TO t\bar t}\right\rvert^2=\frac{3\alpha_{S}^2}{32\pi^2}s^3 c_t^2 \frac{m_t^2}{f_a^2} \left\lvert\sum_q \frac{c_q }{f_a}~   \frac{\frac12+m_q^{2}C_0(p_1,p_2;m_q^2)}{s-m_a^2+im_a\Gamma_a(s)} \right\rvert^2\,,
\label{eq:squareda}
\end{align}
\be
\overline{\sum}\, 2\Re\left(\mathcal{M}_{gg\TO t \bar t}\, \mathcal{M}^{*}_{gg\TO a \TO t\bar t}\right)= \frac{c_t m_t}{f_a}\left(\frac{\alpha_{S}^2  m_t s}{1-\beta^2\cos^2\theta}\right)
{\Re}\left[\sum_q \frac{c_q }{f_a}~ \frac{\frac12 +m_{q}^{2} C_0(p_1,p_2;m_q^2)}{s-m_a^2+im_a\Gamma_a(s)}\right].
\label{eq:inta}
\ee
\begin{figure}[!t]
    \centering
     \includegraphics[width=0.48\textwidth]{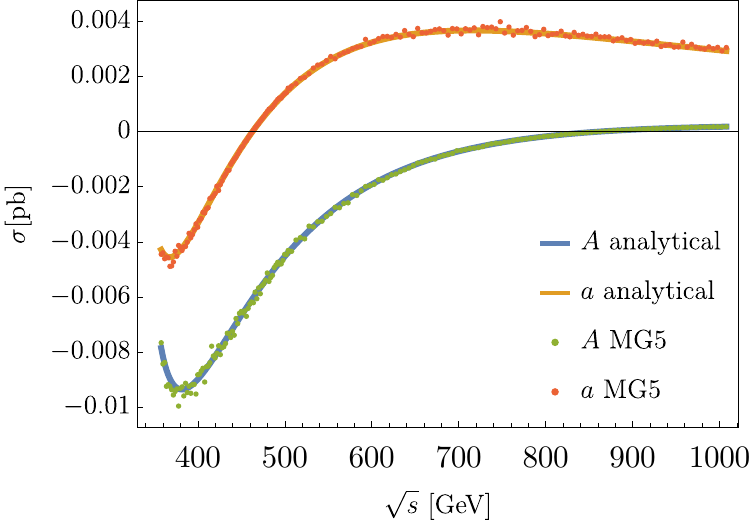}
    \includegraphics[width=0.48\textwidth]{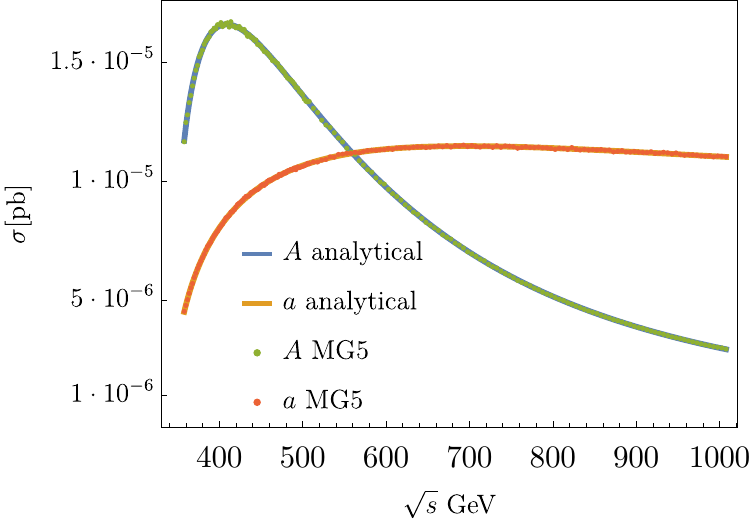}
    \caption{Comparison between analytical calculations and results obtained with  {\aNLO} (MG5 in the plot) for pseudoscalar and ALP contributions to $gg\rightarrow a/A\rightarrow t\bar t$. \textbf{Left}: only the interference term. \textbf{Right}: only the  quadratic term.}
    \label{fig:testggaxtt}
\end{figure}
Comparing the results for a pseudoscalar $A$ with an ALP $a$ we notice the presence of a $\frac12$ in the numerator in the ALP case. In the non-derivative basis, this is due to the contribution from the diagram featuring the $gga$ vertex, see diagram b) in Fig.~\ref{fig:ggatt}. 

We have cross-checked our implementation of the model {\tt ALPTopNLO\_3flav} {\UFO} model presented in Appendix \ref{app:MGModel}, as also done for the $gq \TO aq$ or $gq \TO Aq  $ process and the $qq \TO ag $ or $qq \TO Ag $ process, using the formulas collected in this Appendix. For the  $gg \TO a \TO t\bar t$ and $gg \TO A \TO t\bar t$ process, the comparison is shown in Fig.~\ref{fig:testggaxtt} where   the predictions from the formulae in Eqs.~\ref{eq:ggaxttpseudo}  and Eqs.~\ref{eq:ggaxttALP} are plotted as lines, while the results obtained via {\aNLOs} with the {\tt ALPTopNLO\_3flav} {\UFO} model are shown as dots. 
In the left plot of Fig.~\ref{fig:testggaxtt} we show only the contribution of the interference of the one-loop diagram and the SM QCD one at  tree level,\footnote{In fact, in Sec.~\ref{subsec:ttbar} we have considered only such contributions and not those from squared one-loop diagrams.} Eqs.~\eqref{eq:intA} and \eqref{eq:inta}, while in the right plot we show only the contribution from the former diagram squared, Eqs.~\eqref{eq:squaredA} and \eqref{eq:squareda}. In both plots the cross section is integrated of $\cos\theta$ and the azimuthal angle. As can be seen from the comparison of lines and dots, the results obtained via the {\tt ALPTopNLO\_3flav} {\UFO} model are in agreement with the analytical formulae.  
\medskip

\section{Two-loop effects and relevance}\label{app:dip}

\begin{figure}[!h]
    \centering
\includegraphics[width=0.9\textwidth]{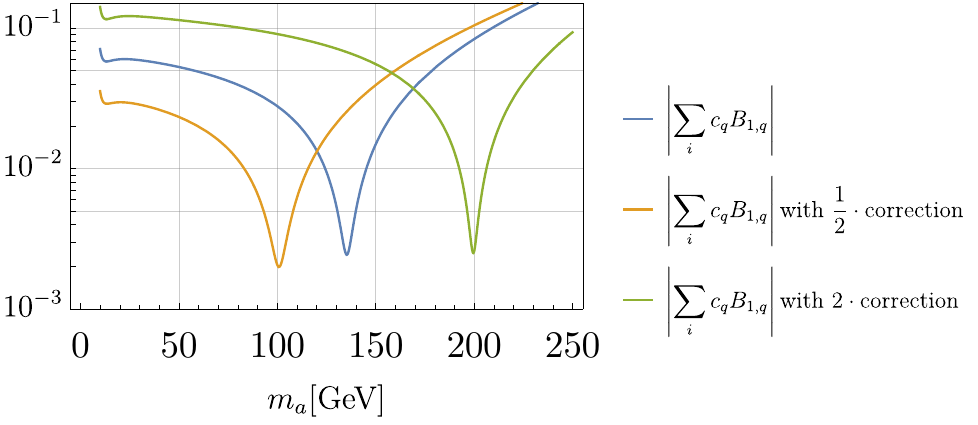}
    \caption{Position of the minimum of the ALP BR into gluons for different values of the prefactor on front of the contribution of the additional quark in the loop. }
    \label{fig:B1functions_scale}
\end{figure}

We now present an estimation of the  uncertainty due to our ignorance of the two-loop effects impacting the ALP branching ratio into gluons Br($a\TO gg$) and $gg\TO a$ production.
In the following we will refer only to $a\TO gg$, but the argument is valid in both cases.
As mentioned in the main body of the paper, complete two-loop calculations for some processes are beyond the scope of our analysis.
Our approach is based on keeping the leading logarithmic terms in the  one-loop induced couplings to other quarks, $c_{f}(m_{t})$, and then employ these effective couplings to estimate the full two loops, using \emph{de facto} a one-loop computation for the $a \to gg$ amplitude.

In order to roughly estimate the 
uncertainty of our approximation, 
we repeat the computation of the branching ratio allowing the two-loop effects to float by a factor of two. 
The impact on the shape of the branching ratio into gluons is given in Fig.~\ref{fig:B1functions_scale}.
The location of the minimum in the branching ratio is due to a precise cancellation between the genuine one-loop and estimated two-loop effects.  
The variations 
significantly shift the position of the minimum. 
For ALP masses much larger than the top mass, the two-loop effect becomes subdominant and hence the branching ratio curves tend to the same value for $m_a \gg m_t$.
For ALP masses much smaller than $m_t$, the two-loops effects are the dominant contributions and the resulting uncertainty leads to $\mathcal {O}(1)$ variations. Nevertheless, we assume  that the overall order of magnitude is correctly captured. 

Finally, we note that the variation by a factor of two  adopted above is equivalent to a variation of the  scale $\Lambda$ of new physics by  a factor of 10, see Eqs.~\eqref{eq:cthird} and \eqref{eq:clight}. This means that also the knowledge of the value of $\Lambda$ has an impact on the 
precise determination of the $a\TO gg$ amplitude with external on-shell particles. See also Eq.~\ref{eq:Cgg2loopintermediate}.

A comment on the validity of our conclusions given the approximations involved is in order.  The processes where the two-loop uncertainty is present are those where the couplings to quarks other than the top (which are themselves one-loop) are included in a loop to obtain the $ gg a $ interaction.  In principle this includes $pp\TO a$,  $pp \TO t\bar t$ (the $s$-channel diagrams)
as well as $pp \TO a+j$.
However, the effectively two-loop contributions to the
$a gg $ vertex are relevant only if the typical momentum of the process is smaller than $2 m_t$, as discussed in Sec. \ref{subsec:gluphot}.
This is not the case in the $pp \TO t\bar t$, as well as in the $pp \TO a+j$ as soon as the $p_{\rm T}$ of the extra jet is sufficiently large (see discussion around
Fig.~\ref{fig:ptaj}.)

Hence the only process that is really affected by the two-loop effects is 
the $pp\TO a$ (since $m_a \ll 2m_t$ in our mass window).
In our study, we have checked that the $pp\TO a$ process does not lead to any relevant bound, not even in the regime $m_a \ll m_t$ or $m_a \gtrsim m_t$, where, as mentioned, our estimate for the two-loops should correctly capture the order of magnitude of the effective coupling. 
We conclude that the two-loop uncertainty estimate does not lead to qualitative modifications of our general conclusions.

\section{Limits from mono-jet searches on the invisible top-philic ALP}\label{app:mono-jet}

\begin{table}[h!]
    \centering
    \begin{tabular}{|c|c|c|}
    \hline
          $m_a~ [\text{GeV}]$ 
         & $f_a/c_t ~[\text{GeV}]$ &  Cut $[\text{GeV}]$ \\ \hline
         10 & 230.6 & $p_{\rm T}>900$
                    \\ \hline
        50 & 230.2&  $p_{\rm T}>900$
         \\ \hline
        100 & 227.7 &  $p_{\rm T}>900$
         \\ \hline
        150 & 225.0&  $p_{\rm T}>900$\\ \hline
        200 & 220.7&  $p_{\rm T}>1200$\\
        \hline
    \end{tabular}
    \caption{Limit on $f_a/ c_t$ coming from mono-jet searches. In the third column, the kinematical region coming from the experimental paper from which the best bound is obtained.}
    \label{tab:mono-jet}
\end{table}

\begin{figure}
\centering
\includegraphics[width=0.8\textwidth]{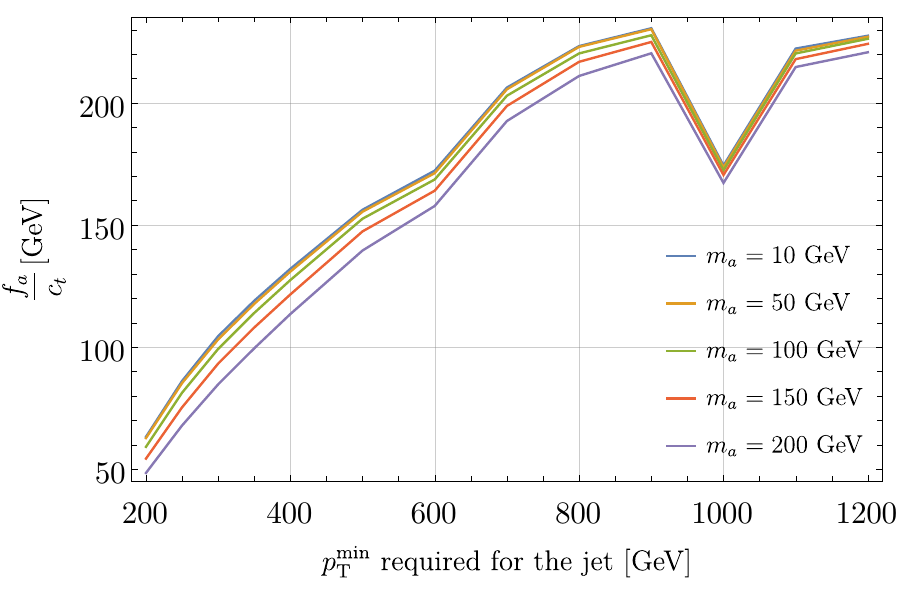}
\captionof{figure}{Bound on $f_a / c_t$ (GeV) for  different $m_a$ (GeV). The strength of the bound depends on the kinematical region considered, modulated on the $x$-axis by the minimum $p_{\rm T}$ required for the jet.}
\label{fig:mono-jetfig}
\end{figure}

In this Appendix we describe how the mono-jet limits in Fig.~\ref{fig:consraints_invisible} were obtained. Using the UFO model {\tt ALPTopNLO\_3flav} discussed in detail in Appendix.~\ref{app:MGModel}, we simulate the production of an ALP in association with one light-jet with the help of  {\aNLO}. The diagrams contributing to this process are illustrated in Figs. \ref{fig:qqag},  \ref{fig:qgqa}, \ref{fig:ggag}. 
When the top-philic ALP decays invisibly  with a 100\% branching ratio (as assumed in Sec.~\ref{sec:DM-CS}), these processes result in a mono-jet signature. 
We can then compare the mono-jet signal of the invisible top-philic ALP with the mono-jet +  $\slashed E_{\rm T}$ search performed by the ATLAS experiment~\cite{ATLAS:2021kxv}. 
The ATLAS collaboration provides model independent 95\% C.L.
upper limits on the BSM mono-jet cross-section, in signal regions defined with increasing minimum transverse momentum ($p_{\rm T}^{\text{min}}$) cuts for the jet.
For a given ALP mass, we compare our cross section prediction as a function of the $p_{\rm T}^{\text{min}}$  with the upper limit table of ATLAS, and then extract a limit on $f_a/c_t$ from every different signal region.
Our results are reported in Fig. \ref{fig:mono-jetfig} for few representative ALP mass cases.

Out of these, we consider for each ALP mass the most stringent bound and report it in Tab.~\ref{tab:mono-jet} and in Fig.~\ref{fig:consraints_invisible}.
As expected from the behaviour of the $pp\TO a+j$  cross-section illustrated in Fig.~\ref{fig:production}, the bound is relatively insensitive to the ALP mass. 
The low $p_{\rm T}^{\text{min}}$ region is not efficient in establishing a meaningful bound as we can see in Fig.~\ref{fig:mono-jetfig}.
This can be understood by looking at how the cross-section of the $pp\TO a+j$ process is distributed {w.r.t.}~the transverse momentum of the jet 
in the top-philic ALP (see Fig. \ref{fig:ptaj}). 
The bound becomes more and more relevant when we increase the $p_{\rm T}^{\text{min}}$ value because the experimental searches become more constraining while the $p_{\rm T}$ differential distribution for $pp\TO a+j$ decreases slowly due to the presence of the gluon contact interaction (note that the dip around $p_{\rm T}=1\tev$ is a fluctuation in the experimental data and is not related to the $pp\TO a+j$ cross-section prediction).
\end{appendices}

\bibliography{}


\end{document}